\documentclass[bibyear]{aa}  

\usepackage{graphicx}
\usepackage{txfonts}
\usepackage{natbib}
\bibpunct{(}{)}{;}{a}{}{,} 
\usepackage{booktabs}
\usepackage{multicol}
\usepackage{graphicx}
\usepackage{fancyhdr}
\usepackage[colorlinks=true,linkcolor=blue,citecolor=blue]{hyperref}
\usepackage{amsmath}	
\usepackage{amssymb}	
\usepackage[inter-unit-product=\cdot]{siunitx}
\usepackage{enumerate}
\usepackage{multirow}
\usepackage{mathtools}
\usepackage{comment}
\usepackage{longtable}
\usepackage{tabularx}
\usepackage{xcolor} 
\usepackage{ulem} 
\usepackage{comment}
\usepackage{cuted}
\usepackage{soul}
\usepackage{longtable}

\newcommand{\fastcluster}[1]{{\sc fastcluster}}
\newcommand{\sevn}[1]{{\sc sevn}}
\newcommand{\pagn}[1]{{\fontfamily{lmtt}\selectfont pAGN}}
\newcommand{\tsunami}[1]{{\sc tsunami}}
\newcommand{\princess}[1]{Princess}
\newcommand{\mcfacts}[1]{{\sc McFacts}}

\newcommand{\orcidicon}[1]{\href{https://orcid.org/#1}{\includegraphics[width=11pt]{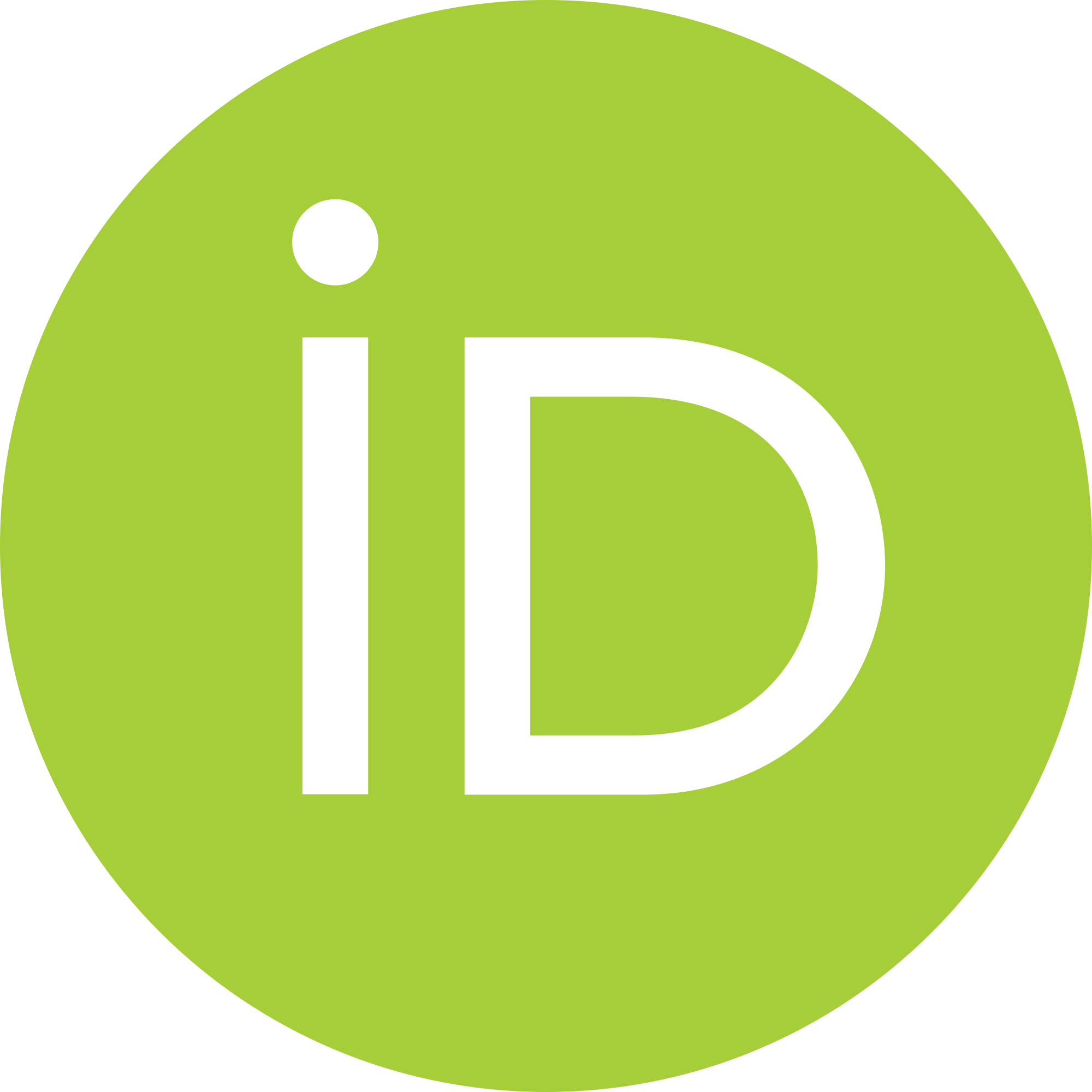}}}
\newcommand{\orcid}[1]{\href{https://orcid.org/#1}{\protect\orcidicon{#1}}}

\DeclareSIUnit\year{yr}
\DeclareSIUnit\au{AU}
\DeclareSIUnit\parsec{pc}
\DeclareSIUnit\erg{erg}

\defcitealias{SG}{SG}
\defcitealias{TQM}{TQM}
\defcitealias{Bellovary_2016}{B16}
\defcitealias{Grishin_2024}{G24}
\defcitealias{Kanagawa_2018}{K18}

\newcommand{\modA}{A} 

\newcommand{\modAfb}{Afb} 

\newcommand{\modAfaOj}{AJ} 

\newcommand{\modAlowZ}{AJlowZ}  

\newcommand{\modBfaOfive}{B} 

\newcommand{\modBfaOj}{BJ} 

\newcommand{\modC}{C} 

\newcommand{\modD}{D} 

\newcommand{\Gaia}{{\it Gaia}}

\begin{document} 

  \title{The role of accretion efficiency, natal kicks, and angular momentum transport in the formation of the \Gaia{} black holes}

   \author{Michela Mapelli
   \thanks{e-mail: \href{mailto:mapelli@uni-heidelberg.de}{mapelli@uni-heidelberg.de}}\inst{1,2,3,4}\orcid{0000-0001-8799-2548},
    Cecilia Sgalletta\inst{1,2}\orcid{0009-0003-7951-4820},
    Johanna M\"uller-Horn\inst{5,6}\orcid{0000-0001-9590-3170},
     Giuliano Iorio\inst{7}\orcid{0000-0003-0293-503X},\\
    Stefano Rinaldi\inst{1,2}\orcid{0000-0001-5799-4155},
    Christian Burt\inst{1,2}\orcid{0009-0008-2061-4946}, 
    Daniel Mar\'in Pina\inst{1,2}\orcid{0000-0001-6482-1842}, Amedeo Romagnolo \inst{1,2,3}\orcid{0000-0001-9583-4339}}
    
    \authorrunning{M. Mapelli et al.}
    \titlerunning{The role of accretion and natal kicks in the formation of \Gaia{} black holes}
    \institute{
      $^{1}$Universit\"at Heidelberg, Zentrum f\"ur Astronomie (ZAH), Institut f\"ur Theoretische Astrophysik, Albert-Ueberle-Str. 2, 69120,\\ $^{\,}$ Heidelberg, Germany\\
    $^{2}$Universit\"at Heidelberg, Interdisziplin\"ares Zentrum f\"ur Wissenschaftliches Rechnen, Heidelberg, Germany\\
    $^{3}$Physics and Astronomy Department Galileo Galilei, University of Padova, Vicolo dell'Osservatorio 3, I--35122, Padova, Italy\\
    $^{4}$INAF - Osservatorio Astronomico di Padova, Vicolo dell’Osservatorio 5, I-35122, Padova, Italy\\
    $^{5}$Max-Planck-Institut für Astronomie, Königstuhl 17, 69117 Heidelberg, Germany \\ 
    $^{6}$Fakultät für Physik und Astronomie, Universität Heidelberg, Im Neuenheimer Feld 226, 69120 Heidelberg, Germany\\
    $^{7}$\textcolor{black}{Departament de Física Quàntica i Astrofísica, Institut de Ciències del Cosmos, Universitat de Barcelona, Martí i Franquès 1, E-08028 Barcelona, Spain}\\
    }

   \date{Received XXX / Accepted YYY}

 
  \abstract{
  \Gaia{} has the potential to deliver  several tens of new dormant black holes (BHs) with low-mass stellar companions (hereafter, \Gaia{} BHs) in the upcoming fourth data release. Three \Gaia{} BHs are already known, but their formation pathways remain uncertain. Here, we perform a large parametric study to explore the formation of \Gaia{} BHs from isolated binary systems with the population-synthesis code \sevn{} and compare our models with the properties of the three already reported \Gaia{} BHs. Specifically, we explore the impact of accretion efficiency, mass transfer stability, natal kicks, angular momentum transport, and core-collapse supernova prescriptions.  We find that models in which stable mass transfer is highly non-conservative and angular momentum is lost as a wind from the donor surface (Jeans mode)    maximize the probability of forming   dormant systems that match the properties of the observed \Gaia{} BHs in terms of both orbital period and eccentricity, because such assumptions prevent the initial orbit from shrinking too much when the BH progenitor fills its Roche lobe.  If we allow for common-envelope evolution, we find that models with common-envelope ejection efficiency $\alpha{} < 1$  predict dormant systems with orbital periods that are too short compared to the observed \textit{Gaia} BHs. The  eccentricity of the observed \Gaia{} BHs, when combined with  information about orbital period and BH mass, favors  relatively large natal kicks, similar to those inferred from Galactic neutron stars. 
  Finally, models in which  BH natal kicks are low -- e.g. because they are modulated by fallback -- result in the formation of a large population of dormant BHs with long orbital periods ($P_{\rm orb}>10^4$ days), 
  which will be tested soon by the fourth \Gaia{} data release.   
  }


   \maketitle
%

\section{Introduction}


Dormant black holes (BHs, i.e. X-ray silent BHs that are members of a binary system with a companion star,  \citealt{zeldovich1966,trimble1969}) represent a unique opportunity to understand the fate of massive stars and provide complementary information to gravitational wave events \citep{abbottGW150914,abbottGWTC4,GWTC4pop} and X-ray  systems \citep{oezel2010,farr2011}. The number of known dormant BHs is steadily increasing thanks to both radial velocity measurements and astrometry.  Dedicated spectroscopic surveys have searched for dormant BHs in globular clusters, resulting in a still  debated candidate in the Large Magellanic Cloud cluster NGC1850 \citep{saracino2022,saracino2023,elbadry2022} and two dormant BHs plus one additional candidate in the Milky Way cluster NGC3201 \citep{giesers2018,giesers2019}.  All these systems are associated with relatively low mass black holes ($4-10$ M$_\odot$) and low-mass companion stars ($0.5-1$ M$_\odot$) while the orbital periods and eccentricities span a wide range of values: from a few days to several hundred days for the former, and from nearly zero to more than 0.6 for the latter.

The two known dormant BHs with O-/B-type companion stars, HD~130298 in the Milky Way \citep{mahy2022} and VFTS243 in the Large Magellanic Cloud \citep{shenar2022} also differ substantially for their eccentricities: VFTS243 is almost exactly a circular system while the strong candidate HD~130298 has a high eccentricity $e = 0.47$. A number of additional spectroscopic dormant BH candidates are still debated \citep{thompson2019,jayasinghe2021,bodensteiner2022,zak2023,chakrabarti2023} and several long-debated candidates were recently found to be BH impostors \citep[e.g.,][]{muellerhorn2026}.

\Gaia{} astrometry holds great promise for the detection of several tens of new dormant systems and candidates with the forthcoming fourth data release \citep[hereafter \Gaia{} DR4,][]{breivik2017,wiktorowicz2020,chawla2022,chawla2025,shikauchi2022,shahaf2023,elbadry2024,nagarajan2025}. 

Currently we know of three \Gaia{} dormant BHs \citep{elbadry2023,elbadry2023b,tanikawa2023,panuzzo2024}. Both \Gaia{} BH1 and BH2 have BH mass $\sim{9}$~M$_\odot{}$, whereas BH3 is the only known electromagnetic stellar-origin BH with mass above 30 M$_\odot$. Interestingly, the companion of BH3 has low metallicity ([Fe/H]=--2.64, \citealt{panuzzo2024}) corroborating the idea that more massive BHs exist in metal-poor systems because of failed supernovae  \citep{woosley2002,mapelli2009,belczynski2010}. Moreover, BH3 is associated with the remnant of a disrupted globular cluster \citep{balbinot2024}, which might indicate a dynamical origin of the system \citep{marin2024,marin2026}, although a primordial binary system is not excluded \citep{elbadry2024BH3,iorio2024} given the relatively long orbital period  ($\sim{11.6}$ years).

\Gaia{} BH1 and BH2 are both a challenge for stellar and binary evolution theory because of their orbital properties. Specifically, their orbital period is 185 and 1277 days, respectively. In both cases, but especially for \Gaia{} BH1, the orbital period is considered to be too long to be explained with a post-common-envelope system and too short to originate from a non-interacting system. If the initial orbital separation was short enough to lead to the onset of a Roche lobe overflow, we indeed expect that the BH1 and BH2 progenitors underwent an unstable mass transfer episode, i.e. a common-envelope (CE) phase \citep{paczynski1968,ge2015}, given the extreme mass ratio at  the time of binary formation ($q=m_1/m_2>20$). According to the traditional $\alpha{}$ formalism for CE evolution, the expected final orbital period would be a few days, unless the CE ejection efficiency parameter $\alpha>10$ \citep{elbadry2023}. Moreover, both \Gaia{} BH1 and BH2 have high orbital eccentricity, which is usually disfavored after a CE episode, unless the BH formed with a large natal kick.

A dynamical formation channel in a stellar cluster \citep{shikauchi2020,rastello2023,dicarlo2024,tanikawa2024,marin2024,fantoccoli2025} or in a hierarchical triple \citep{hayashi2023,generozov2024,li2024,li2026} does not suffer from this problem, but \Gaia{} BH1 and BH2 are currently not inside a cluster: they must have evaporated or been ejected from their parent cluster long ago. Interestingly,  other types of
post-mass-transfer binaries, for example, the systems
detected by \Gaia{} hosting white dwarfs \citep{yamaguchi2025}  or neutron stars \citep{elbadry2024NS,tanikawa2024NS,chattopadhyay2025} share the same interpretation challenges. Dynamical formation is a reasonable option for dormant BHs, given  that dynamical exchanges favor the formation of more massive systems  \citep{hills1980}, but is quite unlikely for white dwarf and neutron star binaries.

Alternative explanations include the possibility that massive star radii evolve in a more compact way than traditionally expected \citep{gilkis2024} or that stellar winds are more effective than currently modeled \citep{kruckow2024,sabhahit2023,boco2025,pauli2026,romagnolo2026}. When considering binary evolution processes,  non-conservative mass transfer can lead to a substantial widening of the orbit, especially if angular momentum is lost as a wind at the surface of the donor star \citep{vinciguerra2020}.  Recently, \cite{olejak2025} demonstrated through binary evolution models with the code \textsc{mesa} \citep{paxton2011,paxton2018} that indeed this kind of non-conservative mass transfer can lead to orbital periods compatible with \Gaia{} BH1 and BH2. Indeed, non-conservative mass transfer is favored for the progenitors of such systems, considering that the accretor (i.e., the companion of the BH progenitor) is a low-mass star, hence with a $\mathcal{O}(10)$ Myr long Kelvin-Helmoltz timescale.

Here, we perform a large parametric study to explore the formation of the \Gaia{} BHs from isolated binary systems with the population-synthesis code \sevn{} \citep{spera2019,mapelli2020,iorio2023}. {Our simulations do not depend on the  stellar-evolution models described by \citet{hurley2000}, but rather are based on up-to-date stellar tracks adopting observationally calibrated core overshooting parameters and wind mass loss \citep{costa2025}. We explore the impact of accretion efficiency, mass transfer stability, angular momentum transport, and core-collapse supernova (CCSN) prescriptions. For the first time in the context of dormant BHs, we explore the possibility that BH natal kicks are not modulated by fallback.} 

We find that the orbital eccentricity and period of \Gaia{} BH1, BH2, and BH3   favor relatively large natal kicks, similar to those expected for  neutron  stars. Models in which the natal kicks are low -- e.g. because they are modulated by fallback -- result in the formation of a large population of dormant BHs {with long orbital periods ($P_{\rm orb}>10^4$ days),} 
which will be tested soon by the fourth \Gaia{} data release.   Our models confirm that the traditional $\alpha$ CE model with CE ejection efficiency $\alpha{}<1$ is a poor description of the evolution of the \Gaia{} BHs. In contrast, highly non-conservative stable mass transfer is an alternative way to explain the orbital periods of \Gaia{} BH1 and BH2, especially if we assume that angular momentum is lost from the surface of the donor star.

\section{Methods}

 

\subsection{Population synthesis with \sevn{}}
The stellar evolution for N-body code \sevn{} evolves stellar properties through on-the-fly interpolation of pre-computed stellar tracks \citep{spera2017,spera2019,mapelli2020,iorio2023}. In this work, we used the stellar tracks evolved with \textsc{parsec} \citep{bressan2012,costa2025} including stellar masses down to 0.7 M$_\odot$ \citep{nguyen2025}. \sevn{} models binary evolution with semi-analytical prescriptions. We refer to \citet{iorio2023} for a
detailed description of the code. Here, we highlight the main differences between the default \sevn{} setup presented by \citet{iorio2023} and our models.

\subsection{Natal kicks}
BHs and neutron stars receive a natal kick at birth. Here, we consider two alternative models for the natal kick. Model DM25 assumes that the natal kicks of both BHs and neutron stars follow a lognormal distribution with mean value $\mu=5.60$ and standard deviation $\sigma{}=0.68$, adopting the fit by \citet[][hereafter DM25]{disberg2025} to the proper motion of young Galactic pulsars \citep{hobbs2005}. This distribution peaks at 150--200~km~s$^{‑1}$. Model DM25fb starts from the same distribution as DM25 but then corrects the kick values by accounting for the fallback parameter,  namely 
$v_k=(1 - f_{\rm fb})\,{}v_{\rm DM25}$, where $v_{\rm DM25}$ is a random number derived from the DM25 lognormal distribution and $f_{\rm fb}$ is the is the fraction (from 0
to 1) of the stellar envelope mass that falls back \citep{fryer2012}. {Previous works always assume that BH natal kicks are modulated by fallback \citep{shikauchi2022,kruckow2024,chawla2025} or that BHs receive no natal kicks \citep{shikauchi2022}, despite evidence that at least a fraction of Galactic BHs are born with substantial kicks \citep{fragos2009,repetto2017,atri2019,nagarajan2025kick}.} 
The properties of our models are summarized in Table~\ref{tab:param} in Appendix~\ref{app:bigtable}.

\subsection{Core-collapse supernova (CCSN) model}

In our analysis, we use three simplified models to remap the final properties of the star at the onset of core collapse to the mass of the compact remnant. The delayed (D) and rapid (R) models are the same as described by \citet{fryer2012}; they are both fitting formulas to one-dimensional semi-analytic CCSN models and depend only on the total mass bound to the star and the carbon-oxygen core mass at the end of carbon burning. The only difference between the two original models is the assumed timescale to revive the shock, which is short ($<250$ ms) in the rapid model and long ($>500$ ms) in the delayed model. The compactness (C) model takes into account the compactness parameter as defined by \cite{oconnor2011} $\xi{}_{2.5}\equiv{}2.5\,{}{\rm M}_\odot\,{}[10^3\,{}{\rm km}/R(2.5\,{}{\rm M}_\odot)]$, where $R(2.5\,{}{\rm M}_\odot{})$ is the radius that encloses a mass of 2.5 M$_\odot$. Here we calculate the compactness at the onset of core collapse by using the fitting formula derived by \citet{mapelli2020}. We assume that stellar models with $\xi{}_{2.5}\ge{}0.3$ ($\xi{}_{2.5}<0.3$) form BHs (neutron stars). In all the considered CCSN models, we correct the final mass of the compact remnant to account for neutrino mass loss as described by \cite{zevin2020}. 

\subsection{Mass transfer and angular momentum transport}

We model wind mass transfer and Roche-lobe overflow mass transfer as described in \cite{iorio2023} and largely based on \cite{hurley2002}. Specifically, in Roche-lobe overflow the mass accretion rate of a non-degenerate accretor is equal to $\dot{M}=f_{\rm a}\,{}|\dot{M}_{\rm d}|$, where $\dot{M}_{\rm d}$ is the rate of mass lost by the donor and $f_{\rm a}\in{}[0,1]$ is a constant efficiency parameter. For degenerate accretors (compact objects), the mass accretion rate is limited by the Eddington value. In our case, since the accretor is a solar-mass star, the Kelvin-Helmoltz timescale is quite long ($\sim{10}$ Myr) and thus we expect $f_{\rm a}$ to be rather small.

If $f_{\rm a}<1$, accretion is non-conservative and angular momentum is lost. Here, we consider two options for angular momentum loss: isotropic re-emission and Jeans mode. In the isotropic re-emission model (Isot), angular momentum is lost as an isotropic wind from the accretion disk. Hence, the loss of orbital angular momentum scales as $\Delta{}J\propto{}\Delta{}M\,{}M_{\rm d}/(M_{\rm a}+M_{\rm d})$, where $\Delta{}M$ is the mass lost, $M_{\rm d}$ is the donor mass and $M_{\rm a}$ is the accretor mass. In the Jeans mode, angular momentum is lost as an isotropic wind from the donor star and $\Delta{}J\propto{}\Delta{}M\,{}M_{\rm a}/(M_{\rm a}+M_{\rm d})$. In our case, since $M_{\rm a}\ll{M_{\rm d}}$, the Isot mode results in a much larger angular momentum loss than the Jeans mode, thus leading to a more efficient orbital shrinking than the latter. Since the donor star is a O-type star with very fast winds ($\sim{1000}$~km~s$^{-1}$), the Jeans mode might be physically closer to the description of a \Gaia{} BH-like progenitor \citep{olejak2025}.

\begin{figure*}
    \centering
    \includegraphics[width=0.8\linewidth]{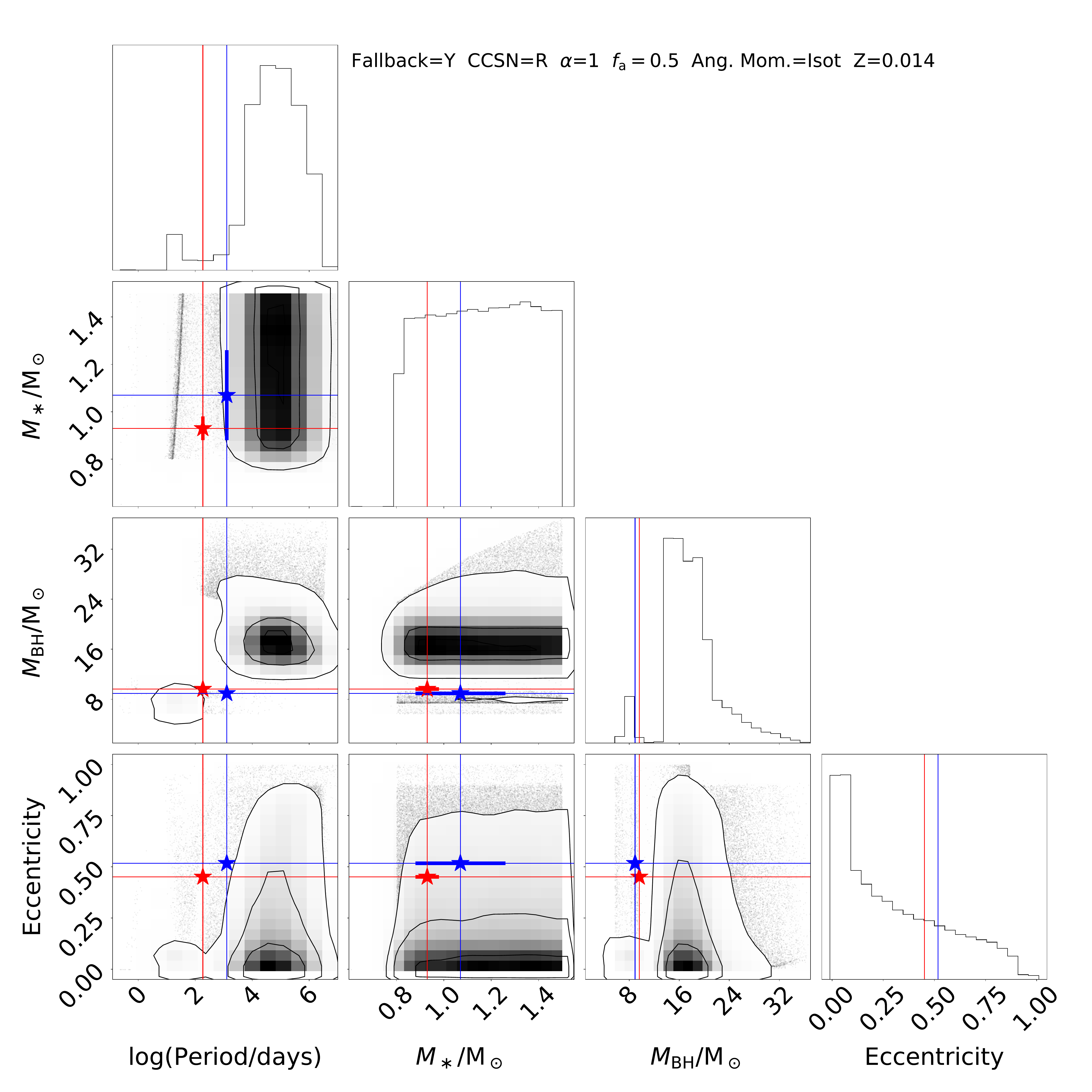}
    \caption{Corner plot showing the orbital period, eccentricity, BH mass ($M_\mathrm{BH}$) and companion star mass ($M_\ast$) in the simulations of model~\modAfb{}. In this model, natal kicks are drawn from DM25 and modulated by fallback, the rapid CCSN model is assumed \citep{fryer2012},  the accretion efficiency is $f_{\rm a}=0.5$, angular momentum transport follows the isotropic re-emission assumption, the critical mass ratios for mass-transfer instability are taken from \cite{hurley2002}, CE evolution assumes $\alpha=1$, and the metallicity is approximately solar ($Z=0.014$). The contour levels of this  and the other corner plots in this work show the 20\%, 50\%, and 90\% probability regions. The red and blue stars show the values for \Gaia{} BH1 \citep{elbadry2023} and BH2 \cite{elbadry2023b} and their corresponding 1$\sigma{}$ uncertainties (smaller than the markers in most cases).}
    \label{fig:modAfb}
\end{figure*}

\begin{figure*}
    \centering
    \includegraphics[width=0.8\linewidth]{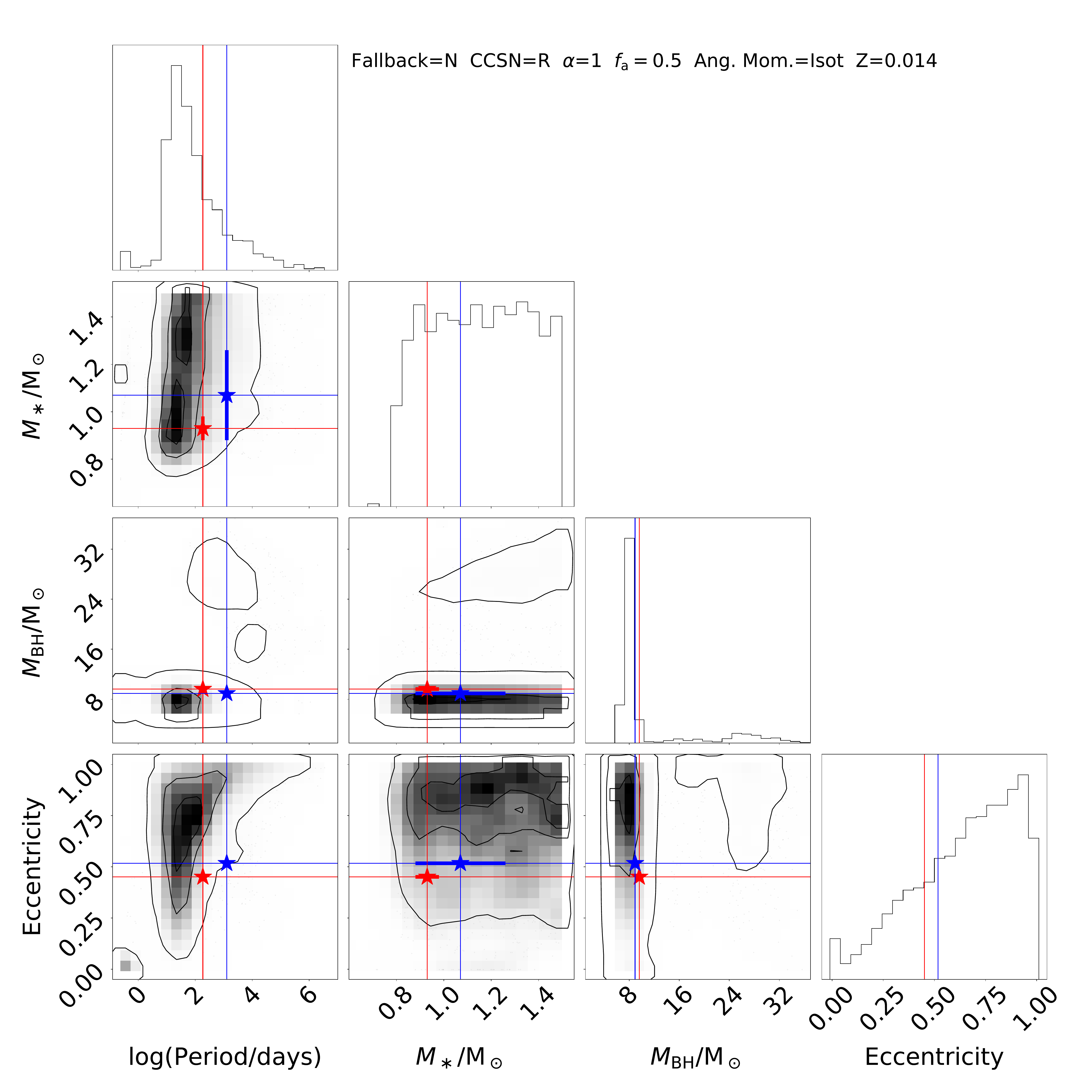}
    \caption{Same as  Fig.~\ref{fig:modAfb}, but for model \modA{}. The only difference between the two models is that model~\modA{} assumes natal kicks from DM25 whereas model~\modAfb{} takes into account the fallback to estimate natal kicks. }
    \label{fig:modA}
\end{figure*}


\subsection{Common-envelope (CE) evolution}
In \sevn{} and other population-synthesis codes, mass transfer is assumed to become unstable when the donor-to-accretor mass ratio is above a given threshold $q_{\rm c}$, whose value depends on the properties of the donor. Here, we assume the same values of $q_{\rm c}$ as in the original paper by \cite{hurley2002}. 
{The $q_{\rm c}$ values are listed in Appendix~\ref{app:alpha}.} When mass transfer is unstable, we use the $\alpha{}$ formalism in the version detailed by \cite{hurley2002} to describe the orbital evolution of the system and the chance of ejecting the envelope. Here, $\alpha{}$ is an efficiency parameter, 
describing what fraction of the orbital energy released during the inspiral of the cores is converted into energy available to unbind the envelope. 
In our models, we consider values of $\alpha{}=0.1,$ 0.5, 1 (fiducial model), 2, 3, 5, 7, 10, 15. Values of  $\alpha{}>1$ are  unphysical according to the original definition of this parameter. They are usually included in population-synthesis simulations with the idea of taking into account additional sources of energy (e.g., recombination, nuclear reactions, outflows), which are not included in the traditional $\alpha{}-$formalism \citep[e.g.,][]{zorotovic2011,ivanova2013,giacobbo2018,fragos2019,elbadry2023b,liz2026}.

The $\alpha{}$ formalism is a simplistic approximation to describe  CE evolution \citep[e.g.,][]{ivanova2013,roepke2023}. For this reason, we also consider models in which we set $q_{\rm c}$ to an unphysically large value ($q_{\rm c}=2000$), as to  prevent the system from starting a CE evolution. With this, we do not claim that the progenitor of the \Gaia{} BHs did not undergo unstable mass transfer. Rather, we want to quantify the differences in the evolution of such systems if we consider  two different models of orbital evolution during mass transfer. We refer to \cite{olejak2025} for an accurate discussion of these two scenarios. In general, using the $\alpha{}$ formalism with $\alpha\leq{}1$ produces an efficient shrinking of the semi-major axis, also by several orders of magnitude. 
Even if we set $q_{\rm c}=2000$, \sevn{} triggers  unstable mass transfer if both stars in the binary fill their Roche lobes or if a collision takes place (i.e., when the pericenter distance between two stars becomes less than the sum of their radii). 

\subsection{Initial conditions}

We sample the masses of the primary star ($M_1$) from a Kroupa initial mass function \citep{kroupa2001}  $\mathcal{F}(M_1) \propto M_{1}^{-2.3}$, in a range between $15$ and $150\,{} \text{M}_\odot$. We do not simulate lower-mass primary stars because we are only interested in BH progenitors.

We draw the secondary star mass ($M_2$) within the range $[0.7, 150]\,{} {\rm M}_{\odot}$ from the distribution $\mathcal{F}(q) \propto q^{-0.1}$, where $q=M_2/M_1$ \citep{sana2012} and 
 $0.01 \leq q \leq 1 $.

The orbital periods and eccentricities are also drawn from the distributions derived by \cite{sana2012}, by fitting the properties of a sample of O-type binary stars in open clusters. Specifically, we randomly sample the orbital periods $P_{\rm orb}$ and the eccentricities $e$ from the distributions $ \mathcal{F}(P_{\rm orb}) \propto (\log P_{\rm orb})^{-0.55}$, 
with $0.15 \leq \log \left(P_{\rm orb}/{\rm d}\right) \leq 5.5$,  
and $\mathcal{F}(e) \propto e^{-0.42}$, with $0\leq e \leq 1-\left(P/2 \ \mathrm{days}\right)^{-2/3}$, following the correction by \cite{moe2017}. {In Appendix~\ref{app:ecc}, we consider a set of runs starting with a thermal distribution for the initial eccentricities $\mathcal{F}(e) \propto e$, for comparison with  \citet{shikauchi2022}.}

We sample $10^{7}$ binaries and use them as initial conditions for each considered run. We consider three different metallicities: $Z=0.014$, 0.017, and $1.4\times{}10^{-4}$. The properties of our models are summarized in Table~\ref{tab:param}.

\section{Results}

\begin{figure*}
    \centering
    \includegraphics[width=0.8\linewidth]{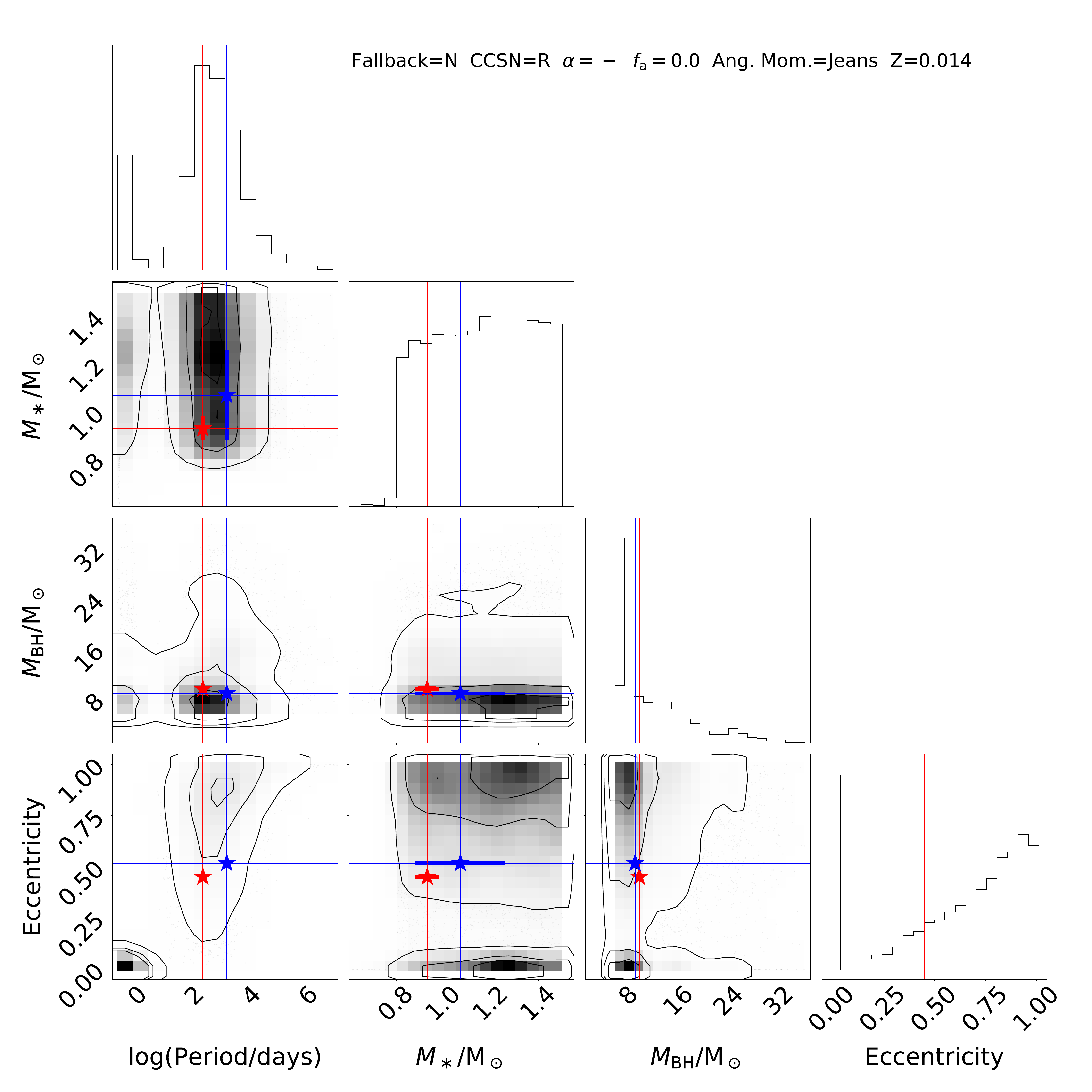}
    \caption{Same as Fig.~\ref{fig:modAfb} but for model \modBfaOj{}, in which we prevent the onset of CE evolution by setting a critical mass ratio $q_{\rm  c}=2000$.  In this model, natal kicks are drawn from DM25 (no fallback), the rapid CCSN model is assumed \citep{fryer2012},  the accretion efficiency is $f_{\rm a}=0.0$, angular momentum transport follows the Jeans mode, 
    and the metallicity is approximately solar ($Z=0.014$). }
    \label{fig:modB}
\end{figure*}

\begin{figure*}
    \centering
    \includegraphics[width=0.8\linewidth]{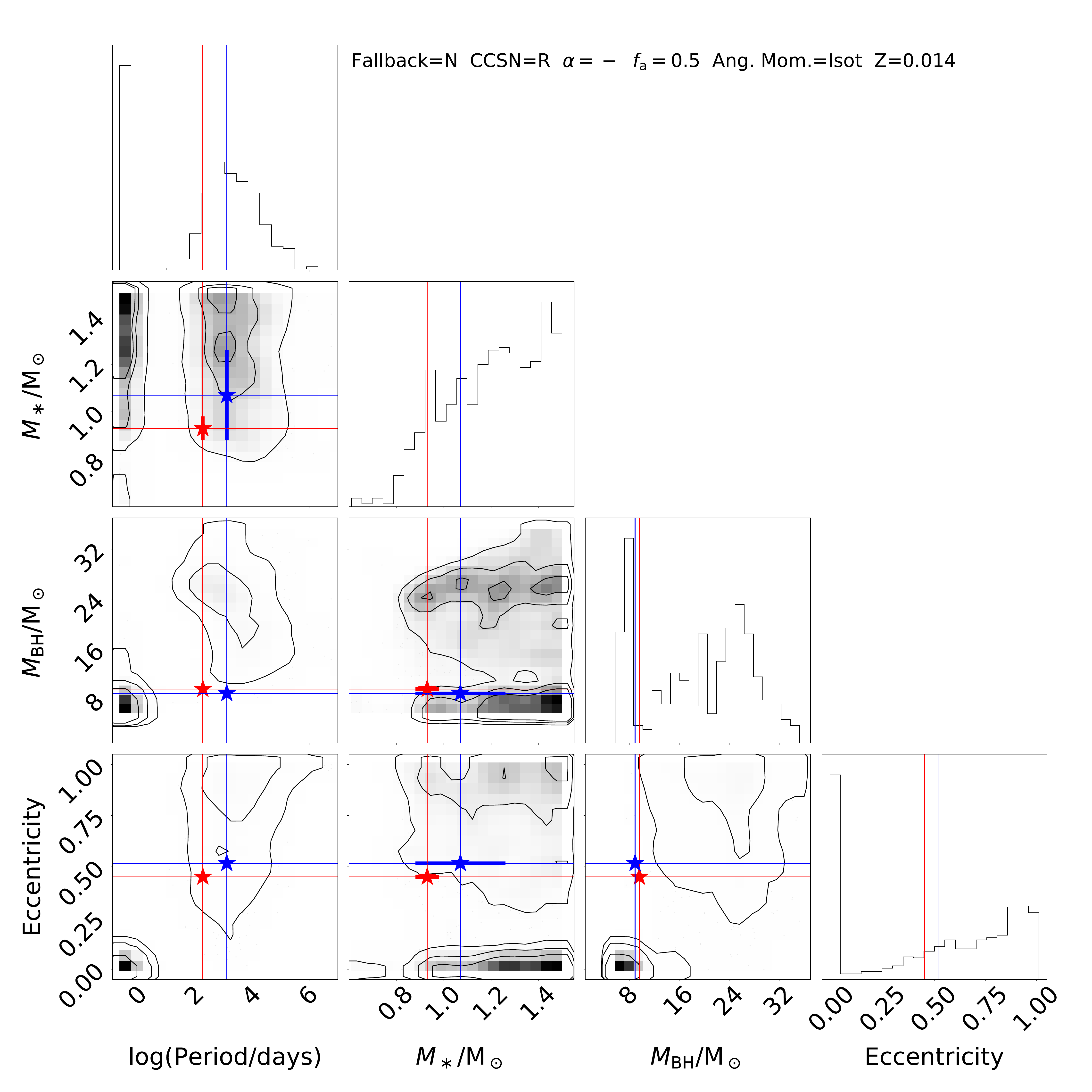}
    \caption{Same as Fig.~\ref{fig:modB} but for model \modBfaOfive{}. This models is the same as model \modBfaOj{} (i.e., mass transfer is always assumed to be stable),  but for two differences:  the accretion efficiency is $f_{\rm a}=0.5$ (instead of $f_{\rm a}=0.0$ as in model \modBfaOj{}), and angular momentum transport follows the isotropic re-emission (instead of Jeans mode as in model \modBfaOj{}).}
    \label{fig:modBfa05}
\end{figure*}

\begin{figure*}
    \centering
    \includegraphics[width=0.8\linewidth]{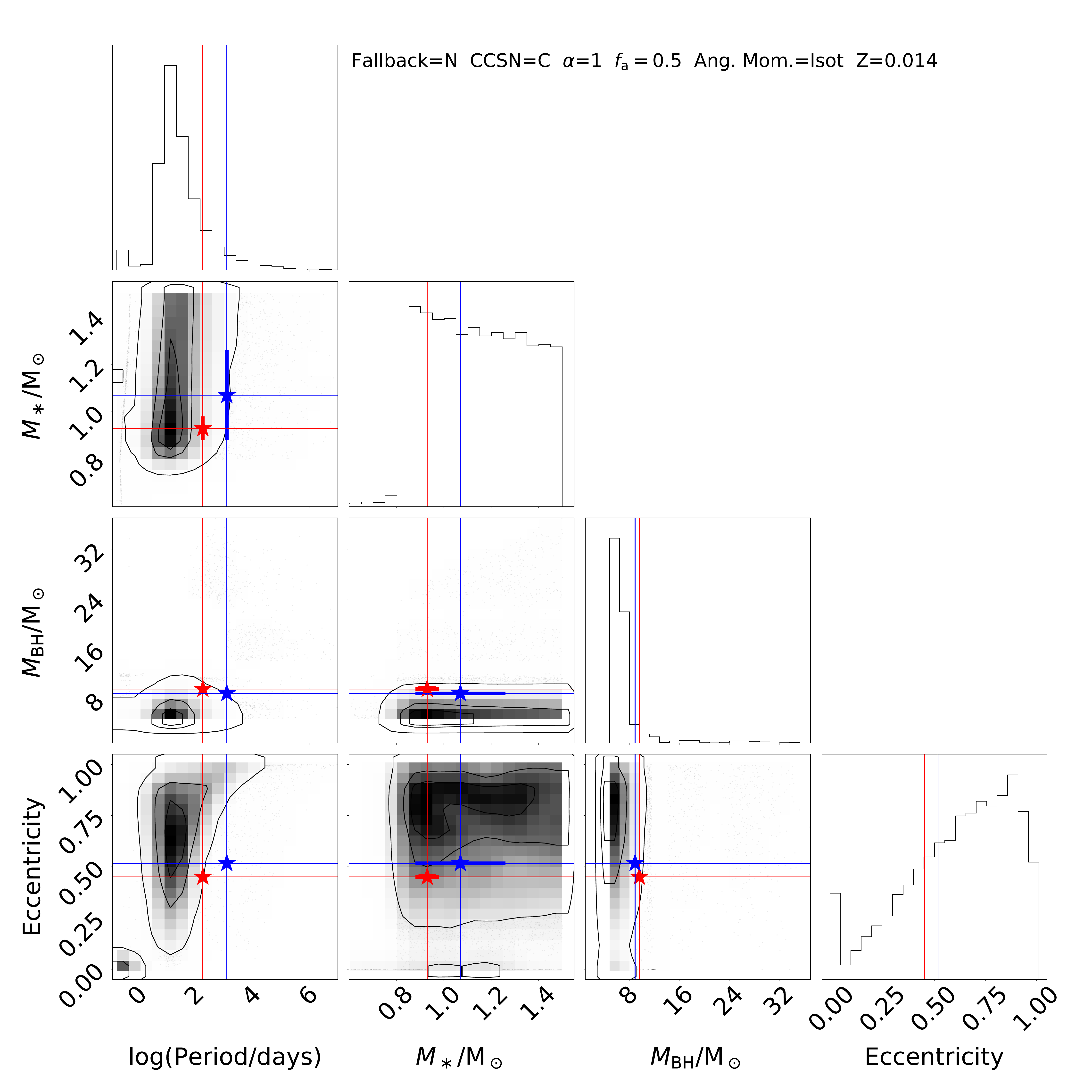}
    \caption{Same as Fig.~\ref{fig:modA} but for model \modC{}. This model is the same as model \modA{} but for one difference: it assumes the compactness criterion by \cite{oconnor2011} for the outcome of a CCSN.}
    \label{fig:modC}
\end{figure*}

\begin{figure*}
    \centering
    \includegraphics[width=0.8\linewidth]{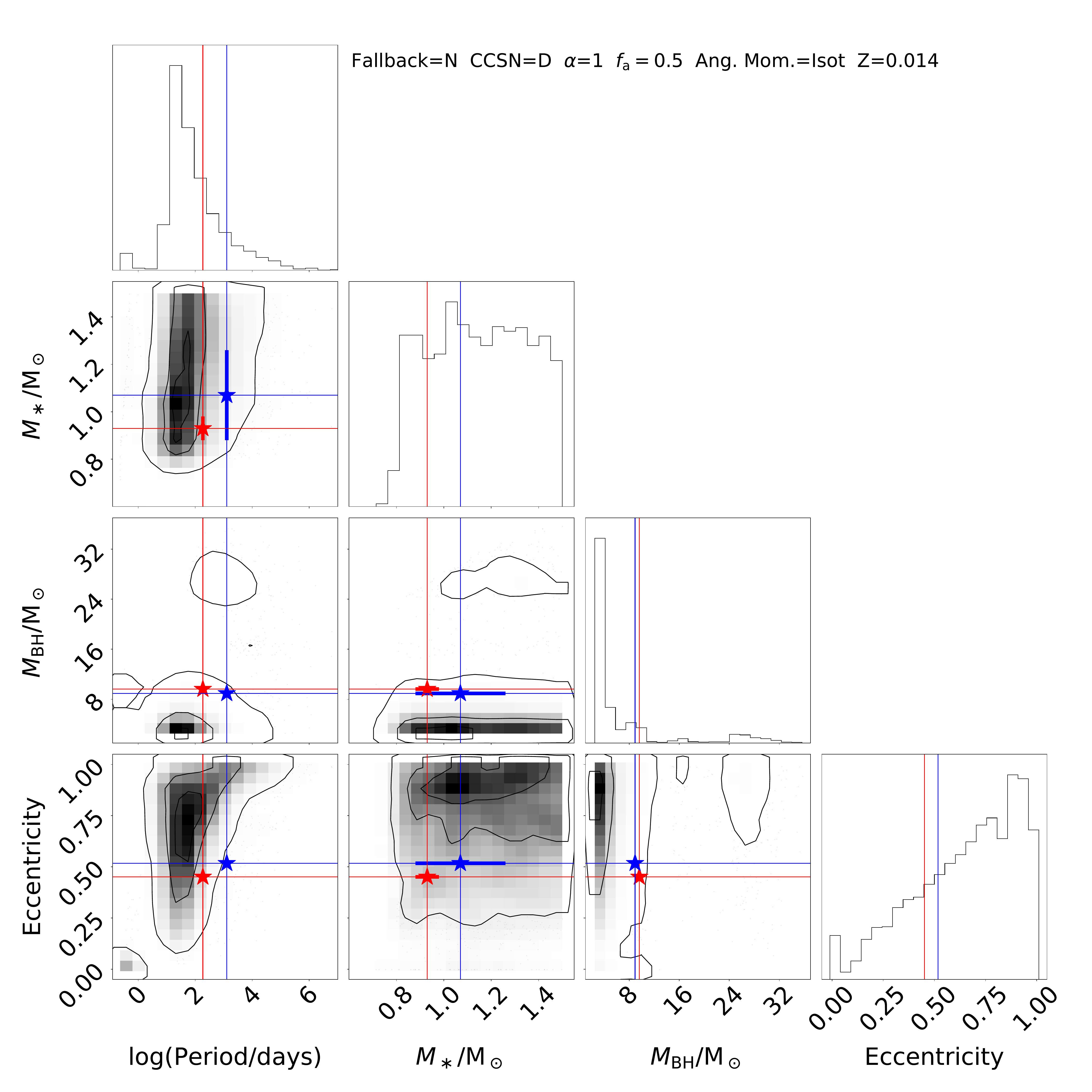}
    \caption{Same as Fig.~\ref{fig:modA} for model \modD{}. This model is the same as model \modA{} but for one difference: it assumes the delayed formalism by \cite{fryer2012} for the outcome of a CCSN.}
    \label{fig:modD}
\end{figure*}

\begin{figure*}
    \centering
    \includegraphics[width=0.8\linewidth]{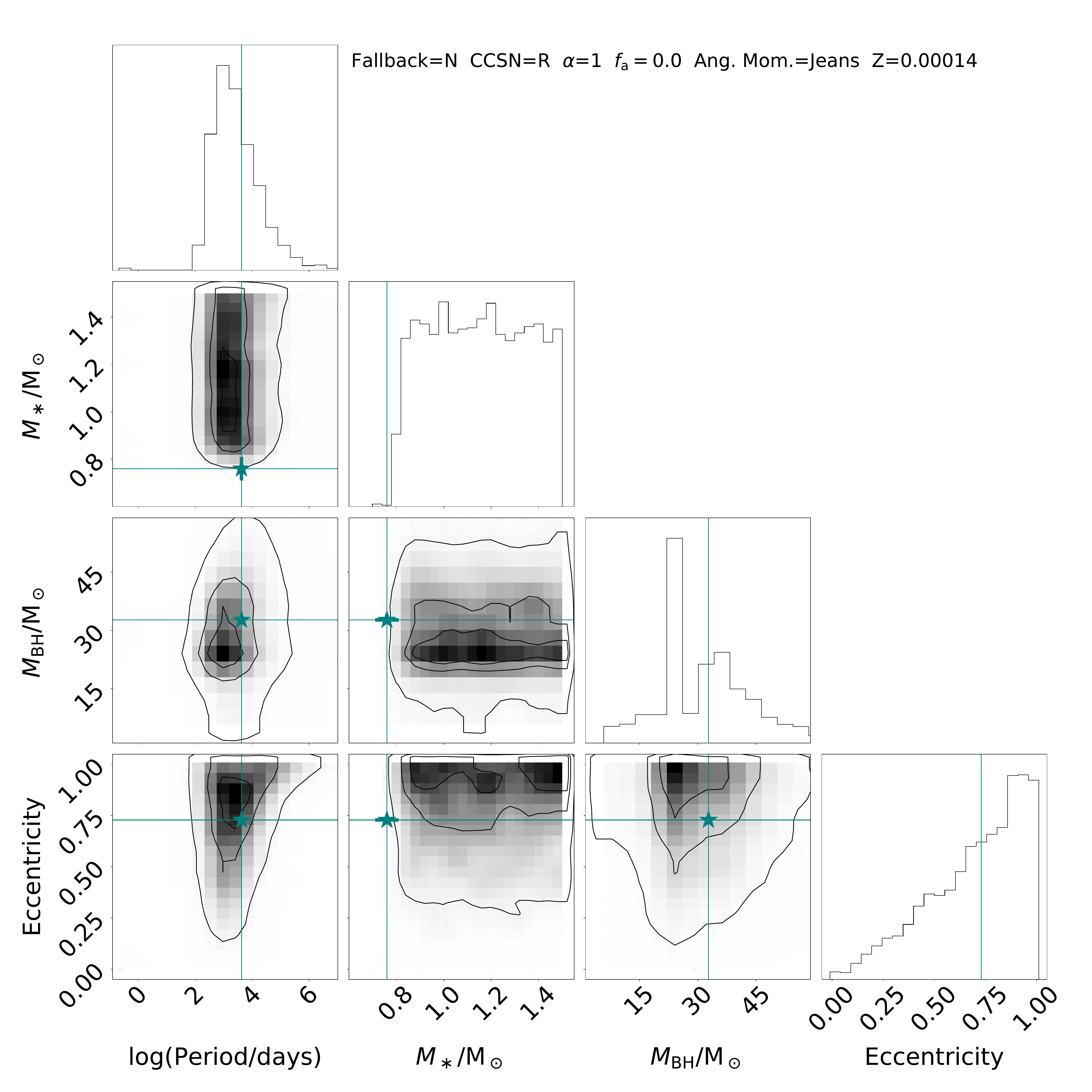}
    \caption{Same as Fig.~\ref{fig:modAfb}  for model \modAlowZ{}. This model is the same as \modAfaOj{} (with CE $\alpha{}=1$, accretion efficiency $f_{\rm a}=0$, Jeans mode for angular-momentum loss, and natal kicks from DM25) but with metallicity $Z=1.4\times{}10^{-4}$. Only \Gaia{} BH3 is shown for comparison (green lines and stars).}
    \label{fig:modBH3}
\end{figure*}

\subsection{Impact of natal kicks}\label{sec:kicks}
Figures~\ref{fig:modAfb} and \ref{fig:modA} show the behavior of orbital period, eccentricity, BH mass ($M_\mathrm{BH}$) and companion mass ($M_\ast$) of the simulated systems in models \modAfb{} and \modA{}, respectively. The only difference between the two models is the formalism for natal kicks. In model \modA{}, we draw the natal kicks from the distribution by DM25, whereas in model \modAfb{} the kicks are reduced taking into account the fallback as described by \cite{fryer2012}. 

The difference is striking: if the kicks are suppressed by fallback (Fig.~\ref{fig:modAfb}), we expect a large population of systems with almost zero eccentricity and long orbital periods ($P>10^3$ days), while a  small fraction ($\sim{0.01-0.02}$) of the simulated dormant systems have orbital period and eccentricity in the range of \Gaia{} BH1 and BH2.  

Instead, if the kicks are drawn from the distribution by DM25, most of the systems with large orbital period are  disrupted by the natal kick.  A higher fraction of the simulated  systems ($\sim{0.1-0.3}$) have orbital period and eccentricity in the range of \Gaia{} BH1 and BH2. 

Moreover, the overall number of dormant BH systems predicted by model \modA{} is a factor of $\approx{60}$ (Table~\ref{tab:param}) lower than that predicted by model \modAfb{}. The reason is that higher natal kicks (model \modA{}) disrupt the systems with initially large orbital periods.

This consideration applies to all the other models we have simulated (Appendices~\ref{app:bigtable},~\ref{app:alpha}, and \ref{app:ecc}): 
if the natal kicks are modulated by fallback, the dormant BH population we recover is extremely numerous and dominated {by systems with long ($P>10^3$ days) orbital periods.} 
Such long-period systems mostly host massive BHs ($>12$ M$_\odot$), because these form via failed supernovae and  are assumed to have small or no natal kick in the fallback formalism. \Gaia{} DR4 will  probe the existence of such class of long-period massive BH systems \citep{nagarajan2025}, yielding  constraints on the natal kicks of BHs.

Qualitatively, models where the BH natal kick does not depend on  fallback and is similar to the distribution inferred for the Galactic young pulsars (DM25) are able to produce systems with the orbital properties of \Gaia{} BH1 and BH2.

{Appendix~\ref{app:ecc} shows that the main takeaway of this Section does not change if we assume large initial orbital eccentricities: natal kicks modulated by fallback result in the formation of a population of non-interacting long-orbital-period ($P>10^3$ days) systems, which is instead completely suppressed if natal kicks are higher. However, the final eccentricities of the long-orbital-period systems formed in the fallback models strongly depend on the initial eccentricity distribution: high (low) initial eccentricities lead to high (low) final eccentricities of the BH systems with long orbital periods, because these do not undergo Roche-lobe overflow during their life. }

\subsection{Stable mass transfer}

When Roche lobe overflow starts in a system like the progenitor of \Gaia{} BH1 and BH2, with a mass ratio $q\gtrsim{}20$ between donor (i.e., the BH progenitor) and accretor, it is thought to lead to unstable mass transfer given the long Kelvin-Helmholtz timescale of the accretor ($\sim{10}$ Myr for a Sun-like star). The description of unstable mass transfer in population-synthesis codes is mostly  implemented through the $\alpha{}$ formalism for CE evolution 
\citep{hurley2002}. For many years, we have been aware that the  $\alpha{}$ formalism for CE evolution is an overly simplified model \citep{ivanova2013,roepke2023}, but we generally use it as a parametric toy model in population synthesis because of its simplicity. In this Section, we will simply make an experiment and assume that the mass transfer of the possible progenitor systems was always stable. 
We use this assumption only as a thought experiment, but we refer to the work by \citet{olejak2025} and \cite{ge2020} for a discussion on the stability of similar systems. Here, we set up \sevn{} so that the mass-ratio threshold for mass transfer to become unstable is never reached (critical mass ratio $q_c=2000$).

Figure~\ref{fig:modB} shows model \modBfaOj{}, where we assume that mass transfer is stable, the accretion efficiency is zero ($f_{\rm a}=0$) and angular momentum is lost from the system as a wind from the donor star (the so-called Jeans mode).  The natal kicks are drawn from DM25. For comparison, Fig.~\ref{fig:modBfa05} shows model \modBfaOfive{}, which differs from \modBfaOj{} only because the  accretion efficiency is $f_{\rm a}=0.5$ and angular momentum is lost according to the isotropic re-emission formalism.

Requesting that mass transfer is always stable instead of allowing for CE evolution helps to produce systems with the period of \Gaia{} BH1 and BH2 especially (but not only) with kicks from DM25. 
{However, many more systems merge or circularize if we assume a large accretion efficiency ($f_{\rm a}\geq{0.5}$) and isotropic re-emission  compared to low accretion efficiency and Jeans mode. For instance,  model B shown in Fig.~\ref{fig:modBfa05} (with $f_{\rm a}={0.5}$ and isotropic re-emission) produces a factor of $\approx{7}$ less BH systems than model BJ (Fig.~\ref{fig:modB}). About 25\% and 14\% of the systems shown in Fig.~\ref{fig:modBfa05} and Fig.~\ref{fig:modB} have circularized at the end of the simulation, respectively.
The reason for these differences} is that a dissipative mass transfer with the Jeans mode keeps the orbital period larger than a more conservative mass transfer with isotropic re-emission \citep{olejak2025}. 
{Hence, many more binaries merge in the isotropic re-emission case, whereas the survivors tend to circularize during mass transfer.}

{Models that assume no CE evolution, accretion efficiency $f_{\rm a}\sim{0}$, and the Jeans mode for angular momentum transport maximize the probability of forming systems like \Gaia{} BH1 and BH2, according to our metrics (Tab.~\ref{tab:param}).}

\subsection{Black hole mass function}

To determine the mass of the BH from the properties of the progenitor at the onset of core collapse, \sevn{} uses several formalisms obtained from 1D CCSN models. Figures~\ref{fig:modA} and \ref{fig:modB}  adopt the rapid model by \cite{fryer2012}. Figures~\ref{fig:modC} and ~\ref{fig:modD} are the same as Fig.~\ref{fig:modA} but for other CCSN models,   based on the compactness parameter \citep{oconnor2011} in Fig.~\ref{fig:modC} (see model~\modC{})  and the delayed model \citep{fryer2012} in Fig.~\ref{fig:modD}. The comparison of such Figures shows that 
with the rapid model we expect a peak in the BH mass range at $\approx{8-10}$ M$_\odot$, similar to the mass of \Gaia{} BH1 and \Gaia{} BH2. In contrast, delayed and compactness models tend to produce many lower-mass systems.

 However, with low kicks (based on the fallback criterion) all the CCSN models predict a large population of massive ($\sim{}16-32$~M$_\odot$) BHs in binary systems with low mass stars. 
 \Gaia{} DR4 will definitely probe the existence of such subpopulation \citep{nagarajan2025}.

\subsection{CE efficiency parameter}

Increasing  the CE efficiency parameter $\alpha{}$ from 0.1 to 15 leads to dormant BHs with progressively longer orbital period and higher eccentricities, especially if only large natal kicks are considered (according to the DM25 distribution without modulation by fallback). This happens because higher values of $\alpha{}$ correspond to more efficient ejection of the envelope with less orbital shrinking. As a result, models with $\alpha{}<1$ cannnot reproduce the orbital properties of the observed \Gaia{} BHs, because they predict systems with too short orbital periods.  In contrast, models with $\alpha{}\ge{}1$ produce  similar results (in terms of orbital period and eccentricity distribution) to our stable mass-transfer models with low accretion efficiency ($f_a\sim{0}$) and Jeans mode for angular momentum loss.

Our result  agrees with previous works finding a preference for $\alpha{}\geq{}1$ \citep{giacobbo2018,mapelli2018,fragos2019,hirai2022,elbadry2023b}. This possibly supports the importance of additional sources of energy for unbinding the CE, or the need for a different formalism to describe orbital evolution during a CE phase \citep{nelemans2000,nelemans2005,fragos2019,hirai2022}.  We further discuss the models with $\alpha{}\neq{1}$ in  Appendix~\ref{app:alpha}.


\subsection{Metallicity and \Gaia{} BH3}

\Gaia{} BH3, an anticipation of \Gaia{} DR4, is substantially different from BH1 and BH2. The metallicity of the companion is well below solar ($\rm [Fe/H]=-2.64$), the mass of the BH is $\approx{33}$ M$_\odot$, and the orbital period is around 4000 days \citep{panuzzo2024}. The orbital eccentricity is also  high ($e\approx{0.75}$). 
Figure~\ref{fig:modBH3} shows model~\modAlowZ{}. 
{This} model has low metallicity $Z=1.4\times{}10^{-4}$. Similar to BH1 and BH2, also BH3 is more likely formed with low accretion efficiency and the Jeans mode for angular momentum transport, but, unlike the other two \Gaia{} BHs, a CE evolution 
is the configuration that maximizes the probability of producing a \Gaia{} BH3-like system (Table~\ref{tab:param}). 

Overall, this result remarks the relatively minor difference between our best CE and non-CE evolution systems: the most important ingredients to form \Gaia{} BH1, BH2, and BH3 seem to be the large natal kicks (DM25), the low accretion efficiency ($f_a=0$) and the Jeans mode for angular momentum transport. The latter keeps the binary orbital period relatively wide, while the kick is essential to produce large eccentricities.

\section{Discussion}

\subsection{Evolutionary channels}

\begin{figure}
    \centering
    \includegraphics[width=\linewidth]{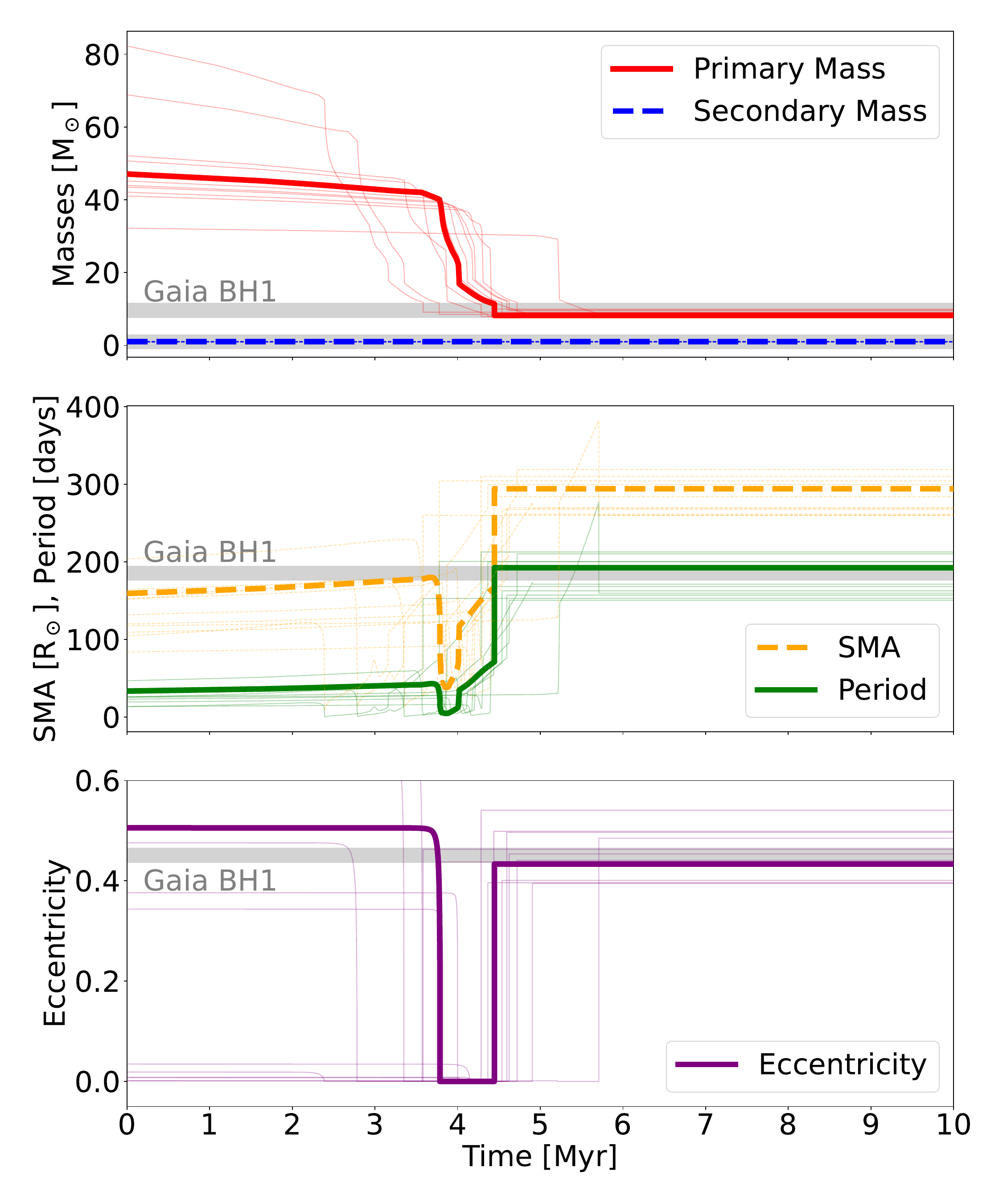} 
    \caption{Evolution of  the best-matching \Gaia{} BH1-like systems through stable mass transfer. We show model \modBfaOj{}{} (same model as Fig.~\ref{fig:modB}), with stable mass transfer ($q_c=2000$), $f_{\rm a}=0$, Jeans mode, and no fallback. From top to bottom: temporal evolution of the primary and secondary mass,  semi-major axis (SMA),  orbital period, and  eccentricity. The thick line shows the best-matching model among all the simulated systems; the thin lines show the next best-matching models.}
    \label{fig:bin6}
\end{figure}

\begin{figure}
    \centering
    \includegraphics[width=\linewidth]{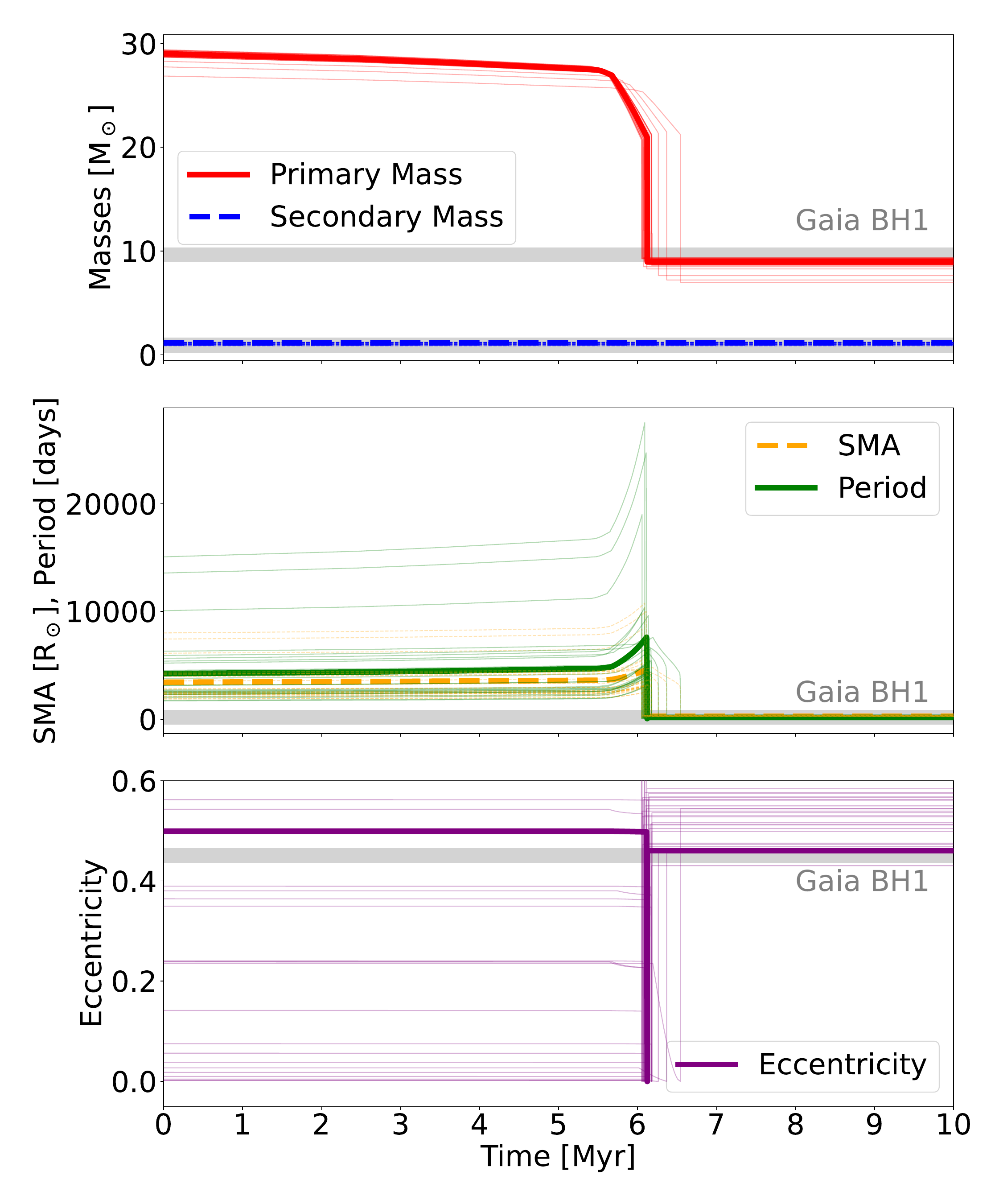} 
    \caption{Same as Fig.~\ref{fig:bin6}, but considering the best-matching \Gaia{} BH1-like systems formed through CE evolution. We show model \modAfb{} (same model as Fig.~\ref{fig:modAfb}), with CE evolution, $\alpha=1$, $f_{\rm a}=0.5$, isotropic re-emission, and fallback.}
    \label{fig:bin3}
\end{figure}

Our simulations demonstrate that there are two main evolutionary channels able to produce \Gaia{} BH-like systems from isolated binaries:  stable mass transfer and CE evolution. Figures ~\ref{fig:bin6} and \ref{fig:bin3} display the temporal evolution of some specific simulations from models \modBfaOj{} and \modAfb{} as examples of stable mass transfer and CE evolution, respectively. 
In both Figures, we show the systems that have the closest final masses, orbital periods, and eccentricities  to the measured values of \Gaia{} BH1.

In the first channel (Fig.~\ref{fig:bin6}, let us call it stable mass transfer channel), the zero-age main sequence mass of the primary star spans from $\approx{30}$ to $\approx{90}$ M$_\odot$ and the initial orbital period $P_{\rm in}$ is shorter than the final one ($P_{\rm in}\approx{10-50}$ days). 
This channel becomes important only if mass transfer remains stable, is highly dissipative ($f_{\rm a}\sim{0}$) and mass/angular momentum are lost as a wind from the donor (Jeans mode). In this case, the orbit  shrinks by a factor of $\lesssim{2}$ during mass transfer. The CCSN explosion happens during stable mass transfer and the natal kick sets both  final period and eccentricity. Without a significant kick ($v_{\rm kick}\approx{}30-100$ km/s), it is not possible to attain the observed eccentricity of the \Gaia{} BHs, because we assume that the orbit circularizes during Roche-lobe overflow \citep[see, e.g.,][]{kotko2024}.

The second formation channel (Fig.~\ref{fig:bin3}, let us call it CE-evolution channel) produces all the \Gaia{} BH-like systems in the runs where  CE evolution is activated. In this case, if $\alpha{}=1$, the primary star has initially a mass of 25--30 M$_\odot$
 (the range is very narrow because of the dependence of the final semi-major axis on the donor mass in the traditional $\alpha{}$ formalism), the initial semi-major axis is at least a factor of 10 higher than the final one, and the orbit shrinks dramatically during the CE phase. As in the other formation channel, the natal kick is crucial to determine the final eccentricity of the system.

\subsection{Percentile Distribution}

To quantitatively assess whether the models presented in this paper are capable of producing BHs with parameters similar to \Gaia{} BH1, BH2, and (for $Z=1.4\times{}10^{-4}\approx{10^{-2}\,{}{\mathrm{Z}_\odot}}$) BH3, we consider the percentiles of the model probability distribution function (PDF) at which the \Gaia{} BHs are found. 
This approach allows us to quickly identify 
models for which the  observations are not a plausible outcome.  The main caveat is that this analysis does not include selection effects.

In practice, for each model we approximated the PDF with a Gaussian mixture model trained on the samples produced by the population synthesis code with \textsc{figaro} 
\citep{rinaldi2024} and evaluate the approximated PDF for each of the available samples. These samples, once ranked according to their PDF, can be used as an approximate one-dimensional percentile grid. 
To account for the measurement uncertainty of the \Gaia{} BHs, we used a Monte Carlo approach and -- for each object -- draw 100 samples from a multivariate Gaussian distribution centered on the measured values and with standard deviation equal to the measurement uncertainty. Here, we neglected  possible correlations between parameters. For each of these samples, we evaluated the PDF and computed the corresponding percentile. 

Table~\ref{tab:param} reports, for each model and object, the median percentile value and the 68\% credible interval. The majority of the models have  percentiles larger than 
95. 
The models highlighted in blue  recover the properties of at least one of the {\it Gaia} BHs within the $\leq{}90$th percentile.

\subsection{Selection effects}
The percentile-based approach described above quantifies the probability of forming \Gaia{} BHs with the observed properties for a given model. However, it does not account for observational selection effects. In particular, systems that are intrinsically rare may still be preferentially detected, while models that efficiently produce \Gaia{} BH-like systems may over or underpredict the total number of detectable sources.

Here, we provide an order-of-magnitude estimate of the number of dormant BHs that would have been detected in the third \Gaia{} data release (\Gaia{} DR3) for our different models. To this end, we adopted a forward-modeling approach. First, we scaled our simulated populations to the Milky Way to estimate the total number of \Gaia{} BH systems. Second, we generated mock observations of these binaries to emulate \Gaia{} measurements. Finally, we applied a model of the \Gaia{} non-single star selection function to estimate the number of systems that would have been detected as binaries in DR3.

{Here, we focus on models \modAfb{} (Fig.~\ref{fig:modAfb}) and \modA{} (Fig.~\ref{fig:modA}) as examples of runs where natal kicks are and are not modulated by fallback, respectively. In our models \modAfb{} and \modA{}}, the number of dormant BHs with a main-sequence or a giant companion in the mass range $0.7-1.5$~M$_\odot$ is {$N_{\rm dorm}=$108347 and 1932}, respectively. Restricting to BH masses in the range $8$--$12\,\mathrm{M}_\odot$ reduces these numbers to {2161 and 576}, respectively.   

The total initial mass of our stellar population is $M_{\rm sim}=2.32\times{}10^9$ M$_\odot$ (after correcting for the low-mass stars that we do not simulate), corresponding to approximately $5\%$ of the stellar mass in the Milky Way \citep{licquia2015,blandhawthorn2016}. However, we assume that all stars are initially members of binary systems, leading to a factor of $\gtrsim{}2$ overestimate of the binary star population. The actual correction factor depends on the mass of the primary star, but here we are just considering an order-of-magnitude estimate \citep{moe2017}. In addition, at solar metallicity stellar evolution reduces the total mass of a stellar population older than $\sim{1}$ Gyr to $\sim50\%$ of its initial value due to winds and supernovae, implying a correction factor $f_{\rm loss}\sim0.5$. We neglect variations in metallicity across the Milky Way for this estimate. 

Combining these factors, we estimate the total number of dormant BH systems in the Milky Way for the model \modA{} as

\begin{eqnarray}
N_{\rm dorm, MW}= {{ 4.16}\times{}10^4}\,{}\left(\frac{N_{\rm dorm}}{ 1932}\right)\,{}\left(\frac{M_\mathrm{MW}}{5\times{}10^{10}\,{}\mathrm{M}_\odot}\right)\nonumber{}\\\left(\frac{2.32\times{}10^9\,{}\mathrm{M}_\odot}{M_\mathrm{sim}}\right)
\,{}\left(\frac{f_{\mathrm{bin}}}{0.5}\right)\,{}\left(\frac{0.5}{f_{\mathrm{loss}}}\right),
\end{eqnarray}
where $M_{\mathrm{MW}}$ is the  current total stellar mass of the Milky Way and $f_\mathrm{bin}=0.5$ the fraction of stars in binary systems.

The number of detected \Gaia{} BHs is related to the total population by detection probability $p_\mathrm{det}$. We estimated $p_\mathrm{det}$ by generating mock observations of the simulated BH binaries, forward modeling the selection effects of \Gaia{}, and counting the fraction of systems that would be detectable. We describe this procedure in Appendix~\ref{app:pdet}. 
{We} estimate \textcolor{black}{{$p_\mathrm{det}={ 2.9}\times10^{-4}$ and ${ 8.0}\times10^{-5}$}} for models \modA{} and \modAfb{}. Combining this with our estimate of the total number of dormant BHs in the Milky Way, we obtain an order-of-magnitude prediction for the number of \Gaia{} BHs detectable in DR3:
\begin{equation}
N_{\rm det}=\mathcolor{black}{{ 12}}\,{}\left(\frac{p_{\rm det}}{ 0.00029}\right)\,{}\left(\frac{N_{\rm dorm, MW}}{\mathcolor{black}{ 41600}}\right)
\end{equation}
for model \modA{}. 
This number raises up to  \textcolor{black}{ 186} for model \modAfb{}.

This estimate exceeds the effectively detected number of \Gaia{} BHs by approximately one (two) order(s) of magnitude for model \modA{} (model \modAfb{}), respectively. We will present a more detailed treatment of selection effects and detectability, including predictions for future \Gaia{} data releases, in a follow-up study (M\"uller-Horn et al., in prep.).

\subsection{Formation efficiency}

Natal kicks have a strong impact on the number of dormant BH systems ($N_{\rm dorm}$) per total initial stellar mass ($M_{\rm sim}=2.32\times{}10^9$~M$_\odot$), i.e. the efficiency parameter:
\begin{equation}
\eta{}_\mathrm{dorm}=\frac{N_{\rm dorm}}{M_{\rm sim}}.\label{eq:eff}
  \end{equation}
  As shown in Table~\ref{tab:param}, models with fallback yield an efficiency parameter $\eta{}_\mathrm{dorm}$ of a few $\times{}10^{-5}$ M$_\odot^{-1}$,  which is 20--70 times higher than models run with the DM25 distribution of natal kicks ($\eta{}_\mathrm{dorm}\sim{}10^{-7}-10^{-6}$ M$_\odot^{-1}$).


The large difference of $\eta{}_\mathrm{dorm}$ is almost entirely due to  the subpopulation of massive dormant BHs ($M_{\rm BH}>12$ M$_\odot$) with long orbital period ($P_{\rm orb}>10^4$ days) that form if natal kicks are modulated by fallback and hence strongly quenched for large BH masses. Large natal kicks ($\gtrsim{}100$ km/s) unbind most long orbital period systems. 

Such long-period systems  ($>10^4$ days) are almost undetectable with \Gaia{} DR3 but will be accessible to \Gaia{} DR4 \citep{nagarajan2025}. Hence, \Gaia{} DR4 will test this scenario and possibly provide indirect constraints on BH natal kicks. 

Current observations of dormant systems  \citep{nagarajan2025kick} and BHs in X-ray binaries \citep{fragos2009,repetto2017,atri2019} indicate that a significant fraction of BHs in the Milky Way received non-negligible ($\approx{100}$ km s$^{-1}$) natal kicks, but at least two systems (V404~Cyg and VFTS~243) have a strong preference for low  ($\lesssim{}10$ km s$^{-1}$) natal kicks \citep{willcox2025,nagarajan2025kick}.

\subsection{Comparison with previous work}\label{sec:comp}

{Here, we compare our approach and results to previous work on \textit{Gaia} BHs. In the literature, only \citet{olejak2025} have explored the Jeans mode with highly non-conservative mass transfer as a possible scenario for the formation of \textit{Gaia} BHs. Similarly to our work, they conclude that stable mass transfer with low accretion efficiency and Jeans angular momentum transport favors the formation of \textit{Gaia} BH1-like systems. They simulate  one specific case with the code \textsc{mesa} \citep{paxton2011,paxton2018}, starting with zero initial eccentricity, and do not model  BH formation and natal kicks. 
They suggest that a population synthesis study exploring mass transfer stability, angular momentum transport, and accretion efficiency is needed, in order to draw statistically significant conclusions about the formation channels of \textit{Gaia} BHs.}

{Here, we perform such parameter-space exploration with our code \textsc{sevn}, which adopts stellar tracks completely different from \textsc{bse} and other codes based on the fitting formulas derived by \citet{hurley2000}. Specifically, the evolutionary models adopted here come from the PARSEC stellar tracks  \citep{iorio2023, nguyen2025, costa2025}. We refer to the original papers for details, but here we mention that these tracks include an accurate modeling of stars close to or above a zero-age main sequence mass of 100 M$_\odot$, a large overshooting parameter ($\Lambda_{\rm ov}=0.5$), resulting in larger massive star cores than \textsc{bse} models, and a radiation-driven wind model including the dependence of mass loss on the Eddington ratio $\Gamma=L/L_{\rm Edd}$ (where $L$ and $L_{\rm Edd}$ are the luminosity and Eddington luminosity, respectively). The inclusion of the mass-loss dependence on the Eddington ratio allows to model stellar winds in the optically thick regime -- if a star is radiation-pressure dominated -- and helps massive stars to avoid the Humphreys-Davidson limit \citep{graefener2008,chen2015}. These features result in a different evolution of the stellar radius compared to \textsc{bse} \citep[e.g., Fig.~7 of][]{iorio2023} and lead to a remarkably different BH mass function even when the same core-collapse supernova model is applied \citep[see][for a detailed discussion and comparison]{iorio2023}. Previous work relies on the fitting formulas by \cite{hurley2000}. For example, \citet{shikauchi2022} use their custom version of \textsc{bse} with wind models derived from \citet{belczynski2010}, whereas \citet{chawla2025} use \textsc{cosmic}, which is also based on \textsc{bse} as for stellar and binary evolution.}

{Here, we consider two models for BH natal kicks, comparing a scenario in which natal kicks are the same as those of Galactic pulsars (DM25) with one in which they are modulated by fallback (DM25fb). Previous works \citep[e.g.,][]{shikauchi2022,shikauchi2023,chawla2022,chawla2025}  assume  zero natal kicks or fallback-modulated kicks for BHs and hence result in the prediction of a large number of long-period \textit{Gaia} BH-like systems \citep[see, e.g., Figure 1 of][]{chawla2022}. Here, we explore also cases with BH kicks drawn from the same distribution as neutron star kicks. The possibility that BHs receive significant kicks at birth is compatible with observations \citep{fragos2009,repetto2017,atri2019,nagarajan2025kick}, and significantly helps reconciling the estimated merger rate density of binary BHs with the constraints inferred from gravitational-wave data \citep[e.g.,][]{boesky2024,sgalletta2025}.}

{The models presented in the main text assume low initial orbital eccentricity, following the observational constraints by \cite{sana2012}.  In contrast, \citet{shikauchi2022} draw the initial orbital eccentricities from a thermal distribution, which is usually adopted for the dynamical scenario in star clusters. In Appendix ~\ref{app:ecc}, we show that starting from a thermal eccentricity distribution has almost no impact on the models with DM25 natal kicks, whereas significantly affects the final eccentricity of the long-orbital period systems in the fallback models. Indeed, since the long-orbital period systems ($P_{\rm orb}>10^3$ days) evolve without Roche-lobe overflow, their final eccentricity largely reflects their initial eccentricity if natal kicks are negligible.}

{As for the treatment of stable and unstable mass transfer, previous population-synthesis works assume the default model of \textsc{bse}, which relies on the isotropic re-emission formalism for angular momentum transport and on a mostly conservative mass transfer formalism \citep{hurley2002}. Our results show that the assumption on mass-accretion efficiency and angular-momentum transport drastically affect the results. CE evolution has  already been thoroughly investigated in previous work with the $\alpha$ formalism, exploring values of the free parameter $\alpha{}$ from 0.1 to 15 \citep[e.g.,][]{shikauchi2022,elbadry2023}. Here, we show that disabling CE evolution has a major impact on the formation of dormant systems, especially with large orbital periods. But we also demonstrate that a non-conservative Jeans mode yields results similar to those of CE evolution with $\alpha\geq{}1$.}

\subsection{Dynamical formation}\label{sec:dyn}
Here, we have not considered the dynamical channel for dormant BH formation \citep{rastello2023,dicarlo2024,marin2024,nagarajan2025}, which we will consider in a follow-up study (Mar\'in Pina et al., in prep.). A non-negligible fraction of massive stars form in clustered environments \citep{lada2003} and we already know that \Gaia{} BH3 formed in a globular cluster \citep{balbinot2024}. \cite{rastello2023}  show that other \Gaia{} BHs might have formed in a star cluster as well, and were then subsequently ejected from it due to dynamical interactions or cluster evaporation.

Dynamics clearly extends the parameter space and introduces more degrees of freedom, because  the orbital period and eccentricity are affected by dynamical encounters in addition to binary evolution and natal kicks. Dynamics can also affect the masses through dynamical exchanges. \textcolor{black}{Interestingly, despite the large parameter space we explored, our models struggle to match the properties of \Gaia{} BH3, which is the only \Gaia{} BH clearly associated with a globular-cluster origin.}

\cite{nagarajan2025} discuss a set of dynamical simulations where the orbital period of dynamical systems only extend up to a few $\times{}10^3$ days. In contrast, \cite{rastello2023}, \cite{marin2024}, {and \cite{marin2026}} show dynamical simulations where the orbital periods of \Gaia{} BH-like systems extend up to $10^7$ days or more. There are several possible explanations for this reason. First, \cite{nagarajan2025} use the dynamical simulations by \cite{dicarlo2020}, which comprise only 6000 star cluster models, whereas \cite{rastello2023} analyse a sample of $>10^5$ star cluster models. Hence, there could be an effect of low statistics. Moreover, dynamical interactions  tend to unbind soft binaries, i.e. binary systems with binding energy smaller than the average kinetic energy of a star in the cluster \citep{heggie1975}. Hence, the maximum orbital period of binary systems in a star cluster depends on the actual threshold between hard and soft binaries, which in turn depends on the velocity dispersion of stars in the cluster.  

Based on such considerations, we conclude that adding dynamics to our models should not change the main take-home of Section~\ref{sec:kicks}: a  large population of dormant BHs with orbital period $>10^4$ days would strongly favor models where BHs are born with low natal kicks, possibly modulated by  fallback. Otherwise, either natal kicks are relatively large for dormant BHs, or most dormant BHs form in massive ($\gg{}10^3$ M$_\odot$) dense clusters, where dormant BH binaries with $P_{\rm orb}>10^4$ days are soft and get destroyed. \Gaia{} DR4 will possibly validate this hypothesis.


\subsection{Additional remarks}

We have adopted the population synthesis code \sevn{} with stellar tracks from \textsc{parsec} \citep{costa2025}. We also reran several models with the \textsc{mist} tracks \citep{Choi+2016} and found no significant differences. Population-synthesis codes like \sevn{} do not model the response of the donor star during mass transfer. This might affect the long-term evolution of a binary system after mass transfer. In a follow-up study, we will run a comparison between our \sevn{} models and \textsc{mesa} to quantify possible deviations.

Alternative treatments for the transport of angular momentum during non-conservative mass transfer including the possibility that mass is lost via the second Lagrangian point \citep{macleod2018} and the possibility that the excess of angular momentum is redistributed back to the orbit via tides \citep{paczynski1991,popham1991}. The latter scenario should lead to results qualitatively similar to the Jeans mode for angular momentum transport, widening the orbit rather than shrinking it. 

 We have found that models with stable mass transfer might provide a better description of \Gaia{} BH1 and BH2 compared to the $\alpha{}$ formalism for CE evolution with $\alpha{}<1$. This result  does not mean that we propose that  mass-transfer  between a $\gtrsim{}30$ M$_\odot$ donor and a $\sim{}0.5-1.0$ M$_\odot$ accretor should be stable. Rather, it confirms that the $\alpha{}$ formalism is not a good-enough description for the orbital evolution of a binary system undergoing unstable mass transfer. Alternative models, such as the  $\gamma{}$ formalism \citep{nelemans2000, nelemans2005} and the recent double-stage CE evolution described by \cite{hirai2022}, might provide final orbital-period distribution in between our stable mass transfer models and the $\alpha{}$ formalism. For example, \cite{hirai2022} describe CE evolution as  a dynamically unstable mass-transfer phase until the convective layers of the envelope are ejected, followed by a stable mass-transfer phase when the accretor reaches the deeper radiative layers of the donor. In most cases, this mechanism results in wider final orbital separations than the $\alpha{}$ formalism.
 
 {Here, we have considered three models to infer the BH mass from the final properties of the stellar progenitor: the  rapid and delayed models from \citet{fryer2012} and the compactness model based on \cite{oconnor2011}. These models provide a very approximate description, as the compact-object mass is still largely an open question in core-collapse supernova physics \citep[see, e.g.,][for a critical discussion of the uncertainties]{burrows2021,maltsev2025}. Hence, our results for the BH mass should be regarded as an approximated phenomenological approach.}



\section{Summary}

We have presented an extensive population synthesis study of the three \Gaia{} BHs with our code \sevn{}. Our main results can be summarized as follows.

\begin{itemize}

\item[$\bullet$]  Natal kicks have a crucial impact on the population of dormant BHs. Kicks modulated by fallback result in a large population of dormant BHs with long orbital periods ($P_{\rm orb}>10^4$ days), 
{and} relatively large BH masses ($>12$ M$_\odot$). 
\Gaia{} DR4 will probe the existence of these {long orbital period 
systems}, providing constraints on the natal kicks of BHs. 

\item[$\bullet$] Models in which the first mass-transfer episode (the one in which the BH progenitor is filling its Roche lobe) is stable, highly non-conservative ($f_{\rm a}\sim{0}$), and the angular momentum is lost as a wind from the donor's surface (Jeans mode) overall maximize the probability of forming \Gaia{} BH-like systems (Table~\ref{tab:param}) in terms of both orbital period and eccentricity, because such assumptions prevent the initial orbit from shrinking too much. This is a common feature for the three \Gaia{} BHs discovered to date.

\item[$\bullet$] Requesting that mass transfer is always stable instead of allowing the onset of CE evolution  helps to produce systems with the orbital periods of \Gaia{} BH1 and BH2 especially --but not only-- if combined with high natal kicks (DM25).

\item[$\bullet$] The high eccentricity  of 
the three \Gaia{} BHs 
is  hard to match, unless we assume \textcolor{black}{relatively large natal kicks (DM25)}, almost zero accretion efficiency,  and Jeans mode for angular momentum transport.

\item[$\bullet$] If we allow for CE evolution described with the traditional $\alpha{}$ formalism, 
values of $\alpha<1$ are  unable to produce dormant systems similar to the observed \Gaia{} BHs. In contrast, values of $\alpha\geq{}1$ (together with DM25 kicks) yield qualitatively similar results to our runs with stable mass transfer, low accretion efficiency ($f_{\rm a}=0$) and Jeans mode for angular momentum loss. 
This result  confirms that the traditional description of  a CE  phase (with $\alpha{}<1$) 
provides a poor representation of the orbital evolution of the progenitors of dormant BHs.

\item[$\bullet$] At near solar metallicity, the mass function of dormant BHs peaks at $8-10$ M$_\odot$  if we assume the rapid CCSN model. In contrast, the delayed CCSN model and a model based on the compactness criterion tend to produce 
many more systems with lower-mass BHs (3--8 M$_\odot$) in the orbital period and eccentricity regime of \Gaia{} BH1 and BH2.


\item[$\bullet$] According to a simple estimate, our models result in a detection rate of \textcolor{black}{$\mathcal{O}(10-100)$} dormant BHs in \Gaia{} DR3. Models in which natal kicks are low result in the largest detection rates and are possibly already in tension with \Gaia{} DR3. We will perform more detailed studies in a follow-up paper.

\end{itemize}

Our results show that \Gaia{} systems might provide key constraints on the formation channels of BHs, complementary to gravitational-wave data. Scenarios with low natal kicks (e.g., modulated by fallback) for BHs will be likely rejected if \Gaia{} DR4 does not include a population of {large orbital period} 
systems. Moreover, the efficiency of mass transfer and the mechanism of angular momentum transport during mass transfer also play a key role in shaping the orbital properties of \Gaia{} BHs. A major question mark is represented by the description of orbital evolution during unstable mass transfer, as our models confirm that the $\alpha{}$ formalism is a poor description of the process. In future work, we will extend our analysis to include dynamical interactions in star clusters, which  drastically increase the parameter space and add further challenges to the numerical modeling of dormant compact objects.


\section*{Data availability}
\textsc{sevn} \citep{iorio2023} is available via GitLab following \href{https://gitlab.com/sevncodes/sevn}{this link}. Here, we have used version 
V2.16 (commit 27d4c4e). 
\textsc{figaro} \citep{rinaldi2024} is {publicly available at \href{https://github.com/sterinaldi/FIGARO}{this link} and via \href{https://pypi.org/project/figaro/}{\texttt{pip}}.} 
The main simulation outputs used in this paper and several additional figures are available in Zenodo at \href{https://doi.org/10.5281/zenodo.20965706}{this link}.

\begin{acknowledgements}

We thank Lucas de S\'a, Laura Hegeler, and Hans-Walter Rix for useful discussions. We acknowledge support from the Deutsche Forschungsgemeinschaft (DFG, German Research Foundation) through project number 546850815 (acronym: DoBlack) and under Germany's Excellence Strategy EXC 2181/1 - 390900948 (the Heidelberg STRUCTURES Excellence Cluster). JMH acknowledges support from the European Research Council for the ERC Advanced Grant [101054731]. CS acknowledges financial support from the Alexander von Humboldt Foundation for the Humboldt Research Fellowship.
GI is supported by a fellowship grant from la Caixa Foundation (ID 100010434). The fellowship code is LCF/BQ/PI24/12040020. The simulations were performed thanks to the support of the state of Baden-W\"urttemberg through bwHPC and the DFG through grants INST 35/1597-1 FUGG and INST 35/1503-1 FUGG. 

This research made use of \texttt{cogsworth} and its dependencies \citep{cogsworth:joss,wagg2025,Breivik+2020,gala}. The \texttt{cogsworth} population used a star formation history model based on the following papers \citep{Wagg+2022,Frankel+2018,Bovy+2016,Bovy+2019,McMillan+2011}. Population observables were estimated using dust maps and MIST isochrones \citep{Dotter+2016,Choi+2016,paxton2011,Paxton+2013,Paxton+2015,paxton2018,bayestar2019}.

This research made use of \textsc{NumPy} \citep{Harris20}, \textsc{SciPy} \citep{SciPy2020}, \textsc{Pandas} \citep{Pandas2024}. Plots were produced using \textsc{Matplotlib} \citep{Hunter2007}.

\end{acknowledgements}

\bibliographystyle{aa} 
\bibliography{bibliography}

@ARTICLE{marin2026,
       author = {{Mar{\'\i}n Pina}, Daniel and {Gieles}, Mark and {Rastello}, Sara and {Garcia-Diago}, Cl{\`a}udia and {Iorio}, Giuliano and {Ard{\`e}vol}, Marc},
        title = "{$N$-body modelling of the ED-2 stream progenitor shows Gaia BH3's formation involved dynamical interactions}",
      journal = {arXiv e-prints},
         year = 2026,
        month = apr,
          eid = {arXiv:2604.24874},
        pages = {arXiv:2604.24874},
          doi = {10.48550/arXiv.2604.24874},
archivePrefix = {arXiv},
       eprint = {2604.24874},
 primaryClass = {astro-ph.GA},
       adsurl = {https://ui.adsabs.harvard.edu/abs/2026arXiv260424874M}
}

@ARTICLE{liz2026,
       author = {{Li}, Zhenwei and {Wei}, Dandan and {Jia}, Shi and {Chen}, Hailiang and {Ge}, Hongwei and {Chen}, Zhuo and {Zhang}, Yangyang and {Chen}, Xuefei and {Han}, Zhanwen},
        title = "{A Path to Constraints on Common Envelope Ejection in Massive Binaries: Full Evolutionary Reconstruction of Three Black Hole X-Ray Binaries}",
      journal = {\apj},
         year = 2026,
        month = jun,
       volume = {1004},
       number = {1},
          eid = {31},
        pages = {31},
          doi = {10.3847/1538-4357/ae66fd},
archivePrefix = {arXiv},
       eprint = {2604.10440},
 primaryClass = {astro-ph.SR},
       adsurl = {https://ui.adsabs.harvard.edu/abs/2026ApJ..1004...31L}
}

@ARTICLE{shikauchi2023,
       author = {{Shikauchi}, Minori and {Tsuna}, Daichi and {Tanikawa}, Ataru and {Kawanaka}, Norita},
        title = "{Spatial and Binary Parameter Distributions of Black Hole Binaries in the Milky Way Detectable with Gaia}",
      journal = {\apj},
         year = 2023,
        month = aug,
       volume = {953},
       number = {1},
          eid = {52},
        pages = {52},
          doi = {10.3847/1538-4357/acd752},
archivePrefix = {arXiv},
       eprint = {2301.07207},
 primaryClass = {astro-ph.GA},
       adsurl = {https://ui.adsabs.harvard.edu/abs/2023ApJ...953...52S}
}

@ARTICLE{chen2015,
       author = {{Chen}, Yang and {Bressan}, Alessandro and {Girardi}, L{\'e}o and {Marigo}, Paola and {Kong}, Xu and {Lanza}, Antonio},
        title = "{PARSEC evolutionary tracks of massive stars up to 350 M$_{☉}$ at metallicities 0.0001 {\ensuremath{\leq}} Z {\ensuremath{\leq}} 0.04}",
      journal = {\mnras},
         year = 2015,
        month = sep,
       volume = {452},
       number = {1},
        pages = {1068-1080},
          doi = {10.1093/mnras/stv1281},
archivePrefix = {arXiv},
       eprint = {1506.01681},
 primaryClass = {astro-ph.SR},
       adsurl = {https://ui.adsabs.harvard.edu/abs/2015MNRAS.452.1068C}
}

@ARTICLE{graefener2008,
       author = {{Gr{\"a}fener}, G. and {Hamann}, W.-R.},
        title = "{Mass loss from late-type WN stars and its Z-dependence. Very massive stars approaching the Eddington limit}",
      journal = {\aap},
         year = 2008,
        month = may,
       volume = {482},
       number = {3},
        pages = {945-960},
          doi = {10.1051/0004-6361:20066176},
archivePrefix = {arXiv},
       eprint = {0803.0866},
 primaryClass = {astro-ph},
       adsurl = {https://ui.adsabs.harvard.edu/abs/2008A&A...482..945G}
}

@ARTICLE{maltsev2025,
       author = {{Maltsev}, K. and {Schneider}, F.~R.~N. and {Mandel}, I. and {M{\"u}ller}, B. and {Heger}, A. and {R{\"o}pke}, F.~K. and {Laplace}, E.},
        title = "{Explodability criteria for the neutrino-driven supernova mechanism}",
      journal = {\aap},
         year = 2025,
        month = aug,
       volume = {700},
          eid = {A20},
        pages = {A20},
          doi = {10.1051/0004-6361/202554931},
archivePrefix = {arXiv},
       eprint = {2503.23856},
 primaryClass = {astro-ph.SR},
       adsurl = {https://ui.adsabs.harvard.edu/abs/2025A&A...700A..20M}
}

@ARTICLE{burrows2021,
       author = {{Burrows}, A. and {Vartanyan}, D.},
        title = "{Core-collapse supernova explosion theory}",
      journal = {\nat},
         year = 2021,
        month = jan,
       volume = {589},
       number = {7840},
        pages = {29-39},
          doi = {10.1038/s41586-020-03059-w},
archivePrefix = {arXiv},
       eprint = {2009.14157},
 primaryClass = {astro-ph.SR},
       adsurl = {https://ui.adsabs.harvard.edu/abs/2021Natur.589...29B}
}

@ARTICLE{repetto2017,
       author = {{Repetto}, Serena and {Igoshev}, Andrei P. and {Nelemans}, Gijs},
        title = "{The Galactic distribution of X-ray binaries and its implications for compact object formation and natal kicks}",
      journal = {\mnras},
         year = 2017,
        month = may,
       volume = {467},
       number = {1},
        pages = {298-310},
          doi = {10.1093/mnras/stx027},
archivePrefix = {arXiv},
       eprint = {1701.01347},
 primaryClass = {astro-ph.HE},
       adsurl = {https://ui.adsabs.harvard.edu/abs/2017MNRAS.467..298R}
}

@ARTICLE{fragos2009,
       author = {{Fragos}, T. and {Willems}, B. and {Kalogera}, V. and {Ivanova}, N. and {Rockefeller}, G. and {Fryer}, C.~L. and {Young}, P.~A.},
        title = "{Understanding Compact Object Formation and Natal Kicks. II. The Case of XTE J1118 + 480}",
      journal = {\apj},
         year = 2009,
        month = jun,
       volume = {697},
       number = {2},
        pages = {1057-1070},
          doi = {10.1088/0004-637X/697/2/1057},
archivePrefix = {arXiv},
       eprint = {0809.1588},
 primaryClass = {astro-ph},
       adsurl = {https://ui.adsabs.harvard.edu/abs/2009ApJ...697.1057F}
}

@ARTICLE{sgalletta2025,
       author = {{Sgalletta}, Cecilia and {Mapelli}, Michela and {Boco}, Lumen and {Santoliquido}, Filippo and {Artale}, M. Celeste and {Iorio}, Giuliano and {Lapi}, Andrea and {Spera}, Mario},
        title = "{The more accurately the metal-dependent star formation rate is modeled, the larger the predicted excess of binary black hole mergers}",
      journal = {\aap},
         year = 2025,
        month = jun,
       volume = {698},
          eid = {A144},
        pages = {A144},
          doi = {10.1051/0004-6361/202452757},
archivePrefix = {arXiv},
       eprint = {2410.21401},
 primaryClass = {astro-ph.HE},
       adsurl = {https://ui.adsabs.harvard.edu/abs/2025A&A...698A.144S}
}

@ARTICLE{boesky2024,
       author = {{Boesky}, Adam P. and {Broekgaarden}, Floor S. and {Berger}, Edo},
        title = "{Investigating the Cosmological Rate of Compact Object Mergers from Isolated Massive Binary Stars}",
      journal = {\apj},
         year = 2024,
        month = nov,
       volume = {976},
       number = {1},
          eid = {24},
        pages = {24},
          doi = {10.3847/1538-4357/ad7fe3},
archivePrefix = {arXiv},
       eprint = {2405.01630},
 primaryClass = {astro-ph.HE},
       adsurl = {https://ui.adsabs.harvard.edu/abs/2024ApJ...976...24B}
}

@ARTICLE{hurley2000,
       author = {{Hurley}, Jarrod R. and {Pols}, Onno R. and {Tout}, Christopher A.},
        title = "{Comprehensive analytic formulae for stellar evolution as a function of mass and metallicity}",
      journal = {\mnras},
         year = 2000,
        month = jul,
       volume = {315},
       number = {3},
        pages = {543-569},
          doi = {10.1046/j.1365-8711.2000.03426.x},
archivePrefix = {arXiv},
       eprint = {astro-ph/0001295},
 primaryClass = {astro-ph},
       adsurl = {https://ui.adsabs.harvard.edu/abs/2000MNRAS.315..543H}
}

@ARTICLE{pauli2026,
       author = {{Pauli}, D. and {Langer}, N. and {Schootemeijer}, A. and {Marchant}, P. and {Jin}, H. and {Ercolino}, A. and {Picco}, A. and {Willcox}, R. and {Sana}, H.},
        title = "{The drastic impact of Eddington-limit induced mass ejections on massive star populations}",
      journal = {\aap},
         year = 2026,
        month = feb,
       volume = {707},
          eid = {A11},
        pages = {A11},
          doi = {10.1051/0004-6361/202557766},
archivePrefix = {arXiv},
       eprint = {2601.08822},
 primaryClass = {astro-ph.SR},
       adsurl = {https://ui.adsabs.harvard.edu/abs/2026A&A...707A..11P}
}

@ARTICLE{romagnolo2026,
       author = {{Romagnolo}, Amedeo and {Broekgaarden}, Floor S. and {Antoniadis}, Konstantinos and {Gormaz-Matamala}, Alex C.},
        title = "{The Stellar Winds Atlas I: Current uncertainties in mass-loss rates}",
      journal = {arXiv e-prints},
         year = 2026,
        month = jan,
          eid = {arXiv:2601.02263},
        pages = {arXiv:2601.02263},
          doi = {10.48550/arXiv.2601.02263},
archivePrefix = {arXiv},
       eprint = {2601.02263},
 primaryClass = {astro-ph.SR},
       adsurl = {https://ui.adsabs.harvard.edu/abs/2026arXiv260102263R}
}

@ARTICLE{boco2025,
       author = {{Boco}, Lumen and {Mapelli}, Michela and {Sander}, Andreas A.~C. and {Mesini}, Sofia and {Ramachandran}, Varsha and {Torniamenti}, Stefano and {Korb}, Erika and {Liu}, Boyuan and {Sabhahit}, Gautham N. and {Vink}, Jorick S.},
        title = "{Metal-poor single Wolf-Rayet stars: The interplay of optically thick winds and rotation}",
      journal = {\aap},
         year = 2025,
        month = nov,
       volume = {703},
          eid = {A243},
        pages = {A243},
          doi = {10.1051/0004-6361/202556187},
archivePrefix = {arXiv},
       eprint = {2507.00137},
 primaryClass = {astro-ph.SR},
       adsurl = {https://ui.adsabs.harvard.edu/abs/2025A&A...703A.243B}
}

@ARTICLE{sabhahit2023,
       author = {{Sabhahit}, Gautham N. and {Vink}, Jorick S. and {Sander}, Andreas A.~C. and {Higgins}, Erin R.},
        title = "{Very massive stars and pair-instability supernovae: mass-loss framework for low metallicity}",
      journal = {\mnras},
         year = 2023,
        month = sep,
       volume = {524},
       number = {1},
        pages = {1529-1546},
          doi = {10.1093/mnras/stad1888},
archivePrefix = {arXiv},
       eprint = {2306.11785},
 primaryClass = {astro-ph.SR},
       adsurl = {https://ui.adsabs.harvard.edu/abs/2023MNRAS.524.1529S}
}

@ARTICLE{zorotovic2011,
       author = {{Zorotovic}, M. and {Schreiber}, M.~R. and {G{\"a}nsicke}, B.~T.},
        title = "{Post common envelope binaries from SDSS. XI. The white dwarf mass distributions of CVs and pre-CVs}",
      journal = {\aap},
         year = 2011,
        month = dec,
       volume = {536},
          eid = {A42},
        pages = {A42},
          doi = {10.1051/0004-6361/201116626},
archivePrefix = {arXiv},
       eprint = {1108.4600},
 primaryClass = {astro-ph.SR},
       adsurl = {https://ui.adsabs.harvard.edu/abs/2011A&A...536A..42Z}
}

@ARTICLE{muellerhorn2026,
       author = {{M{\"u}ller-Horn}, Johanna and {Ramachandran}, Varsha and {El-Badry}, Kareem and {Sander}, Andreas A.~C. and {Bodensteiner}, Julia and {Gies}, Douglas R. and {G{\"o}tberg}, Ylva and {Rivinius}, Thomas and {Shenar}, Tomer and {Sch{\"o}sser}, Elisa C. and {Wang}, Luqian and {Bieryla}, Allyson and {Buchhave}, Lars A. and {Latham}, David W.},
        title = "{Ultraviolet spectroscopy reveals a hot and luminous companion to the Be star+black hole candidate MWC 656}",
      journal = {\aap},
         year = 2026,
        month = apr,
       volume = {708},
          eid = {A187},
        pages = {A187},
          doi = {10.1051/0004-6361/202557960},
archivePrefix = {arXiv},
       eprint = {2601.14403},
 primaryClass = {astro-ph.SR},
       adsurl = {https://ui.adsabs.harvard.edu/abs/2026A&A...708A.187M}
}

@ARTICLE{elbadry2022,
       author = {{El-Badry}, Kareem and {Burdge}, Kevin B.},
        title = "{NGC 1850 BH1 is another stripped-star binary masquerading as a black hole}",
      journal = {\mnras},
         year = 2022,
        month = mar,
       volume = {511},
       number = {1},
        pages = {24-29},
          doi = {10.1093/mnrasl/slab135},
archivePrefix = {arXiv},
       eprint = {2111.07925},
 primaryClass = {astro-ph.SR},
       adsurl = {https://ui.adsabs.harvard.edu/abs/2022MNRAS.511L..24E}
}

@ARTICLE{wagg2025,
       author = {{Wagg}, Tom and {Breivik}, Katelyn and {Renzo}, Mathieu and {Price-Whelan}, Adrian M.},
        title = "{cogsworth: A Gala of COSMIC Proportions Combining Binary Stellar Evolution and Galactic Dynamics}",
      journal = {\apjs},
         year = 2025,
        month = jan,
       volume = {276},
       number = {1},
          eid = {16},
        pages = {16},
          doi = {10.3847/1538-4365/ad8b1f},
archivePrefix = {arXiv},
       eprint = {2409.04543},
 primaryClass = {astro-ph.GA},
       adsurl = {https://ui.adsabs.harvard.edu/abs/2025ApJS..276...16W}
}

@ARTICLE{rinaldi2024,
       author = {{Rinaldi}, Stefano and {Del Pozzo}, Walter},
        title = "{FIGARO: hierarchical non-parametric inference for population studies}",
      journal = {The Journal of Open Source Software},
         year = 2024,
        month = may,
       volume = {9},
       number = {97},
          eid = {6589},
        pages = {6589},
          doi = {10.21105/joss.06589},
       adsurl = {https://ui.adsabs.harvard.edu/abs/2024JOSS....9.6589R}
}

@ARTICLE{mapelli2018,
       author = {{Mapelli}, Michela and {Giacobbo}, Nicola},
        title = "{The cosmic merger rate of neutron stars and black holes}",
      journal = {\mnras},
         year = 2018,
        month = oct,
       volume = {479},
       number = {4},
        pages = {4391-4398},
          doi = {10.1093/mnras/sty1613},
archivePrefix = {arXiv},
       eprint = {1806.04866},
 primaryClass = {astro-ph.HE},
       adsurl = {https://ui.adsabs.harvard.edu/abs/2018MNRAS.479.4391M}
}

@ARTICLE{oconnor2011,
       author = {{O'Connor}, Evan and {Ott}, Christian D.},
        title = "{Black Hole Formation in Failing Core-Collapse Supernovae}",
      journal = {\apj},
         year = 2011,
        month = apr,
       volume = {730},
       number = {2},
          eid = {70},
        pages = {70},
          doi = {10.1088/0004-637X/730/2/70},
archivePrefix = {arXiv},
       eprint = {1010.5550},
 primaryClass = {astro-ph.HE},
       adsurl = {https://ui.adsabs.harvard.edu/abs/2011ApJ...730...70O}
}

@ARTICLE{fragos2019,
       author = {{Fragos}, Tassos and {Andrews}, Jeff J. and {Ramirez-Ruiz}, Enrico and {Meynet}, Georges and {Kalogera}, Vicky and {Taam}, Ronald E. and {Zezas}, Andreas},
        title = "{The Complete Evolution of a Neutron-star Binary through a Common Envelope Phase Using 1D Hydrodynamic Simulations}",
      journal = {\apjl},
         year = 2019,
        month = oct,
       volume = {883},
       number = {2},
          eid = {L45},
        pages = {L45},
          doi = {10.3847/2041-8213/ab40d1},
archivePrefix = {arXiv},
       eprint = {1907.12573},
 primaryClass = {astro-ph.HE},
       adsurl = {https://ui.adsabs.harvard.edu/abs/2019ApJ...883L..45F}
}

@ARTICLE{giacobbo2018,
       author = {{Giacobbo}, Nicola and {Mapelli}, Michela},
        title = "{The progenitors of compact-object binaries: impact of metallicity, common envelope and natal kicks}",
      journal = {\mnras},
         year = 2018,
        month = oct,
       volume = {480},
       number = {2},
        pages = {2011-2030},
          doi = {10.1093/mnras/sty1999},
archivePrefix = {arXiv},
       eprint = {1806.00001},
 primaryClass = {astro-ph.HE},
       adsurl = {https://ui.adsabs.harvard.edu/abs/2018MNRAS.480.2011G}
}

@ARTICLE{generozov2024,
       author = {{Generozov}, A. and {Perets}, H.~B.},
        title = "{A Triple Scenario for the Formation of Wide Black Hole Binaries Such as Gaia BH1}",
      journal = {\apj},
         year = 2024,
        month = mar,
       volume = {964},
       number = {1},
          eid = {83},
        pages = {83},
          doi = {10.3847/1538-4357/ad2356},
archivePrefix = {arXiv},
       eprint = {2312.03066},
 primaryClass = {astro-ph.SR},
       adsurl = {https://ui.adsabs.harvard.edu/abs/2024ApJ...964...83G}
}

@ARTICLE{tanikawa2024NS,
       author = {{Tanikawa}, Ataru and {Wang}, Long and {Fujii}, Michiko S.},
        title = "{Compact Binary Formation in Open Star Clusters II: Difficulty of Gaia NS formation in low-mass star clusters}",
      journal = {The Open Journal of Astrophysics},
         year = 2024,
        month = may,
       volume = {7},
          eid = {39},
        pages = {39},
          doi = {10.33232/001c.117684},
archivePrefix = {arXiv},
       eprint = {2404.01731},
 primaryClass = {astro-ph.SR},
       adsurl = {https://ui.adsabs.harvard.edu/abs/2024OJAp....7E..39T}
}

@ARTICLE{kotko2024,
       author = {{Kotko}, I. and {Banerjee}, S. and {Belczynski}, K.},
        title = "{The enigmatic origin of two dormant BH binaries: Gaia BH1 and Gaia BH2}",
      journal = {\mnras},
         year = 2024,
        month = dec,
       volume = {535},
       number = {4},
        pages = {3577-3594},
          doi = {10.1093/mnras/stae2591},
archivePrefix = {arXiv},
       eprint = {2403.13579},
 primaryClass = {astro-ph.SR},
       adsurl = {https://ui.adsabs.harvard.edu/abs/2024MNRAS.535.3577K}
}

@ARTICLE{fantoccoli2025,
       author = {{Fantoccoli}, Federico and {Barber}, Jordan and {Dosopoulou}, Fani and {Chattopadhyay}, Debatri and {Antonini}, Fabio},
        title = "{Properties of black hole-star binaries formed in N-body simulations of massive star clusters: implications for Gaia black holes}",
      journal = {\mnras},
         year = 2025,
        month = mar,
       volume = {538},
       number = {1},
        pages = {243-257},
          doi = {10.1093/mnras/staf303},
archivePrefix = {arXiv},
       eprint = {2410.17323},
 primaryClass = {astro-ph.GA},
       adsurl = {https://ui.adsabs.harvard.edu/abs/2025MNRAS.538..243F}
}

@ARTICLE{chawla2025,
       author = {{Chawla}, Chirag and {Chatterjee}, Sourav and {Breivik}, Katelyn},
        title = "{Gaia's promise to detect compact-object binaries: where we stand with the third data release}",
      journal = {arXiv e-prints},
         year = 2025,
        month = aug,
          eid = {arXiv:2508.21805},
        pages = {arXiv:2508.21805},
          doi = {10.48550/arXiv.2508.21805},
archivePrefix = {arXiv},
       eprint = {2508.21805},
 primaryClass = {astro-ph.SR},
       adsurl = {https://ui.adsabs.harvard.edu/abs/2025arXiv250821805C}
}

@ARTICLE{willcox2025,
       author = {{Willcox}, R. and {Marchant}, P. and {Vigna-G{\'o}mez}, A. and {Sana}, H. and {Bodensteiner}, J. and {Deshmukh}, K. and {Esseldeurs}, M. and {Fabry}, M. and {H{\'e}nault-Brunet}, V. and {Janssens}, S. and {Mahy}, L. and {Patrick}, L. and {Pauli}, D. and {Renzo}, M. and {Sander}, A.~A.~C. and {Shenar}, T. and {van Son}, L.~A.~C. and {Stoop}, M.},
        title = "{Binarity at LOw Metallicity (BLOeM): Bayesian inference of natal kicks from inert black hole binaries}",
      journal = {\aap},
         year = 2025,
        month = aug,
       volume = {700},
          eid = {A59},
        pages = {A59},
          doi = {10.1051/0004-6361/202555274},
archivePrefix = {arXiv},
       eprint = {2504.16669},
 primaryClass = {astro-ph.SR},
       adsurl = {https://ui.adsabs.harvard.edu/abs/2025A&A...700A..59W}
}

@ARTICLE{wiktorowicz2020,
       author = {{Wiktorowicz}, Grzegorz and {Lu}, Youjun and {Wyrzykowski}, {\L}ukasz and {Zhang}, Haotong and {Liu}, Jifeng and {Justham}, Stephen and {Belczynski}, Krzysztof},
        title = "{Noninteracting Black Hole Binaries with Gaia and LAMOST}",
      journal = {\apj},
         year = 2020,
        month = dec,
       volume = {905},
       number = {2},
          eid = {134},
        pages = {134},
          doi = {10.3847/1538-4357/abc699},
archivePrefix = {arXiv},
       eprint = {2006.08317},
 primaryClass = {astro-ph.HE},
       adsurl = {https://ui.adsabs.harvard.edu/abs/2020ApJ...905..134W}
}

@ARTICLE{hayashi2023,
       author = {{Hayashi}, Toshinori and {Suto}, Yasushi and {Trani}, Alessandro A.},
        title = "{Constraining the Binarity of Black Hole Candidates: A Proof-of-concept Study of Gaia BH1 and Gaia BH2}",
      journal = {\apj},
         year = 2023,
        month = nov,
       volume = {958},
       number = {1},
          eid = {26},
        pages = {26},
          doi = {10.3847/1538-4357/acf4f6},
archivePrefix = {arXiv},
       eprint = {2307.01793},
 primaryClass = {astro-ph.HE},
       adsurl = {https://ui.adsabs.harvard.edu/abs/2023ApJ...958...26H}
}

@ARTICLE{atri2019,
       author = {{Atri}, P. and {Miller-Jones}, J.~C.~A. and {Bahramian}, A. and {Plotkin}, R.~M. and {Jonker}, P.~G. and {Nelemans}, G. and {Maccarone}, T.~J. and {Sivakoff}, G.~R. and {Deller}, A.~T. and {Chaty}, S. and {Torres}, M.~A.~P. and {Horiuchi}, S. and {McCallum}, J. and {Natusch}, T. and {Phillips}, C.~J. and {Stevens}, J. and {Weston}, S.},
        title = "{Potential kick velocity distribution of black hole X-ray binaries and implications for natal kicks}",
      journal = {\mnras},
         year = 2019,
        month = nov,
       volume = {489},
       number = {3},
        pages = {3116-3134},
          doi = {10.1093/mnras/stz2335},
archivePrefix = {arXiv},
       eprint = {1908.07199},
 primaryClass = {astro-ph.HE},
       adsurl = {https://ui.adsabs.harvard.edu/abs/2019MNRAS.489.3116A}
}

@ARTICLE{elbadry2024,
       author = {{El-Badry}, Kareem},
        title = "{Gaia's binary star renaissance}",
      journal = {\nar},
         year = 2024,
        month = jun,
       volume = {98},
          eid = {101694},
        pages = {101694},
          doi = {10.1016/j.newar.2024.101694},
archivePrefix = {arXiv},
       eprint = {2403.12146},
 primaryClass = {astro-ph.SR},
       adsurl = {https://ui.adsabs.harvard.edu/abs/2024NewAR..9801694E}
}

@ARTICLE{breivik2017,
       author = {{Breivik}, Katelyn and {Chatterjee}, Sourav and {Larson}, Shane L.},
        title = "{Revealing Black Holes with Gaia}",
      journal = {\apjl},
         year = 2017,
        month = nov,
       volume = {850},
       number = {1},
          eid = {L13},
        pages = {L13},
          doi = {10.3847/2041-8213/aa97d5},
archivePrefix = {arXiv},
       eprint = {1710.04657},
 primaryClass = {astro-ph.SR},
       adsurl = {https://ui.adsabs.harvard.edu/abs/2017ApJ...850L..13B}
}

@ARTICLE{chawla2022,
       author = {{Chawla}, Chirag and {Chatterjee}, Sourav and {Breivik}, Katelyn and {Moorthy}, Chaithanya Krishna and {Andrews}, Jeff J. and {Sanderson}, Robyn E.},
        title = "{Gaia May Detect Hundreds of Well-characterized Stellar Black Holes}",
      journal = {\apj},
         year = 2022,
        month = jun,
       volume = {931},
       number = {2},
          eid = {107},
        pages = {107},
          doi = {10.3847/1538-4357/ac60a5},
archivePrefix = {arXiv},
       eprint = {2110.05979},
 primaryClass = {astro-ph.GA},
       adsurl = {https://ui.adsabs.harvard.edu/abs/2022ApJ...931..107C}
}

@ARTICLE{shahaf2023,
       author = {{Shahaf}, S. and {Bashi}, D. and {Mazeh}, T. and {Faigler}, S. and {Arenou}, F. and {El-Badry}, K. and {Rix}, H.~W.},
        title = "{Triage of the Gaia DR3 astrometric orbits - I. A sample of binaries with probable compact companions}",
      journal = {\mnras},
         year = 2023,
        month = jan,
       volume = {518},
       number = {2},
        pages = {2991-3003},
          doi = {10.1093/mnras/stac3290},
archivePrefix = {arXiv},
       eprint = {2209.00828},
 primaryClass = {astro-ph.SR},
       adsurl = {https://ui.adsabs.harvard.edu/abs/2023MNRAS.518.2991S}
}

@ARTICLE{shikauchi2020,
       author = {{Shikauchi}, Minori and {Kumamoto}, Jun and {Tanikawa}, Ataru and {Fujii}, Michiko S.},
        title = "{Gaia's detectability of black hole-main sequence star binaries formed in open clusters}",
      journal = {\pasj},
         year = 2020,
        month = jun,
       volume = {72},
       number = {3},
          eid = {45},
        pages = {45},
          doi = {10.1093/pasj/psaa030},
archivePrefix = {arXiv},
       eprint = {2001.11199},
 primaryClass = {astro-ph.HE},
       adsurl = {https://ui.adsabs.harvard.edu/abs/2020PASJ...72...45S}
}

@ARTICLE{shikauchi2022,
       author = {{Shikauchi}, Minori and {Tanikawa}, Ataru and {Kawanaka}, Norita},
        title = "{Detectability of Black Hole Binaries with Gaia: Dependence on Binary Evolution Models}",
      journal = {\apj},
         year = 2022,
        month = mar,
       volume = {928},
       number = {1},
          eid = {13},
        pages = {13},
          doi = {10.3847/1538-4357/ac5329},
archivePrefix = {arXiv},
       eprint = {2112.04798},
 primaryClass = {astro-ph.HE},
       adsurl = {https://ui.adsabs.harvard.edu/abs/2022ApJ...928...13S}
}

@ARTICLE{tanikawa2024,
       author = {{Tanikawa}, Ataru and {Cary}, Savannah and {Shikauchi}, Minori and {Wang}, Long and {Fujii}, Michiko S.},
        title = "{Compact binary formation in open star clusters - I. High formation efficiency of Gaia BHs and their multiplicities}",
      journal = {\mnras},
         year = 2024,
        month = jan,
       volume = {527},
       number = {2},
        pages = {4031-4039},
          doi = {10.1093/mnras/stad3294},
archivePrefix = {arXiv},
       eprint = {2303.05743},
 primaryClass = {astro-ph.GA},
       adsurl = {https://ui.adsabs.harvard.edu/abs/2024MNRAS.527.4031T}
}

@ARTICLE{Harris20,
       author = {{Harris}, Charles R. and {Millman}, K. Jarrod and {van der Walt}, St{\'e}fan J. and {Gommers}, Ralf and {Virtanen}, Pauli and {Cournapeau}, David and {Wieser}, Eric and {Taylor}, Julian and {Berg}, Sebastian and {Smith}, Nathaniel J. and {Kern}, Robert and {Picus}, Matti and {Hoyer}, Stephan and {van Kerkwijk}, Marten H. and {Brett}, Matthew and {Haldane}, Allan and {del R{\'\i}o}, Jaime Fern{\'a}ndez and {Wiebe}, Mark and {Peterson}, Pearu and {G{\'e}rard-Marchant}, Pierre and {Sheppard}, Kevin and {Reddy}, Tyler and {Weckesser}, Warren and {Abbasi}, Hameer and {Gohlke}, Christoph and {Oliphant}, Travis E.},
        title = "{Array programming with NumPy}",
      journal = {\nat},
         year = 2020,
        month = sep,
       volume = {585},
       number = {7825},
        pages = {357-362},
          doi = {10.1038/s41586-020-2649-2},
archivePrefix = {arXiv},
       eprint = {2006.10256},
 primaryClass = {cs.MS},
       adsurl = {https://ui.adsabs.harvard.edu/abs/2020Natur.585..357H}
}

@ARTICLE{SciPy2020,
       author = {{Virtanen}, Pauli and {Gommers}, Ralf and {Oliphant}, Travis E. and {Haberland}, Matt and {Reddy}, Tyler and {Cournapeau}, David and {Burovski}, Evgeni and {Peterson}, Pearu and {Weckesser}, Warren and {Bright}, Jonathan and {van der Walt}, St{\'e}fan J. and {Brett}, Matthew and {Wilson}, Joshua and {Millman}, K. Jarrod and {Mayorov}, Nikolay and {Nelson}, Andrew R.~J. and {Jones}, Eric and {Kern}, Robert and {Larson}, Eric and {Carey}, C.~J. and {Polat}, {\.I}lhan and {Feng}, Yu and {Moore}, Eric W. and {VanderPlas}, Jake and {Laxalde}, Denis and {Perktold}, Josef and {Cimrman}, Robert and {Henriksen}, Ian and {Quintero}, E.~A. and {Harris}, Charles R. and {Archibald}, Anne M. and {Ribeiro}, Ant{\^o}nio H. and {Pedregosa}, Fabian and {van Mulbregt}, Paul and {SciPy 1. 0 Contributors}},
        title = "{SciPy 1.0: fundamental algorithms for scientific computing in Python}",
      journal = {Nature Methods},
         year = 2020,
        month = feb,
       volume = {17},
        pages = {261-272},
          doi = {10.1038/s41592-019-0686-2},
archivePrefix = {arXiv},
       eprint = {1907.10121},
 primaryClass = {cs.MS},
       adsurl = {https://ui.adsabs.harvard.edu/abs/2020NatMe..17..261V}
}

@software{Pandas2024,
       author = {{The pandas development Team}},
        title = "{pandas-dev/pandas: Pandas}",
         year = 2024,
        month = feb,
          eid = {10.5281/zenodo.10697587},
          doi = {10.5281/zenodo.10697587},
      version = {v2.2.1},
    publisher = {Zenodo},
       adsurl = {https://ui.adsabs.harvard.edu/abs/2024zndo..10697587T}
}

@ARTICLE{Hunter2007,
       author = {{Hunter}, John D.},
        title = "{Matplotlib: A 2D Graphics Environment}",
      journal = {Computing in Science and Engineering},
         year = 2007,
        month = may,
       volume = {9},
       number = {3},
        pages = {90-95},
          doi = {10.1109/MCSE.2007.55},
       adsurl = {https://ui.adsabs.harvard.edu/abs/2007CSE.....9...90H}
}

@ARTICLE{hirai2022,
       author = {{Hirai}, Ryosuke and {Mandel}, Ilya},
        title = "{A Two-stage Formalism for Common-envelope Phases of Massive Stars}",
      journal = {\apjl},
         year = 2022,
        month = oct,
       volume = {937},
       number = {2},
          eid = {L42},
        pages = {L42},
          doi = {10.3847/2041-8213/ac9519},
archivePrefix = {arXiv},
       eprint = {2209.05328},
 primaryClass = {astro-ph.SR},
       adsurl = {https://ui.adsabs.harvard.edu/abs/2022ApJ...937L..42H}
}

@ARTICLE{nelemans2000,
       author = {{Nelemans}, G. and {Verbunt}, F. and {Yungelson}, L.~R. and {Portegies Zwart}, Simon F.},
        title = "{Reconstructing the evolution of double helium white dwarfs: envelope loss without spiral-in}",
      journal = {\aap},
         year = 2000,
        month = aug,
       volume = {360},
        pages = {1011-1018},
          doi = {10.48550/arXiv.astro-ph/0006216},
archivePrefix = {arXiv},
       eprint = {astro-ph/0006216},
 primaryClass = {astro-ph},
       adsurl = {https://ui.adsabs.harvard.edu/abs/2000A&A...360.1011N}
}

@ARTICLE{nelemans2005,
       author = {{Nelemans}, G. and {Tout}, C.~A.},
        title = "{Reconstructing the evolution of white dwarf binaries: further evidence for an alternative algorithm for the outcome of the common-envelope phase in close binaries}",
      journal = {\mnras},
         year = 2005,
        month = jan,
       volume = {356},
       number = {2},
        pages = {753-764},
          doi = {10.1111/j.1365-2966.2004.08496.x},
archivePrefix = {arXiv},
       eprint = {astro-ph/0410301},
 primaryClass = {astro-ph},
       adsurl = {https://ui.adsabs.harvard.edu/abs/2005MNRAS.356..753N}
}

@ARTICLE{popham1991,
       author = {{Popham}, Robert and {Narayan}, Ramesh},
        title = "{Does Accretion Cease When a Star Approaches Breakup?}",
      journal = {\apj},
         year = 1991,
        month = apr,
       volume = {370},
        pages = {604},
          doi = {10.1086/169847},
       adsurl = {https://ui.adsabs.harvard.edu/abs/1991ApJ...370..604P}
}

@ARTICLE{paczynski1991,
       author = {{Paczynski}, Bohdan},
        title = "{A Polytropic Model of an Accretion Disk, a Boundary Layer, and a Star}",
      journal = {\apj},
         year = 1991,
        month = apr,
       volume = {370},
        pages = {597},
          doi = {10.1086/169846},
       adsurl = {https://ui.adsabs.harvard.edu/abs/1991ApJ...370..597P}
}

@ARTICLE{macleod2018,
       author = {{MacLeod}, Morgan and {Ostriker}, Eve C. and {Stone}, James M.},
        title = "{Runaway Coalescence at the Onset of Common Envelope Episodes}",
      journal = {\apj},
         year = 2018,
        month = aug,
       volume = {863},
       number = {1},
          eid = {5},
        pages = {5},
          doi = {10.3847/1538-4357/aacf08},
archivePrefix = {arXiv},
       eprint = {1803.03261},
 primaryClass = {astro-ph.SR},
       adsurl = {https://ui.adsabs.harvard.edu/abs/2018ApJ...863....5M}
}

@ARTICLE{heggie1975,
       author = {{Heggie}, D.~C.},
        title = "{Binary evolution in stellar dynamics.}",
      journal = {\mnras},
         year = 1975,
        month = dec,
       volume = {173},
        pages = {729-787},
          doi = {10.1093/mnras/173.3.729},
       adsurl = {https://ui.adsabs.harvard.edu/abs/1975MNRAS.173..729H}
}

@ARTICLE{dicarlo2020,
       author = {{Di Carlo}, Ugo N. and {Mapelli}, Michela and {Giacobbo}, Nicola and {Spera}, Mario and {Bouffanais}, Yann and {Rastello}, Sara and {Santoliquido}, Filippo and {Pasquato}, Mario and {Ballone}, Alessandro and {Trani}, Alessandro A. and {Torniamenti}, Stefano and {Haardt}, Francesco},
        title = "{Binary black holes in young star clusters: the impact of metallicity}",
      journal = {\mnras},
         year = 2020,
        month = oct,
       volume = {498},
       number = {1},
        pages = {495-506},
          doi = {10.1093/mnras/staa2286},
archivePrefix = {arXiv},
       eprint = {2004.09525},
 primaryClass = {astro-ph.HE},
       adsurl = {https://ui.adsabs.harvard.edu/abs/2020MNRAS.498..495D}
}

@ARTICLE{lada2003,
       author = {{Lada}, Charles J. and {Lada}, Elizabeth A.},
        title = "{Embedded Clusters in Molecular Clouds}",
      journal = {\araa},
         year = 2003,
        month = jan,
       volume = {41},
        pages = {57-115},
          doi = {10.1146/annurev.astro.41.011802.094844},
archivePrefix = {arXiv},
       eprint = {astro-ph/0301540},
 primaryClass = {astro-ph},
       adsurl = {https://ui.adsabs.harvard.edu/abs/2003ARA&A..41...57L}
}

@ARTICLE{blandhawthorn2016,
       author = {{Bland-Hawthorn}, Joss and {Gerhard}, Ortwin},
        title = "{The Galaxy in Context: Structural, Kinematic, and Integrated Properties}",
      journal = {\araa},
         year = 2016,
        month = sep,
       volume = {54},
        pages = {529-596},
          doi = {10.1146/annurev-astro-081915-023441},
archivePrefix = {arXiv},
       eprint = {1602.07702},
 primaryClass = {astro-ph.GA},
       adsurl = {https://ui.adsabs.harvard.edu/abs/2016ARA&A..54..529B}
}

@ARTICLE{licquia2015,
       author = {{Licquia}, Timothy C. and {Newman}, Jeffrey A.},
        title = "{Improved Estimates of the Milky Way's Stellar Mass and Star Formation Rate from Hierarchical Bayesian Meta-Analysis}",
      journal = {\apj},
         year = 2015,
        month = jun,
       volume = {806},
       number = {1},
          eid = {96},
        pages = {96},
          doi = {10.1088/0004-637X/806/1/96},
archivePrefix = {arXiv},
       eprint = {1407.1078},
 primaryClass = {astro-ph.GA},
       adsurl = {https://ui.adsabs.harvard.edu/abs/2015ApJ...806...96L}
}

@ARTICLE{ge2020,
       author = {{Ge}, Hongwei and {Webbink}, Ronald F. and {Chen}, Xuefei and {Han}, Zhanwen},
        title = "{Adiabatic Mass Loss in Binary Stars. III. From the Base of the Red Giant Branch to the Tip of the Asymptotic Giant Branch}",
      journal = {\apj},
         year = 2020,
        month = aug,
       volume = {899},
       number = {2},
          eid = {132},
        pages = {132},
          doi = {10.3847/1538-4357/aba7b7},
archivePrefix = {arXiv},
       eprint = {2007.09848},
 primaryClass = {astro-ph.SR},
       adsurl = {https://ui.adsabs.harvard.edu/abs/2020ApJ...899..132G}
}

@ARTICLE{moe2017,
       author = {{Moe}, Maxwell and {Di Stefano}, Rosanne},
        title = "{Mind Your Ps and Qs: The Interrelation between Period (P) and Mass-ratio (Q) Distributions of Binary Stars}",
      journal = {\apjs},
         year = 2017,
        month = jun,
       volume = {230},
       number = {2},
          eid = {15},
        pages = {15},
          doi = {10.3847/1538-4365/aa6fb6},
archivePrefix = {arXiv},
       eprint = {1606.05347},
 primaryClass = {astro-ph.SR},
       adsurl = {https://ui.adsabs.harvard.edu/abs/2017ApJS..230...15M}
}

@ARTICLE{sana2012,
       author = {{Sana}, H. and {de Mink}, S.~E. and {de Koter}, A. and {Langer}, N. and {Evans}, C.~J. and {Gieles}, M. and {Gosset}, E. and {Izzard}, R.~G. and {Le Bouquin}, J.-B. and {Schneider}, F.~R.~N.},
        title = "{Binary Interaction Dominates the Evolution of Massive Stars}",
      journal = {Science},
         year = 2012,
        month = jul,
       volume = {337},
       number = {6093},
        pages = {444},
          doi = {10.1126/science.1223344},
archivePrefix = {arXiv},
       eprint = {1207.6397},
 primaryClass = {astro-ph.SR},
       adsurl = {https://ui.adsabs.harvard.edu/abs/2012Sci...337..444S}
}

@ARTICLE{kroupa2001,
       author = {{Kroupa}, Pavel},
        title = "{On the variation of the initial mass function}",
      journal = {\mnras},
         year = 2001,
        month = apr,
       volume = {322},
       number = {2},
        pages = {231-246},
          doi = {10.1046/j.1365-8711.2001.04022.x},
archivePrefix = {arXiv},
       eprint = {astro-ph/0009005},
 primaryClass = {astro-ph},
       adsurl = {https://ui.adsabs.harvard.edu/abs/2001MNRAS.322..231K}
}

@ARTICLE{ivanova2013,
       author = {{Ivanova}, N. and {Justham}, S. and {Chen}, X. and {De Marco}, O. and {Fryer}, C.~L. and {Gaburov}, E. and {Ge}, H. and {Glebbeek}, E. and {Han}, Z. and {Li}, X.-D. and {Lu}, G. and {Marsh}, T. and {Podsiadlowski}, P. and {Potter}, A. and {Soker}, N. and {Taam}, R. and {Tauris}, T.~M. and {van den Heuvel}, E.~P.~J. and {Webbink}, R.~F.},
        title = "{Common envelope evolution: where we stand and how we can move forward}",
      journal = {\aapr},
         year = 2013,
        month = feb,
       volume = {21},
          eid = {59},
        pages = {59},
          doi = {10.1007/s00159-013-0059-2},
archivePrefix = {arXiv},
       eprint = {1209.4302},
 primaryClass = {astro-ph.HE},
       adsurl = {https://ui.adsabs.harvard.edu/abs/2013A&ARv..21...59I}
}

@ARTICLE{roepke2023,
       author = {{R{\"o}pke}, Friedrich K. and {De Marco}, Orsola},
        title = "{Simulations of common-envelope evolution in binary stellar systems: physical models and numerical techniques}",
      journal = {Living Reviews in Computational Astrophysics},
         year = 2023,
        month = dec,
       volume = {9},
       number = {1},
          eid = {2},
        pages = {2},
          doi = {10.1007/s41115-023-00017-x},
archivePrefix = {arXiv},
       eprint = {2212.07308},
 primaryClass = {astro-ph.SR},
       adsurl = {https://ui.adsabs.harvard.edu/abs/2023LRCA....9....2R}
}

@ARTICLE{hurley2002,
       author = {{Hurley}, Jarrod R. and {Tout}, Christopher A. and {Pols}, Onno R.},
        title = "{Evolution of binary stars and the effect of tides on binary populations}",
      journal = {\mnras},
         year = 2002,
        month = feb,
       volume = {329},
       number = {4},
        pages = {897-928},
          doi = {10.1046/j.1365-8711.2002.05038.x},
archivePrefix = {arXiv},
       eprint = {astro-ph/0201220},
 primaryClass = {astro-ph},
       adsurl = {https://ui.adsabs.harvard.edu/abs/2002MNRAS.329..897H}
}

@ARTICLE{zevin2020,
       author = {{Zevin}, Michael and {Spera}, Mario and {Berry}, Christopher P.~L. and {Kalogera}, Vicky},
        title = "{Exploring the Lower Mass Gap and Unequal Mass Regime in Compact Binary Evolution}",
      journal = {\apjl},
         year = 2020,
        month = aug,
       volume = {899},
       number = {1},
          eid = {L1},
        pages = {L1},
          doi = {10.3847/2041-8213/aba74e},
archivePrefix = {arXiv},
       eprint = {2006.14573},
 primaryClass = {astro-ph.HE},
       adsurl = {https://ui.adsabs.harvard.edu/abs/2020ApJ...899L...1Z}
}

@ARTICLE{fryer2012,
       author = {{Fryer}, Chris L. and {Belczynski}, Krzysztof and {Wiktorowicz}, Grzegorz and {Dominik}, Michal and {Kalogera}, Vicky and {Holz}, Daniel E.},
        title = "{Compact Remnant Mass Function: Dependence on the Explosion Mechanism and Metallicity}",
      journal = {\apj},
         year = 2012,
        month = apr,
       volume = {749},
       number = {1},
          eid = {91},
        pages = {91},
          doi = {10.1088/0004-637X/749/1/91},
archivePrefix = {arXiv},
       eprint = {1110.1726},
 primaryClass = {astro-ph.SR},
       adsurl = {https://ui.adsabs.harvard.edu/abs/2012ApJ...749...91F}
}

@ARTICLE{hobbs2005,
       author = {{Hobbs}, G. and {Lorimer}, D.~R. and {Lyne}, A.~G. and {Kramer}, M.},
        title = "{A statistical study of 233 pulsar proper motions}",
      journal = {\mnras},
         year = 2005,
        month = jul,
       volume = {360},
       number = {3},
        pages = {974-992},
          doi = {10.1111/j.1365-2966.2005.09087.x},
archivePrefix = {arXiv},
       eprint = {astro-ph/0504584},
 primaryClass = {astro-ph},
       adsurl = {https://ui.adsabs.harvard.edu/abs/2005MNRAS.360..974H}
}

@ARTICLE{disberg2025,
       author = {{Disberg}, Paul and {Mandel}, Ilya},
        title = "{The Kick Velocity Distribution of Isolated Neutron Stars}",
      journal = {\apjl},
         year = 2025,
        month = aug,
       volume = {989},
       number = {1},
          eid = {L8},
        pages = {L8},
          doi = {10.3847/2041-8213/adf286},
archivePrefix = {arXiv},
       eprint = {2505.22102},
 primaryClass = {astro-ph.HE},
       adsurl = {https://ui.adsabs.harvard.edu/abs/2025ApJ...989L...8D}
}

@ARTICLE{nguyen2025,
       author = {{Nguyen}, C.~T. and {Costa}, G. and {Bressan}, A. and {Girardi}, L. and {Cescutti}, G. and {Korn}, A.~J. and {Volpato}, G. and {Chen}, Y. and {Pastorelli}, G. and {Trabucchi}, M. and {Shepherd}, K.~G. and {Ettorre}, G. and {Zaggia}, S.},
        title = "{PARSEC V2.0: Rotating tracks and isochrones for seven additional metallicities in the range Z = 0.0001{\textendash}0.03}",
      journal = {\aap},
         year = 2025,
        month = sep,
       volume = {701},
          eid = {A258},
        pages = {A258},
          doi = {10.1051/0004-6361/202556005},
archivePrefix = {arXiv},
       eprint = {2508.02393},
 primaryClass = {astro-ph.SR},
       adsurl = {https://ui.adsabs.harvard.edu/abs/2025A&A...701A.258N}
}

@ARTICLE{bressan2012,
       author = {{Bressan}, Alessandro and {Marigo}, Paola and {Girardi}, L{\'e}o. and {Salasnich}, Bernardo and {Dal Cero}, Claudia and {Rubele}, Stefano and {Nanni}, Ambra},
        title = "{PARSEC: stellar tracks and isochrones with the PAdova and TRieste Stellar Evolution Code}",
      journal = {\mnras},
         year = 2012,
        month = nov,
       volume = {427},
       number = {1},
        pages = {127-145},
          doi = {10.1111/j.1365-2966.2012.21948.x},
archivePrefix = {arXiv},
       eprint = {1208.4498},
 primaryClass = {astro-ph.SR},
       adsurl = {https://ui.adsabs.harvard.edu/abs/2012MNRAS.427..127B}
}

@ARTICLE{costa2025,
       author = {{Costa}, G. and {Shepherd}, K.~G. and {Bressan}, A. and {Addari}, F. and {Chen}, Y. and {Fu}, X. and {Volpato}, G. and {Nguyen}, C.~T. and {Girardi}, L. and {Marigo}, P. and {Mazzi}, A. and {Pastorelli}, G. and {Trabucchi}, M. and {Bossini}, D. and {Zaggia}, S.},
        title = "{Evolutionary tracks, ejecta, and ionizing photons from intermediate-mass to very massive stars with PARSEC}",
      journal = {\aap},
         year = 2025,
        month = feb,
       volume = {694},
          eid = {A193},
        pages = {A193},
          doi = {10.1051/0004-6361/202452573},
archivePrefix = {arXiv},
       eprint = {2501.12917},
 primaryClass = {astro-ph.SR},
       adsurl = {https://ui.adsabs.harvard.edu/abs/2025A&A...694A.193C}
}

@ARTICLE{spera2017,
       author = {{Spera}, Mario and {Mapelli}, Michela},
        title = "{Very massive stars, pair-instability supernovae and intermediate-mass black holes with the sevn code}",
      journal = {\mnras},
         year = 2017,
        month = oct,
       volume = {470},
       number = {4},
        pages = {4739-4749},
          doi = {10.1093/mnras/stx1576},
archivePrefix = {arXiv},
       eprint = {1706.06109},
 primaryClass = {astro-ph.SR},
       adsurl = {https://ui.adsabs.harvard.edu/abs/2017MNRAS.470.4739S}
}

@ARTICLE{iorio2023,
       author = {{Iorio}, Giuliano and {Mapelli}, Michela and {Costa}, Guglielmo and {Spera}, Mario and {Escobar}, Gast{\'o}n J. and {Sgalletta}, Cecilia and {Trani}, Alessandro A. and {Korb}, Erika and {Santoliquido}, Filippo and {Dall'Amico}, Marco and {Gaspari}, Nicola and {Bressan}, Alessandro},
        title = "{Compact object mergers: exploring uncertainties from stellar and binary evolution with SEVN}",
      journal = {\mnras},
         year = 2023,
        month = sep,
       volume = {524},
       number = {1},
        pages = {426-470},
          doi = {10.1093/mnras/stad1630},
archivePrefix = {arXiv},
       eprint = {2211.11774},
 primaryClass = {astro-ph.HE},
       adsurl = {https://ui.adsabs.harvard.edu/abs/2023MNRAS.524..426I}
}

@ARTICLE{mapelli2020,
       author = {{Mapelli}, Michela and {Spera}, Mario and {Montanari}, Enrico and {Limongi}, Marco and {Chieffi}, Alessandro and {Giacobbo}, Nicola and {Bressan}, Alessandro and {Bouffanais}, Yann},
        title = "{Impact of the Rotation and Compactness of Progenitors on the Mass of Black Holes}",
      journal = {\apj},
         year = 2020,
        month = jan,
       volume = {888},
       number = {2},
          eid = {76},
        pages = {76},
          doi = {10.3847/1538-4357/ab584d},
archivePrefix = {arXiv},
       eprint = {1909.01371},
 primaryClass = {astro-ph.HE},
       adsurl = {https://ui.adsabs.harvard.edu/abs/2020ApJ...888...76M}
}

@ARTICLE{spera2019,
       author = {{Spera}, Mario and {Mapelli}, Michela and {Giacobbo}, Nicola and {Trani}, Alessandro A. and {Bressan}, Alessandro and {Costa}, Guglielmo},
        title = "{Merging black hole binaries with the SEVN code}",
      journal = {\mnras},
         year = 2019,
        month = may,
       volume = {485},
       number = {1},
        pages = {889-907},
          doi = {10.1093/mnras/stz359},
archivePrefix = {arXiv},
       eprint = {1809.04605},
 primaryClass = {astro-ph.HE},
       adsurl = {https://ui.adsabs.harvard.edu/abs/2019MNRAS.485..889S}
}

@ARTICLE{paxton2018,
       author = {{Paxton}, Bill and {Schwab}, Josiah and {Bauer}, Evan B. and {Bildsten}, Lars and {Blinnikov}, Sergei and {Duffell}, Paul and {Farmer}, R. and {Goldberg}, Jared A. and {Marchant}, Pablo and {Sorokina}, Elena and {Thoul}, Anne and {Townsend}, Richard H.~D. and {Timmes}, F.~X.},
        title = "{Modules for Experiments in Stellar Astrophysics (MESA): Convective Boundaries, Element Diffusion, and Massive Star Explosions}",
      journal = {\apjs},
         year = 2018,
        month = feb,
       volume = {234},
       number = {2},
          eid = {34},
        pages = {34},
          doi = {10.3847/1538-4365/aaa5a8},
archivePrefix = {arXiv},
       eprint = {1710.08424},
 primaryClass = {astro-ph.SR},
       adsurl = {https://ui.adsabs.harvard.edu/abs/2018ApJS..234...34P}
}

@ARTICLE{paxton2011,
       author = {{Paxton}, Bill and {Bildsten}, Lars and {Dotter}, Aaron and {Herwig}, Falk and {Lesaffre}, Pierre and {Timmes}, Frank},
        title = "{Modules for Experiments in Stellar Astrophysics (MESA)}",
      journal = {\apjs},
         year = 2011,
        month = jan,
       volume = {192},
       number = {1},
          eid = {3},
        pages = {3},
          doi = {10.1088/0067-0049/192/1/3},
archivePrefix = {arXiv},
       eprint = {1009.1622},
 primaryClass = {astro-ph.SR},
       adsurl = {https://ui.adsabs.harvard.edu/abs/2011ApJS..192....3P}
}

@ARTICLE{olejak2025,
       author = {{Olejak}, Aleksandra and {Klencki}, Jakub and {Vigna-Gomez}, Alejandro and {de Mink}, Selma E. and {van Son}, Lieke and {Cehula}, Jakub and {Stegmann}, Jakob and {Ryu}, Taeho and {Hendriks}, David D.},
        title = "{Non-conservative Mass Transfer as a Formation Channel for Gaia Black Hole System}",
      journal = {arXiv e-prints},
         year = 2025,
        month = nov,
          eid = {arXiv:2511.10728},
        pages = {arXiv:2511.10728},
          doi = {10.48550/arXiv.2511.10728},
archivePrefix = {arXiv},
       eprint = {2511.10728},
 primaryClass = {astro-ph.HE},
       adsurl = {https://ui.adsabs.harvard.edu/abs/2025arXiv251110728O}
}

@ARTICLE{vinciguerra2020,
       author = {{Vinciguerra}, Serena and {Neijssel}, Coenraad J. and {Vigna-G{\'o}mez}, Alejandro and {Mandel}, Ilya and {Podsiadlowski}, Philipp and {Maccarone}, Thomas J. and {Nicholl}, Matt and {Kingdon}, Samuel and {Perry}, Alice and {Salemi}, Francesco},
        title = "{Be X-ray binaries in the SMC as indicators of mass-transfer efficiency}",
      journal = {\mnras},
         year = 2020,
        month = nov,
       volume = {498},
       number = {4},
        pages = {4705-4720},
          doi = {10.1093/mnras/staa2177},
archivePrefix = {arXiv},
       eprint = {2003.00195},
 primaryClass = {astro-ph.HE},
       adsurl = {https://ui.adsabs.harvard.edu/abs/2020MNRAS.498.4705V}
}

@ARTICLE{kruckow2024,
       author = {{Kruckow}, Matthias U. and {Andrews}, Jeff J. and {Fragos}, Tassos and {Holl}, Berry and {Bavera}, Simone S. and {Briel}, Max and {Gossage}, Seth and {Kovlakas}, Konstantinos and {Rocha}, Kyle A. and {Sun}, Meng and {Srivastava}, Philipp M. and {Xing}, Zepei and {Zapartas}, Emmanouil},
        title = "{The formation of black holes in non-interacting isolated binaries: Gaia black holes as calibrators of stellar winds from massive stars}",
      journal = {\aap},
         year = 2024,
        month = dec,
       volume = {692},
          eid = {A141},
        pages = {A141},
          doi = {10.1051/0004-6361/202452356},
archivePrefix = {arXiv},
       eprint = {2410.18501},
 primaryClass = {astro-ph.SR},
       adsurl = {https://ui.adsabs.harvard.edu/abs/2024A&A...692A.141K}
}

@ARTICLE{gilkis2024,
       author = {{Gilkis}, A. and {Mazeh}, T.},
        title = "{Gaia BH1 and BH2 - evolutionary models with overshooting of the black hole progenitors within the present-day binary separation}",
      journal = {\mnras},
         year = 2024,
        month = nov,
       volume = {535},
       number = {1},
        pages = {L44-L48},
          doi = {10.1093/mnrasl/slae091},
archivePrefix = {arXiv},
       eprint = {2409.15899},
 primaryClass = {astro-ph.SR},
       adsurl = {https://ui.adsabs.harvard.edu/abs/2024MNRAS.535L..44G}
}

@ARTICLE{hills1980,
       author = {{Hills}, J.~G. and {Fullerton}, L.~W.},
        title = "{Computer simulations of close encounters between single stars and hard binaries}",
      journal = {\aj},
         year = 1980,
        month = sep,
       volume = {85},
        pages = {1281-1291},
          doi = {10.1086/112798},
       adsurl = {https://ui.adsabs.harvard.edu/abs/1980AJ.....85.1281H}
}

@ARTICLE{chattopadhyay2025,
       author = {{Chattopadhyay}, Debatri and {Rocha}, Kyle A. and {Gossage}, Seth and {Kalogera}, Vicky and {El-Badry}, Kareem and {Tchekhovskoy}, Alexander},
        title = "{Modeling the Future of Gaia Neutron Star─Main-sequence Binaries: From Eccentric Orbits to Millisecond Pulsar─White Dwarfs}",
      journal = {\apj},
         year = 2026,
        month = apr,
       volume = {1000},
       number = {2},
          eid = {190},
        pages = {190},
          doi = {10.3847/1538-4357/ae48f5},
archivePrefix = {arXiv},
       eprint = {2510.16201},
 primaryClass = {astro-ph.SR},
       adsurl = {https://ui.adsabs.harvard.edu/abs/2026ApJ..1000..190C}
}

@ARTICLE{elbadry2024NS,
       author = {{El-Badry}, Kareem and {Rix}, Hans-Walter and {Latham}, David W. and {Shahaf}, Sahar and {Mazeh}, Tsevi and {Bieryla}, Allyson and {Buchhave}, Lars A. and {Andrae}, Ren{\'e} and {Yamaguchi}, Natsuko and {Isaacson}, Howard and {Howard}, Andrew W. and {Savino}, Alessandro and {Ilyin}, Ilya V.},
        title = "{A population of neutron star candidates in wide orbits from Gaia astrometry}",
      journal = {The Open Journal of Astrophysics},
         year = 2024,
        month = jul,
       volume = {7},
          eid = {58},
        pages = {58},
          doi = {10.33232/001c.121261},
archivePrefix = {arXiv},
       eprint = {2405.00089},
 primaryClass = {astro-ph.SR},
       adsurl = {https://ui.adsabs.harvard.edu/abs/2024OJAp....7E..58E}
}

@ARTICLE{yamaguchi2025,
       author = {{Yamaguchi}, Natsuko and {El-Badry}, Kareem and {Shahaf}, Sahar},
        title = "{Population Demographics of White Dwarf Binaries with Intermediate Separations: Gaia Constraints on post-AGB Mass Transfer}",
      journal = {\pasp},
         year = 2025,
        month = oct,
       volume = {137},
       number = {10},
          eid = {104205},
        pages = {104205},
          doi = {10.1088/1538-3873/ae0d30},
archivePrefix = {arXiv},
       eprint = {2505.14786},
 primaryClass = {astro-ph.SR},
       adsurl = {https://ui.adsabs.harvard.edu/abs/2025PASP..137j4205Y}
}

@ARTICLE{li2024,
       author = {{Li}, Zhuowen and {Zhu}, Chunhua and {Lu}, Xizhen and {L{\"u}}, Guoliang and {Li}, Lin and {Liu}, Helei and {Guo}, Sufen and {Yu}, Jinlong},
        title = "{A Possible Formation Scenario of the Gaia BH1: Inner Binary Merger in Triple Systems}",
      journal = {\apjl},
         year = 2024,
        month = nov,
       volume = {975},
       number = {1},
          eid = {L8},
        pages = {L8},
          doi = {10.3847/2041-8213/ad8653},
archivePrefix = {arXiv},
       eprint = {2410.10581},
 primaryClass = {astro-ph.SR},
       adsurl = {https://ui.adsabs.harvard.edu/abs/2024ApJ...975L...8L}
}

@ARTICLE{li2026,
       author = {{Li}, Zhuowen and {Lu}, Xizhen and {L{\"u}}, Guoliang and {Zhu}, Chunhua and {Liu}, Helei and {Lei}, Li and {Guo}, Sufen and {He}, Xiaolong and {Beissen}, Nurzada},
        title = "{Formation of the dormant black holes with luminous companions from binary or triple systems}",
      journal = {\aap},
         year = 2026,
        month = feb,
       volume = {706},
          eid = {A105},
        pages = {A105},
          doi = {10.1051/0004-6361/202557437},
archivePrefix = {arXiv},
       eprint = {2512.04774},
 primaryClass = {astro-ph.SR},
       adsurl = {https://ui.adsabs.harvard.edu/abs/2026A&A...706A.105L}
}

@ARTICLE{dicarlo2024,
       author = {{Di Carlo}, Ugo Niccol{\`o} and {Agrawal}, Poojan and {Rodriguez}, Carl L. and {Breivik}, Katelyn},
        title = "{Young Star Clusters Dominate the Production of Detached Black Hole{\textendash}Star Binaries}",
      journal = {\apj},
         year = 2024,
        month = apr,
       volume = {965},
       number = {1},
          eid = {22},
        pages = {22},
          doi = {10.3847/1538-4357/ad2f2c},
archivePrefix = {arXiv},
       eprint = {2306.13121},
 primaryClass = {astro-ph.GA},
       adsurl = {https://ui.adsabs.harvard.edu/abs/2024ApJ...965...22D}
}

@ARTICLE{rastello2023,
       author = {{Rastello}, Sara and {Iorio}, Giuliano and {Mapelli}, Michela and {Arca-Sedda}, Manuel and {Di Carlo}, Ugo N. and {Escobar}, Gast{\'o}n J. and {Shenar}, Tomer and {Torniamenti}, Stefano},
        title = "{Dynamical formation of Gaia BH1 in a young star cluster}",
      journal = {\mnras},
         year = 2023,
        month = nov,
       volume = {526},
       number = {1},
        pages = {740-749},
          doi = {10.1093/mnras/stad2757},
archivePrefix = {arXiv},
       eprint = {2306.14679},
 primaryClass = {astro-ph.SR},
       adsurl = {https://ui.adsabs.harvard.edu/abs/2023MNRAS.526..740R}
}

@ARTICLE{paczynski1968,
       author = {{Paczy{\'n}ski}, B. and {Zi{\'o}{\l}kowski}, J.},
        title = "{On the Origin of Planetary Nebulae and Mira Variables}",
      journal = {\actaa},
         year = 1968,
        month = jan,
       volume = {18},
        pages = {255},
       adsurl = {https://ui.adsabs.harvard.edu/abs/1968AcA....18..255P}
}

@ARTICLE{ge2015,
       author = {{Ge}, Hongwei and {Webbink}, Ronald F. and {Chen}, Xuefei and {Han}, Zhanwen},
        title = "{Adiabatic Mass Loss in Binary Stars. II. From Zero-age Main Sequence to the Base of the Giant Branch}",
      journal = {\apj},
         year = 2015,
        month = oct,
       volume = {812},
       number = {1},
          eid = {40},
        pages = {40},
          doi = {10.1088/0004-637X/812/1/40},
archivePrefix = {arXiv},
       eprint = {1507.04843},
 primaryClass = {astro-ph.SR},
       adsurl = {https://ui.adsabs.harvard.edu/abs/2015ApJ...812...40G}
}

@ARTICLE{iorio2024,
       author = {{Iorio}, Giuliano and {Torniamenti}, Stefano and {Mapelli}, Michela and {Dall'Amico}, Marco and {Trani}, Alessandro A. and {Rastello}, Sara and {Sgalletta}, Cecilia and {Rinaldi}, Stefano and {Costa}, Guglielmo and {Dahl-Lahtinen}, Bera A. and {Escobar}, Gast{\'o}n J. and {Korb}, Erika and {Vaccaro}, M. Paola and {Lacchin}, Elena and {Mestichelli}, Benedetta and {Di Carlo}, Ugo N. and {Spera}, Mario and {Arca Sedda}, Manuel},
        title = "{The boring history of Gaia BH3 from isolated binary evolution}",
      journal = {\aap},
         year = 2024,
        month = oct,
       volume = {690},
          eid = {A144},
        pages = {A144},
          doi = {10.1051/0004-6361/202450531},
archivePrefix = {arXiv},
       eprint = {2404.17568},
 primaryClass = {astro-ph.GA},
       adsurl = {https://ui.adsabs.harvard.edu/abs/2024A&A...690A.144I}
}

@ARTICLE{elbadry2024BH3,
       author = {{El-Badry}, Kareem},
        title = "{On the formation of a 33 solar-mass black hole in a low-metallicity binary}",
      journal = {The Open Journal of Astrophysics},
         year = 2024,
        month = may,
       volume = {7},
          eid = {38},
        pages = {38},
          doi = {10.33232/001c.117652},
archivePrefix = {arXiv},
       eprint = {2404.13047},
 primaryClass = {astro-ph.SR},
       adsurl = {https://ui.adsabs.harvard.edu/abs/2024OJAp....7E..38E}
}

@ARTICLE{marin2024,
       author = {{Mar{\'\i}n Pina}, Daniel and {Rastello}, Sara and {Gieles}, Mark and {Kremer}, Kyle and {Fitzgerald}, Laura and {Rando Forastier}, Bruno},
        title = "{Dynamical formation of Gaia BH3 in the progenitor globular cluster of the ED-2 stream}",
      journal = {\aap},
         year = 2024,
        month = aug,
       volume = {688},
          eid = {L2},
        pages = {L2},
          doi = {10.1051/0004-6361/202450460},
archivePrefix = {arXiv},
       eprint = {2404.13036},
 primaryClass = {astro-ph.GA},
       adsurl = {https://ui.adsabs.harvard.edu/abs/2024A&A...688L...2M}
}

@ARTICLE{balbinot2024,
       author = {{Balbinot}, E. and {Dodd}, E. and {Matsuno}, T. and {Lardo}, C. and {Helmi}, A. and {Panuzzo}, P. and {Mazeh}, T. and {Holl}, B. and {Caffau}, E. and {Jorissen}, A. and {Babusiaux}, C. and {Gavras}, P. and {Wyrzykowski}, {\L}. and {Eyer}, L. and {Leclerc}, N. and {Bombrun}, A. and {Mowlavi}, N. and {Seabroke}, G.~M. and {Cabrera-Ziri}, I. and {Callingham}, T.~M. and {Ruiz-Lara}, T. and {Starkenburg}, E.},
        title = "{The 33 M$_{☉}$ black hole Gaia BH3 is part of the disrupted ED-2 star cluster}",
      journal = {\aap},
         year = 2024,
        month = jul,
       volume = {687},
          eid = {L3},
        pages = {L3},
          doi = {10.1051/0004-6361/202450425},
archivePrefix = {arXiv},
       eprint = {2404.11604},
 primaryClass = {astro-ph.GA},
       adsurl = {https://ui.adsabs.harvard.edu/abs/2024A&A...687L...3B}
}

@ARTICLE{belczynski2010,
       author = {{Belczynski}, Krzysztof and {Bulik}, Tomasz and {Fryer}, Chris L. and {Ruiter}, Ashley and {Valsecchi}, Francesca and {Vink}, Jorick S. and {Hurley}, Jarrod R.},
        title = "{On the Maximum Mass of Stellar Black Holes}",
      journal = {\apj},
         year = 2010,
        month = may,
       volume = {714},
       number = {2},
        pages = {1217-1226},
          doi = {10.1088/0004-637X/714/2/1217},
archivePrefix = {arXiv},
       eprint = {0904.2784},
 primaryClass = {astro-ph.SR},
       adsurl = {https://ui.adsabs.harvard.edu/abs/2010ApJ...714.1217B}
}

@ARTICLE{mapelli2009,
       author = {{Mapelli}, M. and {Colpi}, M. and {Zampieri}, L.},
        title = "{Low metallicity and ultra-luminous X-ray sources in the Cartwheel galaxy}",
      journal = {\mnras},
         year = 2009,
        month = may,
       volume = {395},
       number = {1},
        pages = {L71-L75},
          doi = {10.1111/j.1745-3933.2009.00645.x},
archivePrefix = {arXiv},
       eprint = {0902.3540},
 primaryClass = {astro-ph.HE},
       adsurl = {https://ui.adsabs.harvard.edu/abs/2009MNRAS.395L..71M}
}

@ARTICLE{woosley2002,
       author = {{Woosley}, S.~E. and {Heger}, A. and {Weaver}, T.~A.},
        title = "{The evolution and explosion of massive stars}",
      journal = {Reviews of Modern Physics},
         year = 2002,
        month = nov,
       volume = {74},
       number = {4},
        pages = {1015-1071},
          doi = {10.1103/RevModPhys.74.1015},
       adsurl = {https://ui.adsabs.harvard.edu/abs/2002RvMP...74.1015W}
}

@ARTICLE{tanikawa2023,
       author = {{Tanikawa}, Ataru and {Hattori}, Kohei and {Kawanaka}, Norita and {Kinugawa}, Tomoya and {Shikauchi}, Minori and {Tsuna}, Daichi},
        title = "{Search for a Black Hole Binary in Gaia DR3 Astrometric Binary Stars with Spectroscopic Data}",
      journal = {\apj},
         year = 2023,
        month = apr,
       volume = {946},
       number = {2},
          eid = {79},
        pages = {79},
          doi = {10.3847/1538-4357/acbf36},
archivePrefix = {arXiv},
       eprint = {2209.05632},
 primaryClass = {astro-ph.SR},
       adsurl = {https://ui.adsabs.harvard.edu/abs/2023ApJ...946...79T}
}

@ARTICLE{panuzzo2024,
       author = {{Gaia Collaboration} and {Panuzzo}, P. and {Mazeh}, T. and {Arenou}, F. and {Holl}, B. and {Caffau}, E. and {Jorissen}, A. and {Babusiaux}, C. and {Gavras}, P. and {Sahlmann}, J. and {Bastian}, U. and {Wyrzykowski}, {\L}. and {Eyer}, L. and {Leclerc}, N. and {Bauchet}, N. and {Bombrun}, A. and {Mowlavi}, N. and {Seabroke}, G.~M. and {Teyssier}, D. and {Balbinot}, E. and {Helmi}, A. and {Brown}, A.~G.~A. and {Vallenari}, A. and {Prusti}, T. and {de Bruijne}, J.~H.~J. and {Barbier}, A. and {Biermann}, M. and {Creevey}, O.~L. and {Ducourant}, C. and {Evans}, D.~W. and {Guerra}, R. and {Hutton}, A. and {Jordi}, C. and {Klioner}, S.~A. and {Lammers}, U. and {Lindegren}, L. and {Luri}, X. and {Mignard}, F. and {Nicolas}, C. and {Randich}, S. and {Sartoretti}, P. and {Smiljanic}, R. and {Tanga}, P. and {Walton}, N.~A. and {Aerts}, C. and {Bailer-Jones}, C.~A.~L. and {Cropper}, M. and {Drimmel}, R. and {Jansen}, F. and {Katz}, D. and {Lattanzi}, M.~G. and {Soubiran}, C. and {Th{\'e}venin}, F. and {van Leeuwen}, F. and {Andrae}, R. and {Audard}, M. and {Bakker}, J. and {Blomme}, R. and {Casta{\~n}eda}, J. and {De Angeli}, F. and {Fabricius}, C. and {Fouesneau}, M. and {Fr{\'e}mat}, Y. and {Galluccio}, L. and {Guerrier}, A. and {Heiter}, U. and {Masana}, E. and {Messineo}, R. and {Nienartowicz}, K. and {Pailler}, F. and {Riclet}, F. and {Roux}, W. and {Sordo}, R. and {Gracia-Abril}, G. and {Portell}, J. and {Altmann}, M. and {Benson}, K. and {Berthier}, J. and {Burgess}, P.~W. and {Busonero}, D. and {Busso}, G. and {Cacciari}, C. and {C{\'a}novas}, H. and {Carrasco}, J.~M. and {Carry}, B. and {Cellino}, A. and {Cheek}, N. and {Clementini}, G. and {Damerdji}, Y. and {Davidson}, M. and {de Teodoro}, P. and {Delchambre}, L. and {Dell'Oro}, A. and {Fraile Garcia}, E. and {Garabato}, D. and {Garc{\'\i}a-Lario}, P. and {Haigron}, R. and {Hambly}, N.~C. and {Harrison}, D.~L. and {Hatzidimitriou}, D. and {Hern{\'a}ndez}, J. and {Hestroffer}, D. and {Hodgkin}, S.~T. and {Jamal}, S. and {Jevardat de Fombelle}, G. and {Jordan}, S. and {Krone-Martins}, A. and {Lanzafame}, A.~C. and {L{\"o}ffler}, W. and {Lorca}, A. and {Marchal}, O. and {Marrese}, P.~M. and {Moitinho}, A. and {Muinonen}, K. and {Nu{\~n}ez Campos}, M. and {Oreshina-Slezak}, I. and {Osborne}, P. and {Pancino}, E. and {Pauwels}, T. and {Recio-Blanco}, A. and {Riello}, M. and {Rimoldini}, L. and {Robin}, A.~C. and {Roegiers}, T. and {Sarro}, L.~M. and {Schultheis}, M. and {Smith}, M. and {Sozzetti}, A. and {Utrilla}, E. and {van Leeuwen}, M. and {Weingrill}, K. and {Abbas}, U. and {{\'A}brah{\'a}m}, P. and {Abreu Aramburu}, A. and {Ahmed}, S. and {Altavilla}, G. and {{\'A}lvarez}, M.~A. and {Anders}, F. and {Anderson}, R.~I. and {Anglada Varela}, E. and {Antoja}, T. and {Baig}, S. and {Baines}, D. and {Baker}, S.~G. and {Balaguer-N{\'u}{\~n}ez}, L. and {Balog}, Z. and {Barache}, C. and {Barros}, M. and {Barstow}, M.~A. and {Bartolom{\'e}}, S. and {Bashi}, D. and {Bassilana}, J.-L. and {Baudeau}, N. and {Becciani}, U. and {Bedin}, L.~R. and {Bellas-Velidis}, I. and {Bellazzini}, M. and {Beordo}, W. and {Bernet}, M. and {Bertolotto}, C. and {Bertone}, S. and {Bianchi}, L. and {Binnenfeld}, A. and {Blanco-Cuaresma}, S. and {Bland-Hawthorn}, J. and {Blazere}, A. and {Boch}, T. and {Bossini}, D. and {Bouquillon}, S. and {Bragaglia}, A. and {Braine}, J. and {Bratsolis}, E. and {Breedt}, E. and {Bressan}, A. and {Brouillet}, N. and {Brugaletta}, E. and {Bucciarelli}, B. and {Butkevich}, A.~G. and {Buzzi}, R. and {Camut}, A. and {Cancelliere}, R. and {Cantat-Gaudin}, T. and {Capilla Guilarte}, D. and {Carballo}, R. and {Carlucci}, T. and {Carnerero}, M.~I. and {Carretero}, J. and {Carton}, S. and {Casamiquela}, L. and {Casey}, A. and {Castellani}, M. and {Castro-Ginard}, A. and {Ceraj}, L. and {Cesare}, V. and {Charlot}, P. and {Chaudet}, C. and {Chemin}, L. and {Chiavassa}, A. and {Chornay}, N. and {Chosson}, D.},
        title = "{Discovery of a dormant 33 solar-mass black hole in pre-release Gaia astrometry}",
      journal = {\aap},
         year = 2024,
        month = jun,
       volume = {686},
          eid = {L2},
        pages = {L2},
          doi = {10.1051/0004-6361/202449763},
archivePrefix = {arXiv},
       eprint = {2404.10486},
 primaryClass = {astro-ph.GA},
       adsurl = {https://ui.adsabs.harvard.edu/abs/2024A&A...686L...2G}
}

@ARTICLE{elbadry2023b,
       author = {{El-Badry}, Kareem and {Rix}, Hans-Walter and {Cendes}, Yvette and {Rodriguez}, Antonio C. and {Conroy}, Charlie and {Quataert}, Eliot and {Hawkins}, Keith and {Zari}, Eleonora and {Hobson}, Melissa and {Breivik}, Katelyn and {Rau}, Arne and {Berger}, Edo and {Shahaf}, Sahar and {Seeburger}, Rhys and {Burdge}, Kevin B. and {Latham}, David W. and {Buchhave}, Lars A. and {Bieryla}, Allyson and {Bashi}, Dolev and {Mazeh}, Tsevi and {Faigler}, Simchon},
        title = "{A red giant orbiting a black hole}",
      journal = {\mnras},
         year = 2023,
        month = may,
       volume = {521},
       number = {3},
        pages = {4323-4348},
          doi = {10.1093/mnras/stad799},
archivePrefix = {arXiv},
       eprint = {2302.07880},
 primaryClass = {astro-ph.SR},
       adsurl = {https://ui.adsabs.harvard.edu/abs/2023MNRAS.521.4323E}
}

@ARTICLE{elbadry2023,
       author = {{El-Badry}, Kareem and {Rix}, Hans-Walter and {Quataert}, Eliot and {Howard}, Andrew W. and {Isaacson}, Howard and {Fuller}, Jim and {Hawkins}, Keith and {Breivik}, Katelyn and {Wong}, Kaze W.~K. and {Rodriguez}, Antonio C. and {Conroy}, Charlie and {Shahaf}, Sahar and {Mazeh}, Tsevi and {Arenou}, Fr{\'e}d{\'e}ric and {Burdge}, Kevin B. and {Bashi}, Dolev and {Faigler}, Simchon and {Weisz}, Daniel R. and {Seeburger}, Rhys and {Almada Monter}, Silvia and {Wojno}, Jennifer},
        title = "{A Sun-like star orbiting a black hole}",
      journal = {\mnras},
         year = 2023,
        month = jan,
       volume = {518},
       number = {1},
        pages = {1057-1085},
          doi = {10.1093/mnras/stac3140},
archivePrefix = {arXiv},
       eprint = {2209.06833},
 primaryClass = {astro-ph.SR},
       adsurl = {https://ui.adsabs.harvard.edu/abs/2023MNRAS.518.1057E}
}

@ARTICLE{zak2023,
       author = {{Zak}, J. and {Jones}, D. and {Boffin}, H.~M.~J. and {Beck}, P.~G. and {Klencki}, J. and {Bodensteiner}, J. and {Shenar}, T. and {Van Winckel}, H. and {Skarka}, M. and {Arellano-C{\'o}rdova}, K. and {Viuho}, J. and {Sowicka}, P. and {Guenther}, E.~W. and {Hatzes}, A.},
        title = "{Everything that glitters is not gold: V1315 Cas is not a dormant black hole}",
      journal = {\mnras},
         year = 2023,
        month = oct,
       volume = {524},
       number = {4},
        pages = {5749-5761},
          doi = {10.1093/mnras/stad2137},
archivePrefix = {arXiv},
       eprint = {2307.09594},
 primaryClass = {astro-ph.SR},
       adsurl = {https://ui.adsabs.harvard.edu/abs/2023MNRAS.524.5749Z}
}

@ARTICLE{nagarajan2025,
       author = {{Nagarajan}, Pranav and {El-Badry}, Kareem and {Chawla}, Chirag and {Di Carlo}, Ugo Niccol{\`o} and {Breivik}, Katelyn and {Rodriguez}, Carl L. and {Agrawal}, Poojan and {Delfavero}, Vera and {Chatterjee}, Sourav},
        title = "{Realistic Predictions for Gaia Black Hole Discoveries: Comparison of Isolated Binary and Dynamical Formation Models}",
      journal = {\pasp},
         year = 2025,
        month = apr,
       volume = {137},
       number = {4},
          eid = {044202},
        pages = {044202},
          doi = {10.1088/1538-3873/adc839},
archivePrefix = {arXiv},
       eprint = {2502.03527},
 primaryClass = {astro-ph.GA},
       adsurl = {https://ui.adsabs.harvard.edu/abs/2025PASP..137d4202N}
}

@ARTICLE{nagarajan2025kick,
       author = {{Nagarajan}, Pranav and {El-Badry}, Kareem},
        title = "{Mixed Origins: Strong Natal Kicks for Some Black Holes and None for Others}",
      journal = {\pasp},
         year = 2025,
        month = mar,
       volume = {137},
       number = {3},
          eid = {034203},
        pages = {034203},
          doi = {10.1088/1538-3873/adb6d6},
archivePrefix = {arXiv},
       eprint = {2411.16847},
 primaryClass = {astro-ph.GA},
       adsurl = {https://ui.adsabs.harvard.edu/abs/2025PASP..137c4203N}
}

@ARTICLE{chakrabarti2023,
       author = {{Chakrabarti}, Sukanya and {Simon}, Joshua D. and {Craig}, Peter A. and {Reggiani}, Henrique and {Brandt}, Timothy D. and {Guhathakurta}, Puragra and {Dalba}, Paul A. and {Kirby}, Evan N. and {Chang}, Philip and {Hey}, Daniel R. and {Savino}, Alessandro and {Geha}, Marla and {Thompson}, Ian B.},
        title = "{A Noninteracting Galactic Black Hole Candidate in a Binary System with a Main-sequence Star}",
      journal = {\aj},
         year = 2023,
        month = jul,
       volume = {166},
       number = {1},
          eid = {6},
        pages = {6},
          doi = {10.3847/1538-3881/accf21},
archivePrefix = {arXiv},
       eprint = {2210.05003},
 primaryClass = {astro-ph.GA},
       adsurl = {https://ui.adsabs.harvard.edu/abs/2023AJ....166....6C}
}

@ARTICLE{bodensteiner2022,
       author = {{Bodensteiner}, J. and {Heida}, M. and {Abdul-Masih}, M. and {Baade}, D. and {Banyard}, G. and {Bowman}, D.~M. and {Fabry}, M. and {Frost}, A. and {Mahy}, L. and {Marchant}, P. and {M{\'e}rand}, A. and {Reggiani}, M. and {Rivinius}, T. and {Sana}, H. and {Selman}, F. and {Shenar}, T.},
        title = "{Detecting Stripped Stars While Searching for Quiescent Black Holes}",
      journal = {The Messenger},
         year = 2022,
        month = mar,
       volume = {186},
        pages = {3-9},
          doi = {10.18727/0722-6691/5255},
archivePrefix = {arXiv},
       eprint = {2207.00366},
 primaryClass = {astro-ph.SR},
       adsurl = {https://ui.adsabs.harvard.edu/abs/2022Msngr.186....3B}
}

@ARTICLE{jayasinghe2021,
       author = {{Jayasinghe}, T. and {Stanek}, K.~Z. and {Thompson}, Todd A. and {Kochanek}, C.~S. and {Rowan}, D.~M. and {Vallely}, P.~J. and {Strassmeier}, K.~G. and {Weber}, M. and {Hinkle}, J.~T. and {Hambsch}, F.-J. and {Martin}, D.~V. and {Prieto}, J.~L. and {Pessi}, T. and {Huber}, D. and {Auchettl}, K. and {Lopez}, L.~A. and {Ilyin}, I. and {Badenes}, C. and {Howard}, A.~W. and {Isaacson}, H. and {Murphy}, S.~J.},
        title = "{A unicorn in monoceros: the 3 M$_{☉}$ dark companion to the bright, nearby red giant V723 Mon is a non-interacting, mass-gap black hole candidate}",
      journal = {\mnras},
         year = 2021,
        month = jun,
       volume = {504},
       number = {2},
        pages = {2577-2602},
          doi = {10.1093/mnras/stab907},
archivePrefix = {arXiv},
       eprint = {2101.02212},
 primaryClass = {astro-ph.SR},
       adsurl = {https://ui.adsabs.harvard.edu/abs/2021MNRAS.504.2577J}
}

@ARTICLE{thompson2019,
       author = {{Thompson}, Todd A. and {Kochanek}, Christopher S. and {Stanek}, Krzysztof Z. and {Badenes}, Carles and {Post}, Richard S. and {Jayasinghe}, Tharindu and {Latham}, David W. and {Bieryla}, Allyson and {Esquerdo}, Gilbert A. and {Berlind}, Perry and {Calkins}, Michael L. and {Tayar}, Jamie and {Lindegren}, Lennart and {Johnson}, Jennifer A. and {Holoien}, Thomas W.-S. and {Auchettl}, Katie and {Covey}, Kevin},
        title = "{A noninteracting low-mass black hole-giant star binary system}",
      journal = {Science},
         year = 2019,
        month = nov,
       volume = {366},
       number = {6465},
        pages = {637-640},
          doi = {10.1126/science.aau4005},
archivePrefix = {arXiv},
       eprint = {1806.02751},
 primaryClass = {astro-ph.HE},
       adsurl = {https://ui.adsabs.harvard.edu/abs/2019Sci...366..637T}
}

@ARTICLE{shenar2022,
       author = {{Shenar}, Tomer and {Sana}, Hugues and {Mahy}, Laurent and {El-Badry}, Kareem and {Marchant}, Pablo and {Langer}, Norbert and {Hawcroft}, Calum and {Fabry}, Matthias and {Sen}, Koushik and {Almeida}, Leonardo A. and {Abdul-Masih}, Michael and {Bodensteiner}, Julia and {Crowther}, Paul A. and {Gieles}, Mark and {Gromadzki}, Mariusz and {H{\'e}nault-Brunet}, Vincent and {Herrero}, Artemio and {de Koter}, Alex and {Iwanek}, Patryk and {Koz{\l}owski}, Szymon and {Lennon}, Daniel J. and {Ma{\'\i}z Apell{\'a}niz}, Jes{\'u}s and {Mr{\'o}z}, Przemys{\l}aw and {Moffat}, Anthony F.~J. and {Picco}, Annachiara and {Pietrukowicz}, Pawe{\l} and {Poleski}, Rados{\l}aw and {Rybicki}, Krzysztof and {Schneider}, Fabian R.~N. and {Skowron}, Dorota M. and {Skowron}, Jan and {Soszy{\'n}ski}, Igor and {Szyma{\'n}ski}, Micha{\l} K. and {Toonen}, Silvia and {Udalski}, Andrzej and {Ulaczyk}, Krzysztof and {Vink}, Jorick S. and {Wrona}, Marcin},
        title = "{An X-ray-quiet black hole born with a negligible kick in a massive binary within the Large Magellanic Cloud}",
      journal = {Nature Astronomy},
         year = 2022,
        month = jul,
       volume = {6},
        pages = {1085-1092},
          doi = {10.1038/s41550-022-01730-y},
archivePrefix = {arXiv},
       eprint = {2207.07675},
 primaryClass = {astro-ph.HE},
       adsurl = {https://ui.adsabs.harvard.edu/abs/2022NatAs...6.1085S}
}

@ARTICLE{giesers2019,
       author = {{Giesers}, Benjamin and {Kamann}, Sebastian and {Dreizler}, Stefan and {Husser}, Tim-Oliver and {Askar}, Abbas and {G{\"o}ttgens}, Fabian and {Brinchmann}, Jarle and {Latour}, Marilyn and {Weilbacher}, Peter M. and {Wendt}, Martin and {Roth}, Martin M.},
        title = "{A stellar census in globular clusters with MUSE: Binaries in NGC 3201}",
      journal = {\aap},
         year = 2019,
        month = dec,
       volume = {632},
          eid = {A3},
        pages = {A3},
          doi = {10.1051/0004-6361/201936203},
archivePrefix = {arXiv},
       eprint = {1909.04050},
 primaryClass = {astro-ph.SR},
       adsurl = {https://ui.adsabs.harvard.edu/abs/2019A&A...632A...3G}
}

@ARTICLE{giesers2018,
       author = {{Giesers}, Benjamin and {Dreizler}, Stefan and {Husser}, Tim-Oliver and {Kamann}, Sebastian and {Anglada Escud{\'e}}, Guillem and {Brinchmann}, Jarle and {Carollo}, C. Marcella and {Roth}, Martin M. and {Weilbacher}, Peter M. and {Wisotzki}, Lutz},
        title = "{A detached stellar-mass black hole candidate in the globular cluster NGC 3201}",
      journal = {\mnras},
         year = 2018,
        month = mar,
       volume = {475},
       number = {1},
        pages = {L15-L19},
          doi = {10.1093/mnrasl/slx203},
archivePrefix = {arXiv},
       eprint = {1801.05642},
 primaryClass = {astro-ph.SR},
       adsurl = {https://ui.adsabs.harvard.edu/abs/2018MNRAS.475L..15G}
}

@ARTICLE{saracino2022,
       author = {{Saracino}, S. and {Kamann}, S. and {Guarcello}, M.~G. and {Usher}, C. and {Bastian}, N. and {Cabrera-Ziri}, I. and {Gieles}, M. and {Dreizler}, S. and {Da Costa}, G.~S. and {Husser}, T.-O. and {H{\'e}nault-Brunet}, V.},
        title = "{A black hole detected in the young massive LMC cluster NGC 1850}",
      journal = {\mnras},
         year = 2022,
        month = apr,
       volume = {511},
       number = {2},
        pages = {2914-2924},
          doi = {10.1093/mnras/stab3159},
archivePrefix = {arXiv},
       eprint = {2111.06506},
 primaryClass = {astro-ph.GA},
       adsurl = {https://ui.adsabs.harvard.edu/abs/2022MNRAS.511.2914S}
}

@ARTICLE{GWTC4pop,
       author = {{Abac}, A.~G. and {Abouelfettouh}, I. and {Acernese}, F. and {Ackley}, K. and {Adamcewicz}, C. and {Adhicary}, S. and {Adhikari}, D. and {Adhikari}, N. and {Adhikari}, R.~X. and {Adkins}, V.~K. and {Afroz}, S. and {Agarwal}, D. and {Agathos}, M. and {Aghaei Abchouyeh}, M. and {Aguiar}, O.~D. and {Ahmadzadeh}, S. and {Aiello}, L. and {Ain}, A. and {Ajith}, P. and {Akutsu}, T. and {Albanesi}, S. and {Alfaidi}, R.~A. and {Al-Jodah}, A. and {All{\'e}n{\'e}}, C. and {Allocca}, A. and {Al-Shammari}, S. and {Altin}, P.~A. and {Alvarez-Lopez}, S. and {Amarasinghe}, O. and {Amato}, A. and {Amra}, C. and {Ananyeva}, A. and {Anderson}, S.~B. and {Anderson}, W.~G. and {Andia}, M. and {Ando}, M. and {Andrade}, T. and {Andr{\'e}s-Carcasona}, M. and {Andri{\'c}}, T. and {Anglin}, J. and {Ansoldi}, S. and {Antelis}, J.~M. and {Antier}, S. and {Aoumi}, M. and {Appavuravther}, E.~Z. and {Appert}, S. and {Apple}, S.~K. and {Arai}, K. and {Araya}, A. and {Araya}, M.~C. and {Arca Sedda}, M. and {Areeda}, J.~S. and {Argianas}, L. and {Aritomi}, N. and {Armato}, F. and {Armstrong}, S. and {Arnaud}, N. and {Arogeti}, M. and {Aronson}, S.~M. and {Arun}, K.~G. and {Ashton}, G. and {Aso}, Y. and {Assiduo}, M. and {Assis de Souza Melo}, S. and {Aston}, S.~M. and {Astone}, P. and {Attadio}, F. and {Aubin}, F. and {AultONeal}, K. and {Avallone}, G. and {Babak}, S. and et al.},
        title = "{GWTC-4.0: Population Properties of Merging Compact Binaries}",
      journal = {arXiv e-prints},
         year = 2025,
        month = aug,
          eid = {arXiv:2508.18083},
        pages = {arXiv:2508.18083},
          doi = {10.48550/arXiv.2508.18083},
archivePrefix = {arXiv},
       eprint = {2508.18083},
 primaryClass = {astro-ph.HE},
       adsurl = {https://ui.adsabs.harvard.edu/abs/2025arXiv250818083T}
}

@ARTICLE{farr2011,
       author = {{Farr}, Will M. and {Sravan}, Niharika and {Cantrell}, Andrew and {Kreidberg}, Laura and {Bailyn}, Charles D. and {Mandel}, Ilya and {Kalogera}, Vicky},
        title = "{The Mass Distribution of Stellar-mass Black Holes}",
      journal = {\apj},
         year = 2011,
        month = nov,
       volume = {741},
       number = {2},
          eid = {103},
        pages = {103},
          doi = {10.1088/0004-637X/741/2/103},
archivePrefix = {arXiv},
       eprint = {1011.1459},
 primaryClass = {astro-ph.GA},
       adsurl = {https://ui.adsabs.harvard.edu/abs/2011ApJ...741..103F}
}

@ARTICLE{oezel2010,
       author = {{{\"O}zel}, Feryal and {Psaltis}, Dimitrios and {Narayan}, Ramesh and {McClintock}, Jeffrey E.},
        title = "{The Black Hole Mass Distribution in the Galaxy}",
      journal = {\apj},
         year = 2010,
        month = dec,
       volume = {725},
       number = {2},
        pages = {1918-1927},
          doi = {10.1088/0004-637X/725/2/1918},
archivePrefix = {arXiv},
       eprint = {1006.2834},
 primaryClass = {astro-ph.GA},
       adsurl = {https://ui.adsabs.harvard.edu/abs/2010ApJ...725.1918O}
}

@ARTICLE{abbottGW150914,
       author = {{Abbott}, B.~P. and {Abbott}, R. and {Abbott}, T.~D. and {Abernathy}, M.~R. and {Acernese}, F. and {Ackley}, K. and {Adams}, C. and {Adams}, T. and {Addesso}, P. and {Adhikari}, R.~X. and {Adya}, V.~B. and {Affeldt}, C. and {Agathos}, M. and {Agatsuma}, K. and {Aggarwal}, N. and {Aguiar}, O.~D. and {Aiello}, L. and {Ain}, A. and {Ajith}, P. and {Allen}, B. and {Allocca}, A. and {Altin}, P.~A. and {Anderson}, S.~B. and {Anderson}, W.~G. and {Arai}, K. and {Arain}, M.~A. and {Araya}, M.~C. and {Arceneaux}, C.~C. and {Areeda}, J.~S. and {Arnaud}, N. and {Arun}, K.~G. and {Ascenzi}, S. and {Ashton}, G. and {Ast}, M. and {Aston}, S.~M. and {Astone}, P. and {Aufmuth}, P. and {Aulbert}, C. and {Babak}, S. and {Bacon}, P. and {Bader}, M.~K.~M. and {Baker}, P.~T. and {Baldaccini}, F. and {Ballardin}, G. and {Ballmer}, S.~W. and {Barayoga}, J.~C. and {Barclay}, S.~E. and {Barish}, B.~C. and {Barker}, D. and {Barone}, F. and {Barr}, B. and {Barsotti}, L. and {Barsuglia}, M. and {Barta}, D. and {Bartlett}, J. and {Barton}, M.~A. and {Bartos}, I. and {Bassiri}, R. and {Basti}, A. and {Batch}, J.~C. and {Baune}, C. and {Bavigadda}, V. and {Bazzan}, M. and {Behnke}, B. and {Bejger}, M. and {Belczynski}, C. and {Bell}, A.~S. and {Bell}, C.~J. and {Berger}, B.~K. and {Bergman}, J. and {Bergmann}, G. and {Berry}, C.~P.~L. and {Bersanetti}, D. and {Bertolini}, A. and {Betzwieser}, J. and {Bhagwat}, S. and {Bhandare}, R. and {Bilenko}, I.~A. and {Billingsley}, G. and {Birch}, J. and {Birney}, I.~A. and {Birnholtz}, O. and {Biscans}, S. and {Bisht}, A. and {Bitossi}, M. and {Biwer}, C. and {Bizouard}, M.~A. and {Blackburn}, J.~K. and {Blair}, C.~D. and {Blair}, D.~G. and {Blair}, R.~M. and {Bloemen}, S. and et al.},
        title = "{Observation of Gravitational Waves from a Binary Black Hole Merger}",
      journal = {\prl},
         year = 2016,
        month = feb,
       volume = {116},
       number = {6},
          eid = {061102},
        pages = {061102},
          doi = {10.1103/PhysRevLett.116.061102},
archivePrefix = {arXiv},
       eprint = {1602.03837},
 primaryClass = {gr-qc},
       adsurl = {https://ui.adsabs.harvard.edu/abs/2016PhRvL.116f1102A}
}

@ARTICLE{abbottGWTC4,
       author = {{Abac}, A.~G. and {Abouelfettouh}, I. and {Acernese}, F. and {Ackley}, K. and {Adamcewicz}, C. and {Adhicary}, S. and {Adhikari}, D. and {Adhikari}, N. and {Adhikari}, R.~X. and {Adkins}, V.~K. and {Afroz}, S. and {Agapito}, A. and {Agarwal}, D. and {Agathos}, M. and {Aggarwal}, N. and {Aggarwal}, S. and {Aguiar}, O.~D. and {Ahrend}, I. -L. and {Aiello}, L. and {Ain}, A. and {Ajith}, P. and et al.},
        title = "{GWTC-4.0: Updating the Gravitational-Wave Transient Catalog with Observations from the First Part of the Fourth LIGO-Virgo-KAGRA Observing Run}",
      journal = {arXiv e-prints},
         year = 2025,
        month = aug,
          eid = {arXiv:2508.18082},
        pages = {arXiv:2508.18082},
          doi = {10.48550/arXiv.2508.18082},
archivePrefix = {arXiv},
       eprint = {2508.18082},
 primaryClass = {gr-qc},
       adsurl = {https://ui.adsabs.harvard.edu/abs/2025arXiv250818082T}
}

@ARTICLE{trimble1969,
       author = {{Trimble}, Virginia L. and {Thorne}, Kip S.},
        title = "{Spectroscopic Binaries and Collapsed Stars}",
      journal = {\apj},
         year = 1969,
        month = jun,
       volume = {156},
        pages = {1013},
          doi = {10.1086/150032},
       adsurl = {https://ui.adsabs.harvard.edu/abs/1969ApJ...156.1013T}
}

@ARTICLE{zeldovich1966,
       author = {{Zeldovich}, Ya. B. and {Guseynov}, O.~H.},
        title = "{Collapsed Stars in Binaries}",
      journal = {\apj},
         year = 1966,
        month = may,
       volume = {144},
        pages = {840},
          doi = {10.1086/148672},
       adsurl = {https://ui.adsabs.harvard.edu/abs/1966ApJ...144..840Z}
}

@ARTICLE{saracino2023,
       author = {{Saracino}, S. and {Shenar}, T. and {Kamann}, S. and {Bastian}, N. and {Gieles}, M. and {Usher}, C. and {Bodensteiner}, J. and {Kochoska}, A. and {Orosz}, J.~A. and {Sana}, H.},
        title = "{Updated radial velocities and new constraints on the nature of the unseen source in NGC1850 BH1}",
      journal = {\mnras},
         year = 2023,
        month = may,
       volume = {521},
       number = {2},
        pages = {3162-3171},
          doi = {10.1093/mnras/stad764},
archivePrefix = {arXiv},
       eprint = {2303.07369},
 primaryClass = {astro-ph.GA},
       adsurl = {https://ui.adsabs.harvard.edu/abs/2023MNRAS.521.3162S}
}

@ARTICLE{mahy2022,
       author = {{Mahy}, L. and {Sana}, H. and {Shenar}, T. and {Sen}, K. and {Langer}, N. and {Marchant}, P. and {Abdul-Masih}, M. and {Banyard}, G. and {Bodensteiner}, J. and {Bowman}, D.~M. and {Dsilva}, K. and {Fabry}, M. and {Hawcroft}, C. and {Janssens}, S. and {Van Reeth}, T. and {Eldridge}, C.},
        title = "{Identifying quiescent compact objects in massive Galactic single-lined spectroscopic binaries}",
      journal = {\aap},
         year = 2022,
        month = aug,
       volume = {664},
          eid = {A159},
        pages = {A159},
          doi = {10.1051/0004-6361/202243147},
archivePrefix = {arXiv},
       eprint = {2207.07752},
 primaryClass = {astro-ph.SR},
       adsurl = {https://ui.adsabs.harvard.edu/abs/2022A&A...664A.159M}
}

@article{gala,
        doi = {10.21105/joss.00388},
        url = {https://doi.org/10.21105%2Fjoss.00388},
        year = 2017,
        month = {oct},
        publisher = {The Open Journal},
        volume = {2},
        number = {18},
        author = {Adrian M. Price-Whelan},
        title = {Gala: A Python package for galactic dynamics},
        journal = {The Journal of Open Source Software}}

\begin{appendix}

\section{Summary of the models presented in the main text}\label{app:bigtable}

Table~\ref{tab:param} summarizes the models presented in the main text (i.e., with CE parameter $\alpha=1$ or  stable mass transfer).
\begin{table*}[htbp] 

    \caption{{Summary of the models presented in the main text (with CE parameter $\alpha=1$ or  stable mass transfer).}}
    \renewcommand{\arraystretch}{1.2}
    \begin{tabular}{c cc cc cc c ccc c}
    \hline   
        Model & CE & $\alpha$ & CCSN & Fallback & $f_{\rm a}$ & Ang. Mom. & Z & Perc. BH1 & Perc. BH2 & Perc. BH3 & $\eta_{\rm dorm}$\\
        & && && && & &&& [$\times{}10^{-7}$ M$_\odot{}^{-1}$]\\
    \hline

A   & Y & 1 & R & N & 0.5 & Isot & 0.014 & $99.69_{-5.78}^{+0.26}$ & $99.95_{-0.01}^{+0.01}$ & $99.95_{-0.07}^{+0.01}$ & 8.3 \\
Afb & Y & 1 & R & Y & 0.5 & Isot & 0.014 & $96.54_{-10.98}^{+3.20}$ & $99.78_{-0.37}^{+0.22}$ & $99.74_{-0.79}^{+0.17}$  & 467.4 \\ 
 & Y & 1 & R & N & 0.0 & Isot. & 0.014 & $99.75_{-2.56}^{+0.20}$ & $99.95_{-0.01}^{+0.01}$ & $99.95_{-0.52}^{+0.01}$ & 8.5 \\
 & Y & 1 & R & Y & 0.0 & Isot. & 0.014 & $98.95_{-4.54}^{+0.97}$ & $99.90_{-0.17}^{+0.10}$ & $99.81_{-0.90}^{+0.16}$ & 467.3\\ %
AJ        & Y & 1 & R & N & 0.0 & Jeans & 0.014 & $99.64_{-5.51}^{+0.31}$ & $99.95_{-0.01}^{+0.01}$ & $99.95_{-0.01}^{+0.01}$  & 8.3 \\ 
AJfb & Y & 1 & R & Y & 0.0 & Jeans & 0.014 & $95.76_{-3.65}^{+3.85}$ & $99.76_{-1.05}^{+0.24}$ & $99.58_{-0.75}^{+0.35}$  & 467.3 \\ 
       & Y & 1 & R & N & 1.0 & Isot & 0.014 & $99.75_{-2.31}^{+0.20}$ & $99.95_{-0.10}^{+0.01}$ & $99.92_{-0.73}^{+0.03}$ & 8.6 \\ 
       & Y & 1 & R & Y & 1.0 & Isot & 0.014 & $96.29_{-6.02}^{+3.57}$ & $99.92_{-0.30}^{+0.08}$ & $99.50_{-0.83}^{+0.39}$ & 467.1 \\ 
 & Y & 1 & R & N & 0.5 & Isot & 0.017 & $97.92_{-9.36}^{+1.99}$ & $99.92_{-0.01}^{+0.01}$ & $99.92_{-0.01}^{+0.01}$  & 5.2 \\ 
 & Y & 1 & R & N & 0.0 & Jeans & 0.017 & $96.41_{-10.91}^{+3.51}$ & $99.91_{-0.01}^{+0.01}$ & $99.91_{-0.01}^{+0.01}$ & 4.9 \\ 
 AlowZ       & Y & 1 & R & N & 0.5 & Isot & $1.4\times{}10^{-4}$ & $99.96_{-0.01}^{+0.01}$ & $99.81_{-18.76}^{+0.15}$ & $98.19_{-36.93}^{+1.77}$ & 11.2\\ 
 AJlowZ & Y & 1 & R & N & 0.0 & Jeans & 1.4$\times{}10^{-4}$ & $99.96_{-0.04}^{+0.01}$ & $99.96_{-0.84}^{+0.01}$ & $99.90_{-19.83}^{+0.06}$ &  10.4\vspace{0.2cm}\\ 
 
B & N & -- & R & N & 0.5 & Isot. & 0.014 & $99.86_{-0.01}^{+0.01}$ & ${99.86_{-0.01}^{+0.01}}$ & $99.86_{-2.75}^{+0.01}$ & 3.05 \\ 
Bfb & N & -- & R & Y & 0.5 & Isot. & 0.014 & $99.98_{-0.01}^{+0.01}$ & $99.98_{-0.06}^{+0.02}$ & $99.93_{-1.0}^{+0.06}$ & 459.2 \\

 & N & -- & R & N & 0.0 & Isot. & 0.014 & $99.98_{-0.015}^{+0.01}$ & $99.98_{-0.20}^{+0.02}$ & $99.98_{-0.97}^{+0.01}$  & 27.9 \\
 & N & -- & R & Y & 0.0 & Isot. & 0.014 & $99.94_{-0.03}^{+0.02}$ & $99.95_{-0.06}^{+0.05}$ & $99.27_{-1.17}^{+0.54}$  & 488.7 \\
BJ & N & -- & R & N & 0.0 & Jeans & 0.014 & $\mathbf{\mathcolor{blue}{86.39_{-29.16}^{+13.59}}}$ & $\mathbf{\mathcolor{blue}{88.06_{-24.08}^{+11.92}}}$ & $99.98_{-0.13}^{+0.01}$  &  22.7 \\ 
BJfb & N & -- & R & Y & 0.0 & Jeans & 0.014 & $\mathbf{\mathcolor{blue}{83.58_{-12.69}^{+10.51}}}$ & $\mathbf{\mathcolor{blue}{88.63_{-10.59}^{+11.37}}}$ & $99.57_{-1.18}^{+0.38}$ & 743.0 \\
 
 & N & -- & R & N & 0.0 & Jeans & 0.017 & $91.29_{-27.46}^{+8.69}$ & $92.85_{-19.49}^{+7.12}$ & $99.98_{-0.01}^{+0.01}$ & 17.8 \\
 & N & -- & R & N & 0.0 & Jeans & 1.4$\times{}10^{-4}$ & $96.84_{-5.11}^{+3.02}$ & $98.99_{-4.24}^{+1.00}$ & $96.14_{-4.10}^{+3.69}$ & 252.1 \vspace{0.2cm}\\

C & Y & 1 & C & N & 0.5 & Isot. & 0.014 & $99.27_{-0.48}^{+0.53}$ & $99.97_{-0.07}^{+0.02}$ & $99.99_{-0.38}^{+0.01}$ & 47.2 \\
       & Y & 1 & C & Y & 0.5 & Isot & 0.014 &  $99.99_{-0.018}^{+0.01}$ & $99.99_{0.01}^{+0.01}$ & $99.97_{-0.48}^{+0.03}$ & 1267.7 \\  
 & N & -- & C & N & 0.5 & Isot & 0.014 & $99.98_{-0.01}^{+0.01}$ & $99.98_{-0.08}^{+0.01}$ & $99.98_{-0.18}^{+0.01}$ & 22.7 \\
 & N & -- & C & Y & 0.5 & Isot & 0.014 & $99.92_{-0.03}^{+0.05}$ & $99.88_{-0.14}^{+0.12}$ & $99.95_{-1.08}^{+0.05}$ & 1164.3 \vspace{0.2cm}\\

 D & Y & 1 & D & N & 0.5 & Isot. & 0.014 & $99.96_{-4.36}^{+0.01}$ & $99.96_{-0.01}^{+0.01}$ & $99.96_{-0.71}^{+0.01}$  & 11.0 \\
  & Y & 1 & D & Y & 0.5 & Isot. & 0.014 & $96.53_{-5.88}^{+3.11}$ & $99.76_{-0.32}^{+0.24}$ & $99.63_{-1.70}^{+0.34}$ & 382.5 \\
 & Y & 1 & D & N & 0.0 & Isot. & 0.014 & $99.96_{-0.44}^{+0.01}$ & $99.96_{-0.01}^{+0.01}$ & $99.88_{-0.28}^{+0.08}$ & 10.9 \\
 & Y & 1 & D & Y & 0.0 & Isot. & 0.014 & $97.77_{-11.38}^{+2.08}$ & $99.84_{-0.40}^{+0.16}$ & $99.77_{-1.15}^{+0.20}$ & 382.9 \\
 & Y & 1 & D & N & 0.0 & Jeans & 0.014 & $99.90_{-4.82}^{+0.06}$ & $99.96_{-0.01}^{+0.01}$ & $99.96_{-0.74}^{+0.01}$ &  10.8\\
 & Y & 1 & D & Y & 0.0 & Jeans & 0.014 & $96.54_{-8.32}^{+3.29}$ & $99.59_{-0.30}^{+0.41}$ & $99.92_{-0.14}^{+0.06}$ & 382.9 \\
        & Y & 1 & D & N & 1.0 & Isot & 0.014 & $99.96_{-1.94}^{+0.01}$ & $99.96_{-0.01}^{+0.01}$ & $99.96_{-0.75}^{+0.01}$ & 6.69 \\ 
        & Y & 1 & D & Y & 1.0 & Isot & 0.014 & $97.42_{-6.64}^{+2.20}$ & $99.74_{-0.25}^{+0.26}$ & $99.93_{-1.08}^{+0.06}$ & 383.0 \\
 & Y & 1 & D & N & 0.5 & Isot. & 0.017 & $99.75_{-7.88}^{+0.18}$ & $99.93_{-0.01}^{+0.01}$ & $99.93_{-0.01}^{+0.01}$ & 6.1 \\
 & Y & 1 & D & N & 0.5 & Isot & 1.4$\times{}10^{-4}$ & $99.96_{-2.81}^{+0.01}$ & $97.92_{-21.77}^{+2.04}$ & $97.02_{-23.28}^{+2.94}$ & 10.5 \vspace{0.2cm}\\
        
 & N & -- & D & N & 0.5 & Isot. & 0.014 & $99.87_{-0.015}^{+0.02}$ & $99.80_{-0.49}^{+0.19}$  & $99.68_{-0.62}^{+0.20}$  & 84.3\\
 & N & -- & D & Y & 0.5 & Isot. & 0.014 & $99.92_{-0.02}^{+0.01}$ & $99.90_{-0.05}^{+0.10}$ & $99.86_{-0.35}^{+0.05}$  & 472.0 \\
 & N & -- & D & N & 0.0 & Isot. & 0.014 & $99.94_{-0.04}^{+0.05}$ & $99.89_{-0.33}^{+0.093}$ & $99.85_{-0.54}^{+0.14}$ & 34.9 \\
 & N & -- & D & Y & 0.0 & Isot. & 0.014 & $99.89_{-0.09}^{+0.08}$ & $99.88_{-0.12}^{+0.12}$ & $99.83_{-0.19}^{+0.12}$ & 415.1 \\ 
 & N & -- & D & N & 0.0 & Jeans & 0.014 & $\mathbf{\textcolor{blue}{69.33_{-31.19}^{+30.55}}}$ & $92.74_{-25.44}^{7.23}$ & $99.98_{-0.06}^{+0.02}$ & 21.4 \\ 

 & N & -- & D & Y & 0.0 & Jeans & 0.014 & $\mathbf{\textcolor{blue}{80.27_{-8.75}^{+17.94}}}$ & $\mathbf{\mathcolor{blue}{88.19_{-17.31}^{+11.80}}}$ & $99.61_{-1.61}^{+0.31}$  & 533.6 \\
 & N & -- & D & N & 1.0 & Isot. & 0.014 & $99.86_{-0.06}^{+0.06}$ & $99.96_{-0.10}^{+0.04}$ & $99.93_{-0.15}^{+0.04}$ & 122.7\\ 
 & N & -- & D & Y & 1.0 & Isot. & 0.014 & $99.91_{-0.07}^{+0.02}$ & $99.91_{-0.12}^{+0.08}$ & $99.78_{-0.50}^{+0.13}$  & 527.1\\
 
 & N & -- & D & N & 0.5 & Isot. & 0.017 & $99.92_{-0.07}^{+0.03}$ & $99.62_{-0.95}^{+0.38}$ & $99.97_{-0.02}^{+0.02}$ & 73.3 \\ 
 
 & N & -- & D & N & 0.5 & Isot & 1.4$\times{}10^{-4}$ & $\mathbf{\textcolor{blue}{78.56_{-23.59}^{+20.36}}}$ & $94.60_{-19.68}^{+5.40}$ & $97.58_{-26.37}^{+2.41}$ &  70.8 \vspace{0.2cm}\\

\hline 

    \end{tabular}
\renewcommand{\arraystretch}{1.0}
   \label{tab:param}
    \vspace{0.0cm}\footnotesize{Column 1: model name (only the main models discussed in the text and in the Figures have a name for faster reference); column 2: Y (N) means that the $\alpha{}$ formalism for CE  evolution is (is not) activated; column 3: value of the CE $\alpha{}$ parameter; column 4: CCSN  model -- R for rapid, D for delayed, C for compactness criterion; column 5: the natal kick is modulated by fallback (Y) or not (N); column 6: accretion efficiency ($f_{\rm a}$); column 7: angular momentum transport mode (isotropic re-emission or Jeans mode); column 8: metallicity $Z$; column 8-10: percentile value of \Gaia{} BH1, BH2, and BH3 (median values and 68\% credible intervals); column 11: formation efficiency $\eta_{\rm dorm}$ of BH binaries with a 0.5--1.5 M$_\odot$ stellar companion (equation~\ref{eq:eff}).}
\end{table*}
\section{CE evolution and efficiency parameter}\label{app:alpha}

\begin{figure*}[h]
    \centering
    \includegraphics[width=0.8\linewidth]{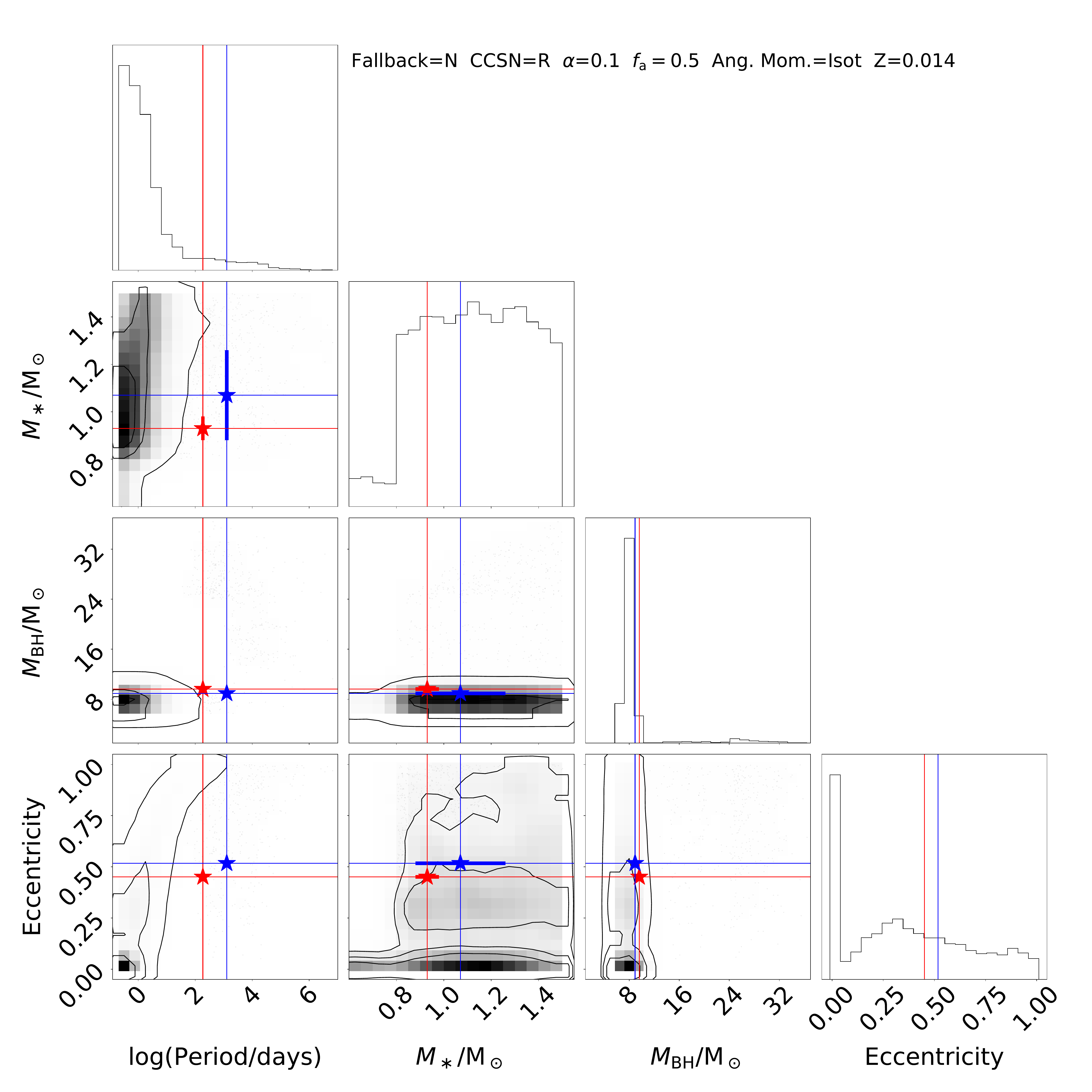}
    \caption{Corner plot showing the orbital period, eccentricity, BH mass ($M_\mathrm{BH}$) and companion star mass ($M_\ast$) in the simulations of model~A$_{\alpha{}01}$ with $\alpha{}=0.1$. The red and blue stars show the values for \Gaia{} BH1 and BH2.}
    \label{fig:modA01}
\end{figure*}
\begin{figure*}[h]
    \centering
    \includegraphics[width=0.8\linewidth]{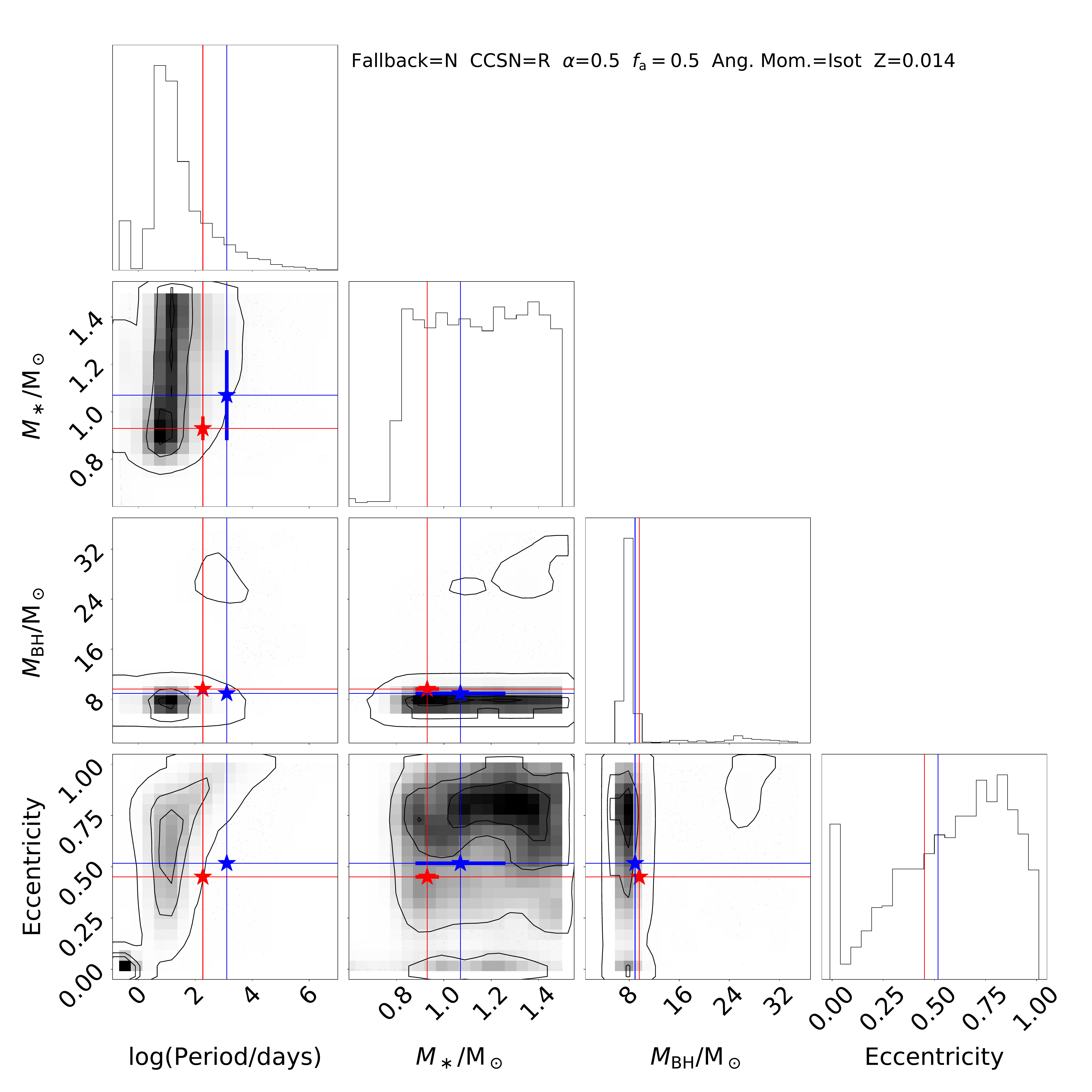}
    \caption{Same as Fig.~\ref{fig:modA01} but for model~A$_{\alpha{}05}$  with $\alpha{}=0.5$.}
    \label{fig:modA05}
\end{figure*}
\begin{figure*}[h]
    \centering
    \includegraphics[width=0.8\linewidth]{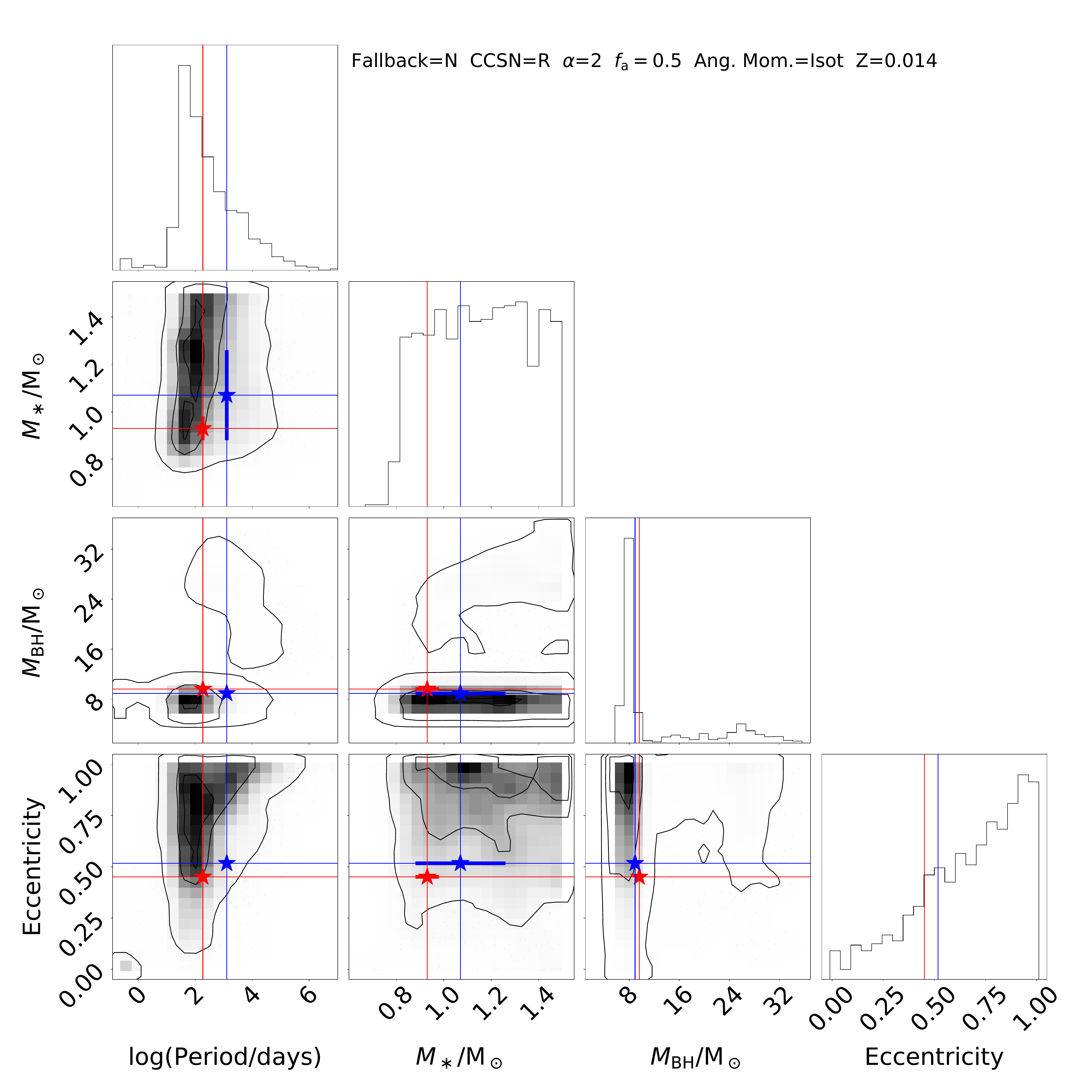}
    \caption{Same as Fig.~\ref{fig:modA01} but for model~A$_{\alpha{}2}$  with $\alpha{}=2$.}
    \label{fig:modA2}
\end{figure*}
\begin{figure*}[h]
    \centering
    \includegraphics[width=0.8\linewidth]{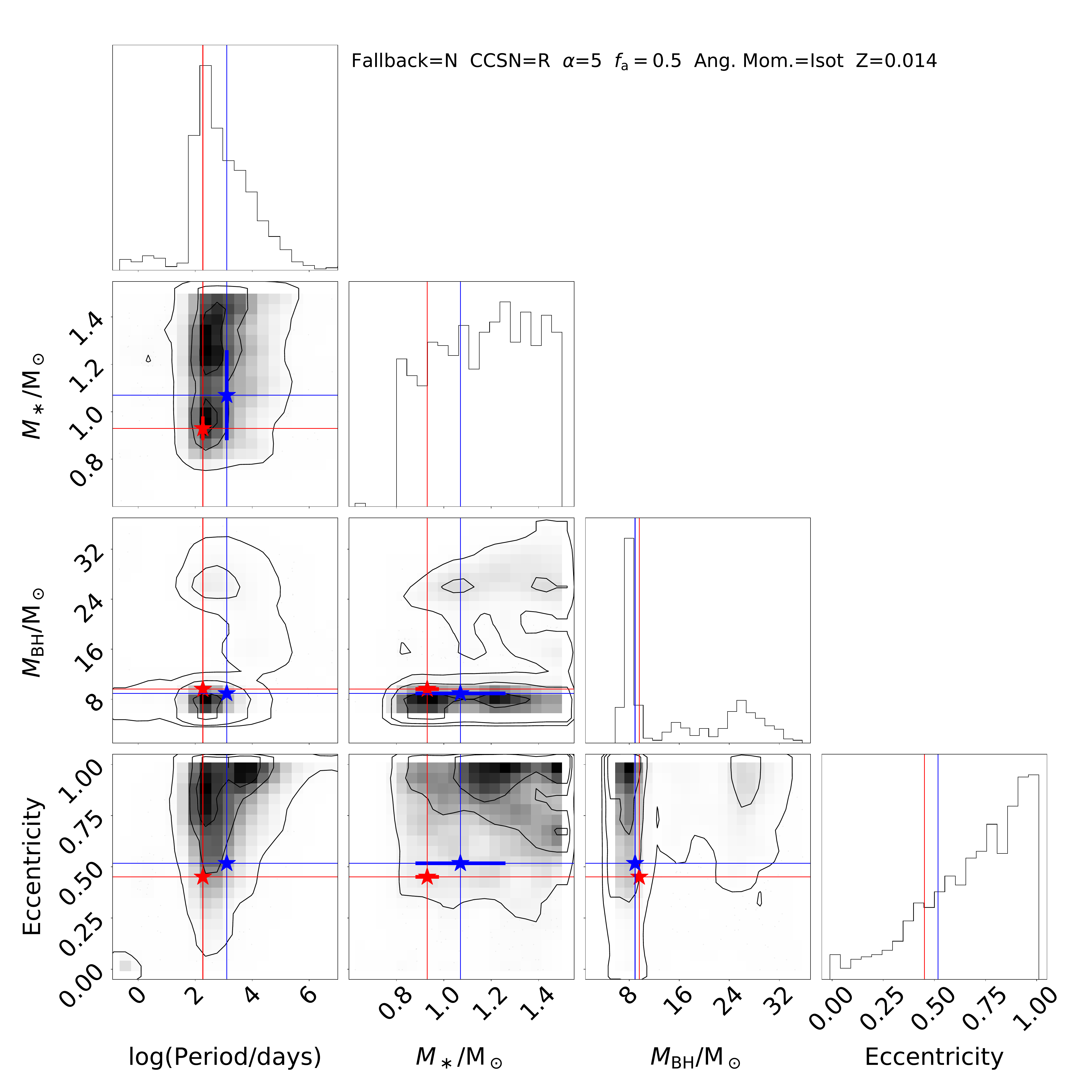}
    \caption{Same as Fig.~\ref{fig:modA01} but for model~A$_{\alpha{}5}$  with $\alpha{}=5$.}
    \label{fig:modA5}
\end{figure*}
\begin{figure*}[h]
    \centering
    \includegraphics[width=0.8\linewidth]{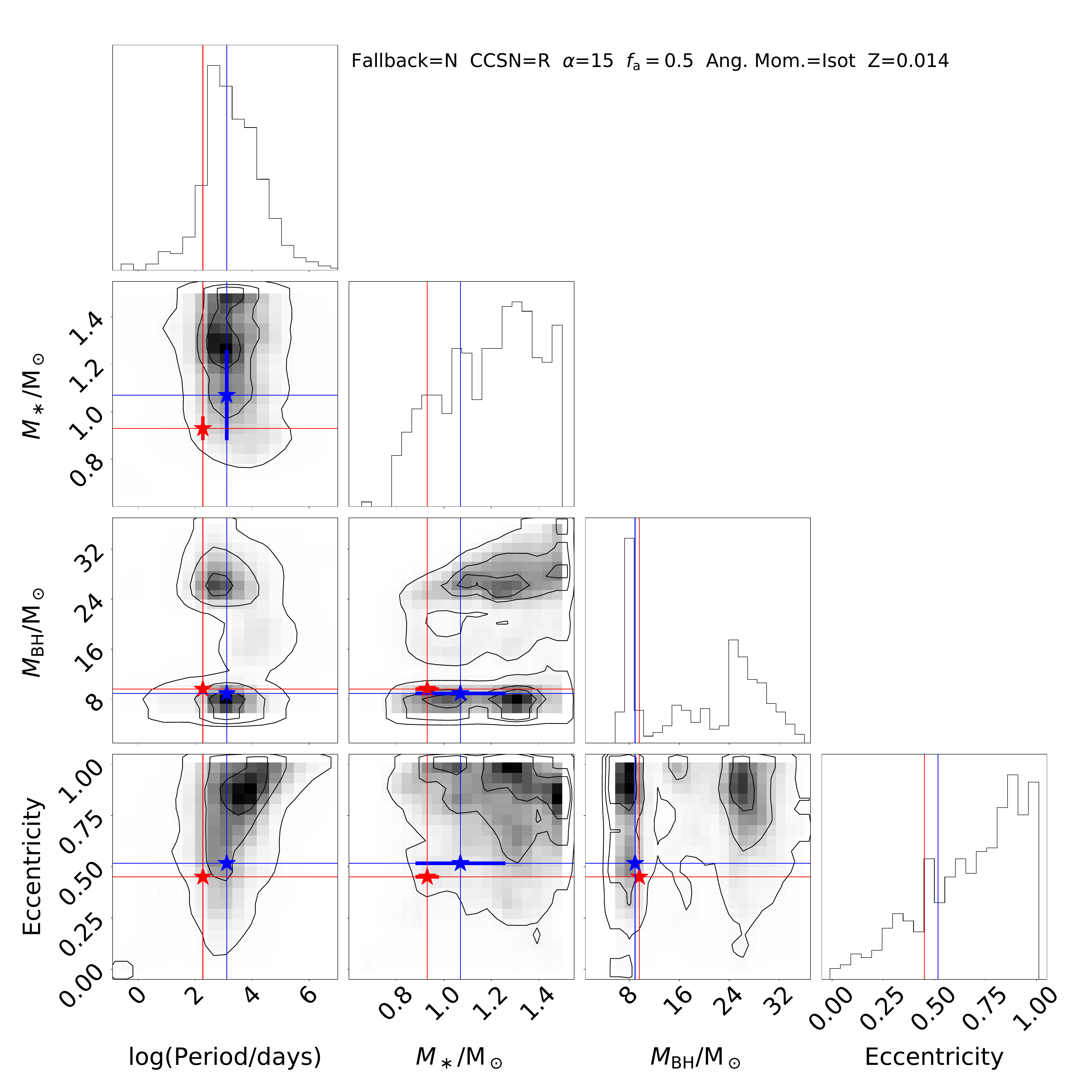}
    \caption{Same as Fig.~\ref{fig:modA01} but for model~A$_{\alpha{}15}$  with $\alpha{}=15$.}
    \label{fig:modA15}
\end{figure*}

{In the runs where CE evolution is allowed, we assume that mass transfer is unstable when the mass ratio $q$ between donor and accretor exceeds the list of critical values $q_c$ indicated in Table~\ref{tab:qc}.}

\begin{table}[htbp] 

    \caption{Summary of the adopted $q_c$ values.}
    \renewcommand{\arraystretch}{1.2}
    \begin{tabular}{lc}
    \hline   
        Evolutionary Stage of the Donor & $q_c$ \\
    \hline
    Main Sequence & 3.0 \\
    Hertzsprung Gap & 4.0 \\
    First Giant Branch & Eq.~57, \citet{hurley2002}\\
    Core He Burning & 3.0\\
        Asymptotic Giant Branch & Eq.~57, \citet{hurley2002}\\

    He Main Sequence & 3.0 \\
    He Hertzsprung Gap & 0.784\\
    He Giant Branch & 0.784 \\
    White Dwarf & 0.628\\
    Neutron Star/Black Hole & Merger \\
    \hline
\end{tabular}
\label{tab:qc}
\end{table}

\begin{table*}[htbp] 

    \caption{ Summary of the models with CE  parameter $\alpha{}\neq{}1$. The description of the columns is the same as in Table~\ref{tab:param}.}
    \renewcommand{\arraystretch}{1.2}
    \begin{tabular}{c cc cc cc c ccc c}
    \hline   
        Model & CE & $\alpha$ & CCSN & Fallback & $f_{\rm a}$ & Ang. Mom. & Z & Perc. BH1 & Perc. BH2 & Perc. BH3 & $\eta_{\rm dorm}$\\
        & && && && & &&& [$\times{}10^{-7}$ M$_\odot{}^{-1}$]\\
    \hline

A$_{\alpha{}01}$   & Y & 0.1 & R & N & 0.5 & Isot & 0.014  & $99.98_{-0.01}^{+0.01}$ & $99.98_{-0.01}^{+0.01}$ & $99.95_{-0.42}^{+0.03}$  & 22.1 \\
AJ$_{\alpha{}01}$   & Y & 0.1 & R & N & 0.0 & Jeans & 0.014  & $99.98_{-0.01}^{+0.01}$ & $99.98_{-0.01}^{+0.01}$ & $99.96_{-0.11}^{+0.02}$ & 21.2  
\vspace{0.1cm}\\
A$_{\alpha{}05}$   & Y & 0.5 & R & N & 0.5 & Isot & 0.014 & $99.96_{-0.18}^{+0.01}$ & $99.96_{-0.01}^{+0.01}$ & $99.91_{-0.60}^{+0.05}$ &  11.9 \\
AJ$_{\alpha{}05}$   & Y & 0.5 & R & N & 0.0 & Jeans & 0.014 & $99.96_{-0.01}^{+0.01}$ & $99.96_{-0.01}^{+0.01}$ & $99.96_{-0.45}^{+0.01}$  & 11.9  
\vspace{0.1cm}\\
A$_{\alpha{}2}$   & Y & 2 & R & N & 0.5 & Isot & 0.014 & $97.56_{-23.92}^{+2.37}$ & $99.93_{-0.01}^{+0.01}$ & $99.93_{-0.01}^{+0.01}$ & 5.8 \\
AJ$_{\alpha{}2}$   & Y & 2 & R & N & 0.0 & Jeans & 0.014 & $98.30_{-7.22}^{+1.55}$ & $99.93_{-0.01}^{+0.01}$ & $99.93_{-0.01}^{+0.01}$ & 5.8  
\vspace{0.1cm}\\
A$_{\alpha{}3}$   & Y & 3 & R & N & 0.5 & Isot & 0.014 & $97.14_{-14.65}^{+2.77}$ & $99.91_{-0.01}^{+0.01}$ & $99.91_{-0.47}^{+0.01}$ & 4.8 \\
AJ$_{\alpha{}3}$   & Y & 3 & R & N & 0.0 & Jeans & 0.014  & $90.77_{-26.51}^{+9.14}$ & $99.86_{-2.63}^{+0.05}$ & $99.91_{-0.66}^{+0.01}$ & 4.7  
\vspace{0.1cm}\\
A$_{\alpha{}5}$   & Y & 5 & R & N & 0.5 & Isot & 0.014  &  $96.26_{-17.51}^{+3.51}$ & $98.04_{-11.64}^{+1.84}$ & $99.87_{-1.39}^{+0.01}$ & 3.4\\
AJ$_{\alpha{}5}$   & Y & 5 & R & N & 0.0 & Jeans & 0.014 & $96.40_{-12.98}^{+3.48}$ & $99.70_{-13.87}^{+0.18}$ & $99.88_{-1.60}^{+0.01}$ &  3.6  
\vspace{0.1cm}\\
A$_{\alpha{}7}$   & Y & 7 & R & N & 0.5 & Isot & 0.014 & $99.55_{-14.03}^{+0.28}$ & $99.82_{-9.65}^{+0.0}$ & $99.82_{-1.80}^{+0.01}$  & 2.4\\
AJ$_{\alpha{}7}$   & Y & 7 & R & N & 0.0 & Jeans & 0.014 & $99.30_{-6.34}^{+0.53}$ & $99.82_{-0.70}^{+0.01}$ & $99.82_{-0.01}^{+0.01}$  &2.5 
\vspace{0.1cm}\\
A$_{\alpha{}10}$   & Y & 10 & R & N & 0.5 & Isot & 0.014 & $98.83_{-13.17}^{+1.01}$ & $95.64_{-28.81}^{+4.19}$ & $99.83_{-0.19}^{+0.01}$  & 2.6\\
AJ$_{\alpha{}10}$   & Y & 10 & R & N & 0.0 & Jeans & 0.014 & $99.19_{-6.51}^{+0.65}$ & $\mathbf{\mathcolor{blue}{78.18_{-37.78}^{+21.66}}}$ & $99.84_{-0.01}^{+0.01}$ & 2.6  
\vspace{0.1cm}\\
A$_{\alpha{}15}$   & Y & 15 & R & N & 0.5 & Isot & 0.014 & $99.80_{-0.40}^{+0.01}$ & $99.80_{-29.07}^{+0.01}$ & $99.80_{-0.23}^{+0.01}$ & 2.2\\
AJ$_{\alpha{}15}$   & Y & 15 & R & N & 0.0 & Jeans & 0.014 &  $99.81_{-2.72}^{+0.01}$ & $99.81_{-34.78}^{+0.01}$ & $99.81_{-0.01}^{+0.01}$ & 2.2  
\vspace{0.1cm}\\

\hline 

    \end{tabular}
\renewcommand{\arraystretch}{1.0}
   \label{tab:paramalpha}
\end{table*}

{In the main text, we assume that CE evolution is described with efficiency parameter $\alpha{}=1$. Here}, we show models with CE ejection efficiency parameter $\alpha{}\neq{}1$. Table~\ref{tab:paramalpha} summarizes the models we considered with $\alpha=0.1,$ 0.5, 2, 3, 5, 7, 10, and 15. Figures~\ref{fig:modA01}, \ref{fig:modA05}, \ref{fig:modA2}, 
\ref{fig:modA5}, and \ref{fig:modA15} show the main properties of models with $\alpha{}=0.1$, 0.5, 2, 
5, and 15. The figures clearly show that the orbital period distribution of dormant BHs shifts to large values with increasing $\alpha{}$. This happens because higher values of $\alpha{}$ imply that less orbital energy is requested to eject the CE; hence the orbital separation shrinks less. The same trend can be seen in both models with isotropic re-emission and $f_{\rm a}=0.5$ and models with Jeans mode and $f_{\rm a}=0.0$. 

Overall, the properties of Gaia~BH1 can be matched by models with $\alpha{}\in[1,7]$, as  shown by the Figures and  by the percentile analysis in Table~\ref{tab:paramalpha}.

Table~\ref{tab:paramalpha} also shows that the efficiency of compact-object binary formation is maximum for small values of $\alpha{}$ and minimum for larger values. This happens because small values of $\alpha{}$ shrink  more efficiently the orbit during CE evolution, thus allowing more systems to remain bound during the CCSN explosion. Some of the systems with the shorter orbital separations (less than a few days) might be observable as X-ray binary systems, but here we do not introduce any emission models to distinguish dormant and X-ray active systems. We will investigate this aspect in a follow-up work. Systems matching the properties of \Gaia{} BH1, BH2, and BH3 are not Roche-lobe filling, thus they do not require that we make a detailed distinction between dormant and X-ray active systems for the purpose of this work.

\section{Thermal eccentricity distribution}\label{app:ecc}

\begin{figure*}
    \centering
    \includegraphics[width=0.8\linewidth]{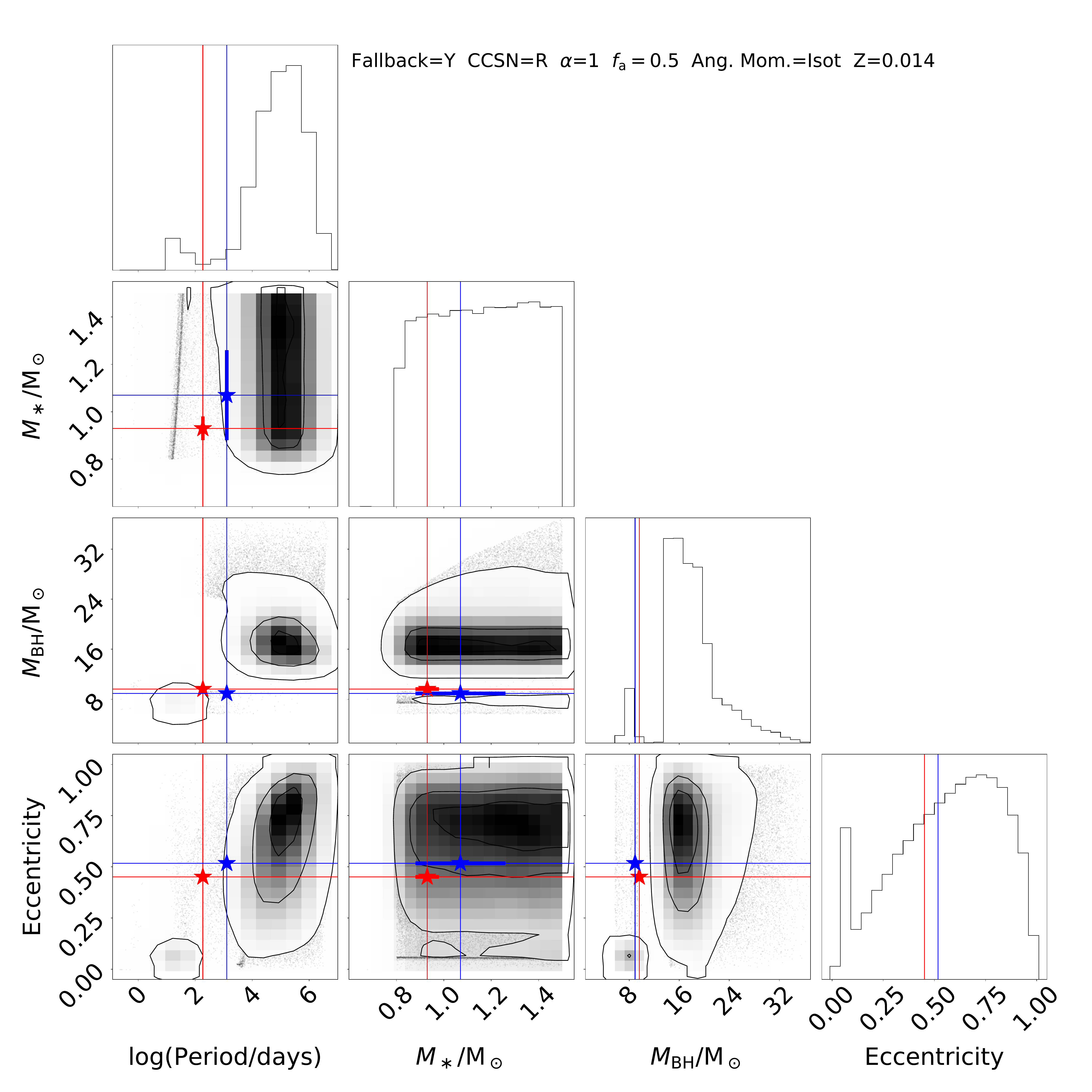}
    \caption{ Same as Fig.~\ref{fig:modAfb} but assuming a thermal distribution $\mathcal{F}(e)\propto{}e$ for the initial orbital eccentricity.}
    \label{fig:modAfb_thermal}
\end{figure*}

\begin{figure*}
    \centering
    \includegraphics[width=0.8\linewidth]{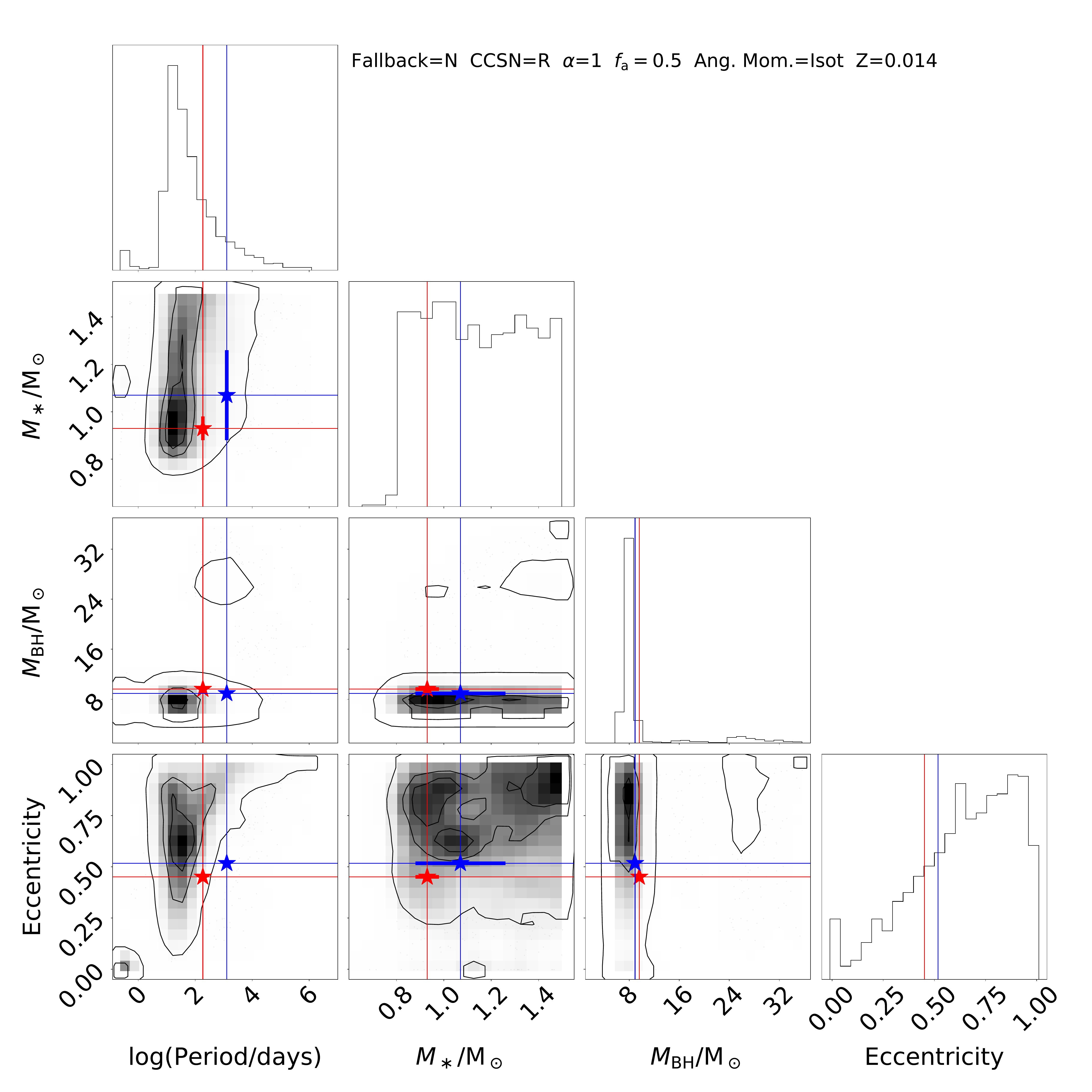}
    \caption{ Same as  Fig.~\ref{fig:modA},  but assuming a thermal distribution $\mathcal{F}(e)\propto{}e$ for the initial orbital eccentricity. }
    \label{fig:modA_thermal}
\end{figure*}

\begin{table*}[htbp] 

    \caption{ Summary of the models that assume a thermal distribution for the initial orbital eccentricity. The description of the columns is the same as in Table~\ref{tab:param}.}
    \renewcommand{\arraystretch}{1.2}
    \begin{tabular}{c cc cc cc c ccc c}
    \hline   
        Model & CE & $\alpha$ & CCSN & Fallback & $f_{\rm a}$ & Ang. Mom. & Z & Perc. BH1 & Perc. BH2 & Perc. BH3 & $\eta_{\rm dorm}$\\
        & && && && & &&& [$\times{}10^{-7}$ M$_\odot{}^{-1}$]\\
    \hline
A\_th   & Y & 1 & R & N & 0.5 & Isot & 0.014  & $99.94_{-0.26}^{+0.01}$ & $99.94_{-0.01}^{+0.01}$  & $99.94_{-1.10}^{+0.01}$ & 6.8  \\
Afb\_th  & Y & 1 & R & Y & 0.5 & Isot & 0.014  & $91.67_{-2.57}^{+5.34}$ & $99.20_{-0.40}^{+0.80}$ & $99.71_{-0.97}^{+0.27}$ & 358.5 \\
       & Y & 1 & R & N & 0.0 & Jeans & 0.014  & $99.94_{-0.26}^{+0.01}$ & $99.94_{-0.01}^{+0.01}$  & $99.94_{-0.01}^{+0.01}$ & 6.7 \\
       & Y & 1 & R & Y & 0.0 & Jeans & 0.014  & $95.16_{-12.41}^{+3.52}$ & $99.83_{-0.44}^{+0.17}$ & $99.73_{-2.02}^{+0.26}$ & 358.5 \\
\hline
\end{tabular}\label{tab:paramecc}
\end{table*}

{In this Section, we report the results of a subset of runs performed assuming a thermal distribution $\mathcal{F}(e)\propto{}e$ for the initial orbital eccentricities. Models in which the natal kicks are modulated by fallback show two main differences. Firstly and most importantly, the long orbital period systems that survive bound until they become \textit{Gaia} BH-like systems tend to have higher eccentricities (Fig.~\ref{fig:modAfb_thermal}). Indeed, these systems do not undergo Roche-lobe overflow and their final eccentricities retain memory of the initial distribution. Secondly, a slightly larger number of systems are disrupted by natal kicks, because of the larger eccentricity. The difference is by a factor of $\approx{1.3}$ and it is visible by comparing the formation efficiency $\eta_{\rm dorm}$ in Table~\ref{tab:paramecc} with the values in Table~\ref{tab:param}.}

{In contrast, when DM25 natal kicks are assumed (Fig.~\ref{fig:modA_thermal}), there is negligible difference between a thermal and a  \protect{\citet{sana2012}} initial eccentricity distribution. This is because the long orbital period systems are disrupted by natal kicks, whereas the short orbital period systems interact via Roche-lobe overflow and circularize before the natal kick: the final evolution of these systems is completely determined by mass transfer and natal kicks.}

\section{Mock observations}\label{app:pdet}

To construct the mock observations, we matched each simulated dormant BH binary to a synthetic Milky Way stellar population generated with the python package \texttt{cogsworth} \citep{wagg2025}. Each binary was assigned to a randomly sampled star with the same luminous companion mass, adopting its 3D Galactocentric position. We generated the \texttt{cogsworth} population using the default Milky Way parameters and solar metallicity. We sampled only single stars, as we have already simulated binary interactions with \sevn{}.
To improve computational efficiency, we proceeded in two steps. First, we estimated the probability that a system lies within an effective search volume of $10\,\mathrm{kpc}$, beyond which the astrometric signal is typically too small for detection in \Gaia{} DR3. We find that $\sim60\%$ of the systems fall within this volume. Second, we evaluated the probability of detection for systems within this volume.

Within the $10\,\mathrm{kpc}$ volume, we resampled stars from the \texttt{cogsworth} population and matched them to the simulated BH binaries by companion mass. We generated 50 realizations per system to increase the statistical robustness. For each realization, we computed observable quantities (sky position, proper motion and distance) with \texttt{cogsworth} and simulated extinction and apparent magnitudes in the \Gaia{} $G$ band using the Bayestar dust map \citep{bayestar2019}. We randomly sampled orbital orientations and phases, while intrinsic binary parameters (BH mass, orbital period, and eccentricity) remained fixed to the values simulated with \sevn{}.

We then processed the mock binaries with the \texttt{gaiamock} pipeline \citep{elbadry2024}, which simulates \Gaia{} epoch astrometry based on the scanning law and empirically calibrated uncertainties. \texttt{gaiamock} fits the resulting data with the full astrometric fitting cascade, assigning single-star, acceleration, or orbital solutions following the DR3 data model. We classify systems as detected if they meet the DR3 non-single star quality cuts \citep{Halbwachs+2023} and would have most likely received a 12-parameter orbital solution. We 
calculated the average detection probability across the population and obtained $p_\mathrm{det}={ 2.9}\times10^{-4}$ and ${ 8.0}\times10^{-5}$ for models \modA{} and 
\modAfb{}, respectively (including the factor of 0.6 for the effective search volume).

\end{appendix}

\end{document}